\documentclass[10pt,journal,compsoc]{IEEEtran}
\usepackage{tikz}
\usetikzlibrary{arrows,automata}
\usepackage{url}
\usepackage{graphicx}
\usepackage{algorithm}
\usepackage{algorithmic}
\usepackage{amsfonts}
\usepackage{balance}
\usepackage{amsmath}
\usepackage{amsthm}
\theoremstyle{definition}
\newtheorem{definition}{Definition}[section]
\newtheorem{theorem}{Theorem}
\newtheorem{corollary}{Corollary}[theorem]
\newtheorem{lemma}[theorem]{Lemma}

\usepackage{amssymb}
\usepackage{multirow}
\usepackage{color}
\newcommand{\enext}[1]{\textbf{EX}{#1}}
\newcommand{\euntil}[2]{\mathit{\textbf{E(}{#1}\textbf{\ U\ }{#2}\textbf{)}}}
\newcommand{\eglobally}[1]{\mathit{\textbf{EG}{#1}}}

\newcommand{\auntil}[2]{\mathit{\textbf{A(}{#1}\textbf{U}{#2}\textbf{)}}}

\newcommand{\pre}[2]{\mathit{pre({#1},{#2})}}
\newcommand{\post}[2]{\mathit{post({#1},{#2})}}
\newcommand{\postStar}[2]{\mathit{post^*({#1},{#2})}}
\newcommand{\preStar}[2]{\mathit{pre^*({#1},{#2})}}
\newcommand{\preStarName}{\mathit{pre^*}}
\newcommand{\postStarName}{\mathit{post^*}}

\newcommand{\prek}[2]{\mathit{pre^k({#1},{#2})_{(k)}}}

\newcommand{\refineSystem}[2]{\mathit{refineSystem({#1},{#2})}}
\newcommand{\sat}[3]{\mathit{SAT({#1},{#2},{#3})}}

\newcommand{\transition}[4]{{#1}\xrightarrow{{#2},{#3}}{#4}}
\newcommand{\traces}[2]{\mathit{traces({#1},{#2})}}
\newcommand{\tracesn}[3]{\mathit{traces({#1},{#2},{#3})}}

\newcommand{\phiinit}{\phi^{\mathit{init}}}

\newcommand{\ctl}{\mathit{CTL}}

\newcommand{\eg}{\mathit{\textbf{EG}}}
\newcommand{\phireach}{\ensuremath{\phi^\mathit{reach}}}
\newcommand{\approximation}{\mathit{approx}}
\newcommand{\computePreStarName}{\mathit{preStar}}
\newcommand{\prekName}{\mathit{compute\_pre^k}}

\newcommand{\computePreStar}[3]{\ensuremath{\computePreStarName({#1},{#2},{#3})}}

\newcommand{\precise}{\mathit{precise}}
\newcommand{\under}{\mathit{under}}
\newcommand{\overapprox}{\mathit{over}}

\newcommand{\join}{\sqcup}

\newcommand{\checkflatteningName}{\mathit{isTraceFlattening}}
\newcommand{\checkflattening}[3]{\checkflatteningName({#1},{#2},{#3})}
\newcommand{\tru}{\mathit{true}}
\newcommand{\fls}{\mathit{false}}

\newcommand{\computeUntil}[4]{\computeUntilName({#1},{#2},{#3},{#4})}
\newcommand{\computeUntilName}{\mathit{computeUntil}}
\newcommand{\computeGlobal}[3]{\computeGlobalName({#1},{#2},{#3})}
\newcommand{\computeGlobalName}{\mathit{computeGlobal}}

\newcommand{\computeGlobalOverNew}[3]{\computeGlobalOverName({#1},{#2},{#3})}

\newcommand{\computeGlobalOverName}{\mathit{computeGlobalOver}}
\newcommand{\latticeApprox}{\mathit{Approx}}

\newcommand{\enumlabel}{\mathit{label}}
\newcommand{\timeout}{(TO)}
 \newcommand{\order}[2]{\mathit{order}{_{#1}}{#2}}
 
 \newcommand{\Yinit}{Y_{\mathit{init}}}

\newcommand{\qstart}{q^\mathit{start}} 
\newcommand{\phiteqchk}{\phi_\mathit{check}}                                    

\newcommand{\egloballyName}{\mathit{\textbf{EG}}}

\newcommand{\TODO}[1]{\textbf{TODO {#1}}}
\newcommand{\growone}{\mathit{grow}_1}
\newcommand{\growtwo}{\mathit{grow}_2}
\newcommand{\growtwoa}{\mathit{grow}_{2a}}
\newcommand{\growtwob}{\mathit{grow}_{2b}}
\newcommand{\asoY}{\phi_\mathit{atmost\_one\_succ\_outside\_Y}}
\newcommand{\inval}{\mathit{invalid}}
\newcommand{\vald}{\mathit{valid}}

\newcommand{\dirt}{\mathit{dirty}}
\newcommand{\extend}[1]{\mathit{extend}{(#1)}}

\usepackage{subfigure}
\usepackage{booktabs}
\usepackage{array}
\usetikzlibrary{arrows,automata,fit}

\renewcommand{\vec}[1]{\mathbf{#1}}
\begin{document}
 \title{Checking Temporal Properties of Presburger Counter Systems using
 Reachability Analysis}
  %
 \author{
     \IEEEauthorblockN{Aravind Acharya}\\
     \IEEEauthorblockA{Department of Computer Science and Automation\\ Indian
     Institute of Science\\ Bangalore, India 560012\\ Email:
 aravind@iisc.ac.in}\\
 \and
     \IEEEauthorblockN{K Vasanta Lakshmi}\\
     \IEEEauthorblockA{Department of Computer Science and Automation\\ Indian
     Institute of Science\\ Bangalore, India 560012\\ Email:
 kvasanta@csa.iisc.ernet.in}\\
 \and
     \IEEEauthorblockN{Raghavan Komondoor}\\
     \IEEEauthorblockA{Department of Computer Science and Automation\\ Indian
     Institute of Science\\ Bangalore, India 560012\\ Email:
 raghavan@iisc.ac.in}
 }
 \IEEEtitleabstractindextext{%
     \begin{abstract}
         Counter systems are a well-known and powerful modeling notation for
specifying infinite-state systems.  In this paper we target the problem of
checking temporal properties of counter systems. 
Given a counter system, and a CTL temporal property, our approach computes
a formula that encodes
the set of concrete states of the counter system that satisfy the given
property.  A novel aspect of our approach is that it uses reachability
analysis techniques, which are well studied in the literature, as a black
box. One advantage of this approach is that it is  able to compute precise answers on a wider class
of systems than previous approaches for the same problem.  Secondly, it
computes results by iterative expansion and contraction, and hence
permits an over-approximated or under-approximated solution to be obtained at any point.  We state the
formal properties of our framework, and also provide experimental results
using standard benchmarks to show the usefulness of our framework.

     \end{abstract}
     \begin{IEEEkeywords}
         Counter Systems, CTL properties, Accelerations, Reachability,
     \end{IEEEkeywords}
 }
\maketitle
\section{Introduction}
\label{sec:intro}

\emph{Counter systems} are a class of
infinite state systems that are
equivalent to simple looping programs that use integer variables, 
without arrays and pointers.
A counter system has a finite set of \emph{control states} and a finite set of \emph{counters}, with each
counter taking values from the infinite domain of integers. There are transitions between control states, with each transition being \emph{guarded} by a predicate on the counters, and having an \emph{action}, which indicates the updated values of the counters in terms of the old values.  \emph{Presburger} logic is the decidable first-order theory of natural numbers. Presburger formulas use variables, constants, addition and subtraction, comparisons, and quantification. 
The class of counter systems where the guards as well as actions are
represented using Presburger formulas are called Presburger counter systems.
Presburger counter systems have been shown to be applicable in various settings~\cite{fast}, such as the analysis of the
TTP protocol, different broadcast protocols, as well as cache coherence protocols. In the rest of this paper we use ``counter system'' or even just ``system'' to refer to Presburger counter systems. 

Verification of properties of counter systems has been an important topic in the research literature. While problems such as reachability analysis and temporal property checking are decidable for infinite systems such as pushdown systems and petri-nets~\cite{Bouajjani97,esparza1998decidability}, these problems are in general undecidable on counter systems because of their greater expressive power.  This said, various interesting subclasses of counter systems have been identified on which reachability analysis is decidable~\cite{Com1998,Ibarra02,finkel02,trex_fast_accelerations05,iteratingOctagons}.
When it comes to \emph{CTL}~\cite{book_clarke} temporal property checking, researchers have shown
decidability of this problem on significantly narrower classes~\cite{dem2006,gaporder12}.  In this paper, we try to bridge this gap to an extent by proposing a CTL property-checking approach that uses  (black-box) reachability queries as subroutines.  It will become clear later in this paper that we succeed in this; at the same time, it is noteworthy that there still exist systems on which reachability analyses terminate but our approach does not.

\subsection{Problem statement}

\begin{figure}
\centering
	\makebox[1.5cm]{
		\subfigure[]{\begin{tikzpicture}[->,>=stealth',shorten >=1pt,auto,node distance=2.8cm,
    semithick,/tikz/initial text=, transform shape, scale =0.8]
    \tikzstyle{every state}=[fill=white,draw=black,text=black]
    \node[state, initial, initial text={\small$x=0$}] (A) at (0,0)  {$q_0$};

    \path (A) edge [loop] node [above,text width=2.7cm, align=left] {\small$x \geq 0 \wedge x<100$/ \\ $x^\prime = x+1$}  node [below] {$t_0$}(A);
    \path (A) edge [out=-40, in =-130,loop] node [below,text width=2.7cm, align=left] {\small$x > 0 \wedge x<5$/ \\ $x^\prime = x-1$}  node [above] {$t_1$}(A);
%
\end{tikzpicture}
%
}
	}\hspace{0.2cm}
	\makebox[3.5cm]{
		\subfigure[]{\begin{tikzpicture}[->,>=stealth',shorten >=1pt,auto,node distance=2.8cm,
    semithick,/tikz/initial text=, transform shape, scale =0.8]
    \tikzstyle{every state}=[fill=white,draw=black,text=black]
    \node[state, initial, initial text={\small$x=0$}] (A) at (0,0)  {$q_0$};

    \path (A) edge [loop] node [above,text width=3.7cm, align=center] {\small$x \geq 0 \wedge x<100\wedge x<10$/ \\ $x^\prime = x+1$}  node [below] {$t_0$}(A);
    \path (A) edge [out=-40, in =-130,loop] node [below,text width=3.7cm, align=center] {\small$x > 0 \wedge x<5 \wedge x<10$/ \\ $x^\prime = x-1$}  node [above] {$t_1$}(A);
%
\end{tikzpicture}
%
}
 	}
	\makebox[3.5cm]{
		\subfigure[]{\begin{tikzpicture}[->,>=stealth',shorten >=1pt,auto,node distance=2.8cm,
    semithick,/tikz/initial text=, transform shape, scale =0.8]
    \tikzstyle{every state}=[fill=white,draw=black,text=black]
    \node[state, initial, initial text={\small$x=0$}] (A) at (0,0)  {$q_0$};

    \path (A) edge [loop] node [above,text width=3.7cm, align=center] {\small$x \geq 0 \wedge x<100\wedge x<10$/ \\ $x^\prime = x+1$}  node [below] {$t_0$}(A);
%
\end{tikzpicture}
%
}
 	}
	\makebox[3.5cm]{
		\subfigure[]{\begin{tikzpicture}[->,>=stealth',shorten >=1pt,auto,node distance=2.8cm,
    semithick,/tikz/initial text=, transform shape, scale =0.8]
    \tikzstyle{every state}=[fill=white,draw=black,text=black]
    \node[state, initial, initial text={\small$x=0$}] (A) at (0,0)  {$q_0$};

    \path (A) edge [out=-40, in =-130,loop] node [below,text width=3.7cm, align=center] {\small$x > 0 \wedge x<5 \wedge x<10$/ \\ $x^\prime = x-1$}  node [above] {$t_1$}(A);
%
\end{tikzpicture}
%
}
 	}
	\makebox[4.5cm]{
		\subfigure[]{\begin{tikzpicture}[->,>=stealth',shorten >=1pt,auto,node distance=2.8cm,
    semithick,/tikz/initial text=, transform shape, scale =0.8]
    \tikzstyle{every state}=[fill=white,draw=black,text=black]
    \node[state, initial, initial text={\small$x=0$}] (A) at (0,0)  {$q_0$};
    \node[state, initial, initial text={\small$x=0$}] (B) at (3,0)  {$q_0$};

    \path (A) edge [out=40, in=140] node [above,text width=3.7cm, align=center] {\small$x \geq 0 \wedge x<100\wedge x<10$/ \\ $x^\prime = x+1$}  node [below]{$t_{01}$}(B);
    \path (B) edge [out=-140, in =-40] node [below,text width=3.7cm, align=center] {\small$x > 0 \wedge x<5 \wedge x<10$/ \\ $x^\prime = x-1$}  node [above] {$t_{11}$}(A);
%
\end{tikzpicture}
%
}
 	}
	\caption{(a) A counter system $M$. \quad(b) Refinement $M_1$ of $M$
		w.r.t $(x<10)$.
        (c)-(e) Different flattenings of $M_1$. Each transition $t_{ij}$ in
		a flattening has the same guard and action as transition $t_i$ of $M_1$.}
  	\label{fig:example_intro}
\end{figure}
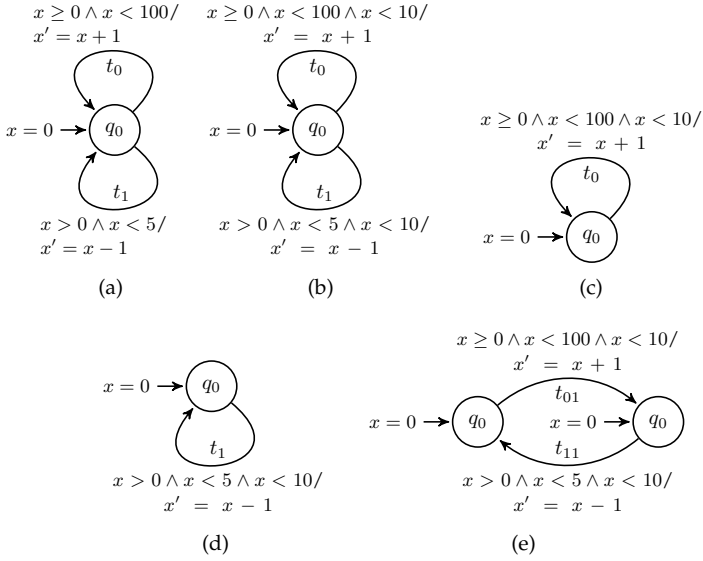

\subsubsection{Counter systems.}
Figure~\ref{fig:example_intro}(a) shows a simple Presburger counter system that we will use as a running example throughout this paper. It has a single control state, $q_0$, a single counter, $x$, and two transitions $t_0$ and $t_1$. The guard of each transition is the part that is depicted before the `/', while the action is the part that is depicted after the `/'. In this example, transition $t_0$ can trigger whenever $x$ has a value between 0 and 99 (both extremes inclusive), and increments the value of $x$. Similarly, transition $t_1$ can trigger whenever $x$ has a value between 1 and 4 (both extremes inclusive), and decrements the value of $x$. A \emph{state} of the system is a tuple consisting of a control state and a valuation to the counters. In our example system, each value for $x$ (i.e., each integer) is a state. Clearly, any system induces a state-transition relation among the set of all states.
Notice that our example system is non-deterministic; e.g., from $x=1$ one can transition via $t_0$ to $x=2$ or via $t_1$ to $x=0$. The initial state of the example system is $x=0$ (indicated using the incoming arrow). We define a \emph{trace} $t$ of a system $M$ as a (finite or infinite) sequence of states such that each state in $t$ is a predecessor of the next state in $t$ as per the state-transition relation induced by $M$.

\subsubsection{CTL in existential normal form.}
In this paper we focus on CTL properties expressed in \emph{existential normal form} (ENF). In this normal form, any  CTL property $\psi$ is either a simple Presburger formula $\phi$ (e.g., $x>0$), 
or is of the form $\enext{\psi_1}$, $\euntil{\psi_1}{\psi_2}$, or $\eglobally{\psi_1}$, or is a conjunction or disjunction of two properties, or is a negation of a property.

The solution of any CTL property $\psi$ wrt a given system $M$ is a Presburger formula that encodes the set of states of $M$ that satisfy the property $\psi$.  A state $\vec{s}$ is said to satisfy a given property $\psi$ in a system $M$ as per the following definition:

\begin{itemize}
\item $\vec{s}$ satisfies $\enext{\psi_1}$ in $M$ if some successor state of $\vec{s}$ in the state-transition relation of $M$ satisfies $\psi_1$.

\item $\vec{s}$ satisfies $\euntil{\psi_1}{\psi_2}$ in $M$ if $\vec{s}$ satisfies $\psi_2$, or there  exists a trace $t$ of $M$ starting from $\vec{s}$ such that (a)  $t$ reaches some state $\vec{s_2}$ that satisfies $\psi_2$, and (b) all states in $t$ until $\vec{s_2}$ (including $\vec{s}$) satisfy $\psi_1$. 

\item $\vec{s}$ satisfies $\eglobally{\psi_1}$ in $M$ if there exists an infinitely long trace $t$ of $M$ starting from $\vec{s}$ such that each state in $t$ satisfies $\psi_1$.

\item $\vec{s}$ satisfies $\psi_1 \vee \psi_2$, $\psi_1 \wedge \psi_2$, or $\neg \psi_1$ as per the usual semantics.
\end{itemize}

In our example system, the solution to $\enext{(x=1)}$ is $x=0 \vee x=2$,  the solution to $\euntil{x\geq 0}{x=0}$ is $x\geq 0 \wedge x < 5$, while the solution to $\eglobally{(x \geq 2)}$ is $x\geq 2 \wedge x < 5$. 

The problem that we address in this paper is to find the solution of a given CTL property $\psi$ wrt a given counter system $M$.
In the sections below we provide an informal overview of our approach. In this introductory section we focus only on \emph{non-nested} properties. These are properties of the form $\enext{\phi_1}$, $\euntil{\phi_1}{\phi_2}$, or $\eglobally{\phi_1}$, where the $\phi_i$'s are simple Presburger formulas. We describe how we extend our approach to solve for general nested properties in Section~\ref{sec:algo}.

\subsection{``Next'' (\textbf{EX}) and ``Until'' (\textbf{EU}) properties}

Solving for a property $\enext{\phi}$ on a system $M$ is straightforward. This just involves finding the (immediate) predecessors of states in $\phi$ as the state-transition relation induced by $M$.
It is always decidable to compute set of predecessor states when guards 
and actions are Presburger formulas. This follows from decidability of 
Presburger arithmetic.

Solving for $\euntil{\phi_1}{\phi_2}$ is more interesting. The first step here is to conjunct the guard of each transition in $M$ with $\phi_1$. We call this operation ``refinement'' of $M$ wrt $\phi_1$. Let the resulting counter system be called $M_1$.  Effectively, the state-transition relation induced by $M_1$ is the same as the one induced by $M$, except that all pairs of the form $(\vec{s_1}, \vec{s_2})$ are pruned away where $\vec{s_1}$ does \emph{not} satisfy $\phi_1$.  We then identify all states from which states in $\phi_2$ can be reached in the state-transition relation induced by $M_1$. We do this using a reachability subroutine as a black-box. It is easy to see that the states thus identified are exactly the ones that satisfy $\euntil{\phi_1}{\phi_2}$ in $M$.

\subsection{``Globally'' (\textbf{EG}) properties}
\label{ssec:intro:globally}

The $\eglobally{\phi_1}$ property is the most interesting one. Our primary contribution in is proposing an approach to solve for this kind of property. The closest related work to our work is the approach of Demri et al.~\cite{dem2006}. Their approach addresses the class of \emph{flat} and \emph{trace flattable} systems. A flat counter system is one in which no two distinct cycles share a common control-state. A trace-flattable counter system $M$ is one such that there exists a flat system $M'$ such that every trace in $M$ is present in $M'$ and vice versa. In this approach a Presburger formula is constructed that encodes all traces in the system, and another Presburger formula is constructed that represents the given temporal property. Finally, it is checked whether the first formula implies the second. 
This approach can model check complete class of $\ctl$ properties for 
flat counter systems. But in case of trace-flattable systems their approach can
model check only non-nested $\ctl$ properties.

Intuitively, our approach consists of two simultaneously executing iterative components. One component, which we call the ``under-approximation'' component, computes a growing under-approximation of the set of states that satisfy $\eglobally{\phi_1}$. The other component, which we call the ``over-approximation'' component, computes a growing under-approximation of the set of states that \emph{do not} satisfy $\eglobally{\phi}$ in $M$.  Both components work on the system $M_1$ that is obtained by refining $M$ wrt $\phi_1$. It is easy to see that the set of states that satisfy $\eglobally{\phi_1}$ in $M_1$ is exactly the same as the set of states that satisfy $\eglobally{\phi_1}$ in $M$. 
When either one of the two components terminates 
our approach yields a precise result. There exists systems on which neither of
the two components terminate but our approach that combines these two components
terminates. We discuss such systems in detail in Section~\ref{sssec:combined-does-better}.
In cases where our approach does not terminate either component can be 
terminated at any time to result in an approximate solution in the 
corresponding direction. 

\TODO{In later sections we need to sync the ``variant'' terminology with the ``component'' terminology used above.}

\subsubsection{The under-approximation component.}

This component works by generating ``flattenings'' of the refined system $M_1$. A flattening is intuitively an unrolled flat version of the system $M_1$, in which a single control-state may be duplicated more than once.  For an illustration, consider the system $M$ in Figure~\ref{fig:example_intro}(a), and the property $\eglobally{(x < 10)}$. Part~(b) of the figure shows the refinement of $M$ wrt the formula $x < 10$. Part~(c)-(e) show three different flattenings of $M_1$. Any system with one or more cycles has an infinite number of flattenings. It is guaranteed that each flattening of a system exhibits a subset of traces as the original system. Also, if the original system is not trace-flattable, then by definition there will exist no flattening (or set of flattenings) that exhibit all the traces that are exhibited by  the original system.

A state $\vec{s}$ satisfies $\eglobally{\phi_1}$ in a flattening if there is at least one infinitely long trace from this state in this flattening.
It is relatively straightforward to obtain the set of such states from any flattening.  For instance, this could be done using the approach of Demri et al. (although we propose our own simpler technique for this). This set, in general, will be a subset of the set of all states that satisfy $\eglobally{\phi_1}$ in $M_1$ (because the flattening exhibits a subset of traces that $M_1$ exhibits). Therefore, we iteratively generate different flattenings of $M_1$, in a systematic manner (e.g., in increasing order of number of transitions contained). From each flattening we determine the set of states that satisfy $\eglobally{\phi_1}$ in it, and add these states to a set of states $X$ that we maintain. This set is the growing under-approximation that we referred to earlier. It is noteworthy that a state $\vec{s}$ gets added to $X$ if there is even one infinitely long trace from it in some flattening that gets generated. That is, the flattenings need \emph{not} exhibit all traces from this state that are present in $M_1$.

The stopping condition for the algorithm occurs when a flattening $M'$ is found such that after adding states that satisfy $\eglobally{\phi_1}$ in $M'$ it to $X$, the approach detects that all traces in $M_1$ starting from all states that are currently not in $X$ are exhibited by $M'$. At this point it becomes clear that $X$ contains all states that satisfy $\eglobally{\phi_1}$ in $M_1$.

Note that the stopping condition can be reached even if $M_1$ is not trace flattable. $M_1$ only needs to possess a flattening such that all traces in $M_1$ from states that \emph{do not} satisfy $\eglobally{\phi_1}$ are preserved in this flattening. Also, clearly, if $M_1$ is trace flattable, then our algorithm necessarily terminate eventually.

To illustrate this approach, we revert back to the example system $M_1$ in Figure~\ref{fig:example_intro}(b).  Since there are arbitrarily long traces in $M_1$ that alternate between the transitions $t_0$ and $t_1$, no flattening (or set of flattenings) can preserve all traces.  In other words, $M_1$ is not trace flattable. However, our approach is able to terminate after generating the three flattenings shown in parts~(c)-(e) of the figure. The correct solution to the property $\eglobally{(x < 10)}$ is the set of states $x \geq 0 \wedge x < 5$ (from each state in this set there is an infinitely long path that alternates between the two transitions). 

\subsubsection{The over-approximation component.}

The under-approximation component described above did not use reachability analysis approaches as a subroutine. The over-approximation component, which we describe here, uses this as a subroutine.

This component maintains a growing set $Y$, which is always a under-approximation of the set of states in $M_1$ that do not satisfy $\eglobally{\phi}$. This set is initialized to the set of ``stuck'' states in $M_1$; these are the states from which there is no outgoing transition. Then, iteratively, in each step of the approach, the set of all states $\vec{s}'$ that satisfy the following properties are identified and added to $Y$:

\begin{itemize}
\item Some state in $Y$ is reachable from $\vec{s}'$ (this is where reachability analysis is used as a subroutine)

\item Ignoring states that are already in $Y$, there is a \emph{single} outgoing trace from $\vec{s}'$, which ends in a state that is already in $Y$. We will present later the details of how we check this property.
\end{itemize}

It is easy to see that an invariant of the approach is that any state that ever gets added to $Y$ is such that all traces from the state reach a stuck state. That is, the set $Y$ is a growing under-approximation of $\eglobally{\phi}$. In other words, $\phi - Y$ is a shrinking over-approximation of $\eglobally{\phi}$.

We now illustrate this approach on the refined system $M_1$ depicted in Figure~\ref{fig:example_intro}(b), for the property $\eglobally{(x < 10)}$. $Y$ gets initialized to the stuck states in this system, which are described by the formula $x < 0 \vee x \geq 10$ (these states do not satisfy the guard of either of the two transitions). In the first iteration, the states $x \geq 5 \wedge x < 9$ get added to $Y$. This happens because from each of these states there is a single trace, which reaches the state $x=10$, which is already in $Y$ (this single trace follows transition $t_0$ one or more times). Therefore, $Y$ becomes equal to $x < 0 \vee x \geq 5$.  Note that from each state in $x \geq 0 \wedge x < 5$ there are multiple traces that do not go through states in $Y$.  These traces follow different interleavings of the two transitions. Therefore, no more states get added to $Y$ in the second iteration, which results in termination. The precise solution is $\phi - Y$, which is $(x < 10) - (x < 0 \vee x \geq 5)$, which is $x \geq 0 \wedge x < 5$. Note that on this example the under-approximation component and the over-approximation component both were able to terminate with the precise solution.

We are able to prove that the over-approximation component also terminates on all flat and trace flattable systems. This proof is quite involved in nature. In general, the termination class of the over-approximation component is incomparable with the termination class of the under-approximation component.

We also describe in this paper an integration of the two components, such that they assist each other during their (simultaneous) iterations. The integrated approach necessarily terminates on all cases where either one of the individual components would have terminated.
There are two advantages of integrating these two approaches:
\begin{itemize}
\item Due to the mutual assistance provided by the two components in 
each iteration, the interated approach terminates on some cases where 
neither of the components terminate.
\item  Users can apply the integrated approach directly on a given 
system, without having to decide upfront which component to try 
first, or without having to try both components simultaneously.
\end{itemize}

\subsection{Contributions}

Our primary contribution is a novel approach for CTL model checking of counter systems. The key novelty of our approach over previous CTL model-checking
approaches~\cite{dem2006,gaporder12} are as follows:

\begin{itemize}
\item Our approach uses reachability analysis
black-boxes as subroutines. As a result, 
the class of systems on which our approach terminates
(with precise solutions) is arguably wider than the
subclass addressed by Demri et al.~\cite{dem2006} (and potentially
incomparable with that of Bozzelli et al.~\cite{gaporder12}).

\item Our approach supports approximations in cases where termination is not guaranteed. This is a useful feature that is not a part of previous approaches.

\item 
We have implemented our approach, and have evaluated it on a
set of 44 real-life counter systems that come with the FAST~\cite{fast}
toolkit. These systems are all outside the class of systems addressed by the approach of
Demri et al.~\cite{dem2006} and Bozelli et al.~\cite{gaporder12}. Our
implementation terminated (with precise results) on 42 of these systems within 1
second of running time.
\end{itemize} 

This paper is an extended version of a conference paper that we wrote earlier~\cite{kvasantafme2014}. The contributions that are new in this paper over the conference paper are as follows:

\begin{itemize}
\item In the conference submission the two components that we introduced earlier in this section for solving for ``global'' properties were presented as independent approaches. In this paper we present an integration of these two approaches into a single approach. The advantages of this integration were mentioned earlier in this section.

\item We describe an implementation and evaluation of the integrated approach. 

\item We give a complete proof of correctness of our integrated approach for solving for ``global'' properties. In the conference paper we had stated but not proved the correctness of the two underlying components. (Note that the proof of correctness of our routines for solving for ``next'' and ``until'' properties are relatively easy.)

\item We present complete proofs for termination of the integrated approach  on certain classes of counter systems. We also show that these classes are wider than the class of flat and trace-flattable systems. (Again, showing classes of systems on which our routines for ``next'' and ``until'' properties terminate is easy.)

\item We also give an example of a system on which neither of the two components
terminate but the integrated approach terminates.

\item We present the complete details of an inductive approach to solve for general (i.e., arbitrarily nested) CTL properties. This approach uses as subroutines our approaches that were introduced above for solving for ``next'', ``until'', and ``global'' properties. This inductive approach was alluded to but not formally presented in the conference submission (due to space restrictions).

\end{itemize}

The remainder  of this paper is organized as follows: In Section~\ref{sec:prelim} we introduce some of the preliminary notions and terminology  that underlies our approaches. 
In Section~\ref{sec:until}  we present our approach to answer until properties. In
Section~\ref{sec:global}, we summarize the
results from our previous work~\cite{kvasantafme2014} along with some extensions to handle global
properties. We discuss the correctness and termination of the routines that handle global properties
in Section~\ref{sec:global-correctness} and Section~\ref{sec:termination}. In Section~\ref{sec:algo} we discuss our 
inductive algorithm for answering CTL properties. 
Section~\ref{ssec:implementation} contains the discussion on our implementation 
and experimental results, while Section~\ref{sec:related_work} discusses related work. We formally state and prove the 
theoretical claims of our Algorithms in Appendix~\ref{app:proofs}.

\section{Notation and Terminology}
\label{sec:prelim}
We had presented many of the key background concepts for our work in
Section~\ref{sec:intro}. In this section we present these definitions and terminology more formally for the sake of completeness.

\subsection{Counter systems and Presburger formulas}

\begin{definition}[Counter System]
 A counter system $M$ is represented by the tuple $M=\langle Q, C,\Sigma,
\phiinit,G,F\rangle$
where $Q$ is a finite set of natural numbers that encode the control states, $C$ is a finite set of $m$ counters,
$\phiinit$ is a Presburger formula that represents the initial states of the system, $\Sigma$ is a finite alphabet representing 
the set of transitions in $M$,
such that for each $b\in \Sigma$ there exists a Presburger formula
$g_b \in G$ and
a Presburger formula $f_b \in F$ that are the guard and action of the transition $b$, respectively.
\end{definition}

In the counter system that was shown in Figure~\ref{fig:example_intro}(a),
$Q = \{q_0\}$ (encoded as the natural number zero), $C = \{x\}$, $\Sigma
= \{t_0, t_1\}$, $\phiinit = (x = 0)$ (shown as the incoming arrow into the
system), $g_{t_0} \equiv (x\geq0)\wedge(x<100)$, $g_{t_1} \equiv
(x>0)\wedge(x<5)$, $f_{t_0} \equiv (x^\prime = x +1)$, $f_{t_1} \equiv
(x^\prime = x - 1)$.

A \emph{state} (denoted by $\vec{s}$, $\vec{s_1}$, $\vec{s'}$ etc.) in a system is a column vector $\vec{v} \in \mathbb{N}^{m+1}$.
The first element $\vec{v}[0]$ represents  the control state, while the values of rest of the elements 
$v[1],\dots,v[m]$ represent the values of the counters $C$. 
We sometimes use the term \emph{concrete state} to refer to a state.  

Presburger formulas that we use in this paper are of two kinds. In the
first kind of formula, the names of the counters, as well
the \emph{control-state variable} $q$ (which refers to the first element
$v_0$ of a state as mentioned above), occur as free variables. Throughout
this paper we use symbols $\phi$, $\phi_i$, etc., to denote Presburger
formulas of the first kind. Since the example systems we use for
illustrations have only a single control-state, we omit the control-state
variable $q$ from the guards, actions, and formulas that we show (it will
always be constrained to be zero).  Also, sometimes we wish to use extra
free variables (on top of the counter names and $q$) in a formula. Our
notation in this case is as follows: $\phi_{(k)}$ is a Presburger formula
with an additional free variable $k$.

A state $\vec{s}$ is said to satisfy a formula $\phi$, denoted as
$\vec{s}\models \phi$, if the formula $\phi$ evaluates to true when the
free variables in $\phi$ are substituted with the corresponding values in
$\vec{s}$. For this reason, we often refer to a formula as a ``set of
states'', by which we mean the states that satisfy the formula.

We use Presburger formulas of the first kind for multiple purposes. The
first use is as guards of transitions. The second use is as part of the
input property given to the approach.  For instance, consider the input
property $\eglobally{(x < 10)}$; the Presburger formula $x < 10$ within
this property indicates that the approach ought to identify the states of
the system from which infinitely long traces exist that pass through only
states in the set $x < 10$. The solution to a given property from our
approach is also represented as a Presburger formula. For instance, for the
input property mentioned above, for the system shown in
Figure~\ref{fig:example_intro}(a), the solution can represented as the
Presburger formula $x \geq 0 \wedge x < 5$.

The second kind of Presburger formula uses unprimed and primed versions of the counter names as well as $q, q'$ as free variables. A pair of states $(\vec{s}, \vec{s^\prime})$ is said to satisfy a formula of this kind if the formula evaluates to $\tru$ when the unprimed free variables in the formula are substituted with the corresponding values in $\vec{s}$ and the primed free variables in the formula are substituted with the corresponding values in $\vec{s^\prime}$. This kind of Presburger formula is used only to represent the transition $f_b$ of each transition $b$ of a counter system. 

The semantics of a counter system is as follows. A \emph{concrete transition}  $\transition{\vec{s}}{M}{b}{\vec{{s^\prime}}}$ is said to be possible in system $M$ due to transition $b$ if $\vec{s}$ satisfies $g_b$ and $(\vec{s}, \vec{s^\prime})$ satisfies $f_b$. In this case we say that $\vec{s}$  is a  predecessor of $\vec{s^\prime}$, and that $\vec{s^\prime}$ is a successor of $\vec{s}$. A counter system can be \emph{non-deterministic}; i.e., a state could have multiple successor states, either by the action of a single transition itself, or due to different transitions out of a control-state with overlapping guards. However, we assume that systems exhibit \emph{finite branching}; i.e., every state has a finite number of  successors. 

\subsection{Traces and flattenings}
\label{ssec:prel:traces-flattenings}

Given a counter system $M$,
a \emph{trace} $t$ ``in'' $M$
\emph{starting from} a state $\vec{s_0}$ is any sequence of states
$\vec{s_0},\vec{s_1}, \ldots, \vec{s_n}$, $n \geq 0$, such that there is a concrete transition in $M$ from each state in the sequence to the next state (if any) in the sequence. This definition also generalizes in a natural way
to infinite traces. 
If $t$ is a trace in $M$ we also say that $M$ \emph{exhibits} $t$. We use the notation $\traces{M}{q}$ to represent the set  of all traces
in $M$ that start from states $\vec{s}$ such that $\vec{s}[0] = q$; i.e., intuitively, these are the traces that start from the control state $q$. 
$\tracesn{M}{q}{\phi}$ represents the subset of  $\traces{M}{q}$ consisting of traces whose
starting states satisfy the given formula $\phi$. 
We define the set of traces starting from states that satisfy $\phi$ as 
$\traces{M}{\phi} = \underset{q\in Q}{\bigcup}\tracesn{M}{q}{\phi}$.

A system $M_1$ is said to be a \emph{refinement} of a system $M$ wrt to a
formula $\phi$, written as $M_1$ $\equiv$ $\refineSystem{M}{\phi}$, if $M_1$ is
identical to $M$ in every way except that the guard of each transition in $M_1$
is the corresponding guard in $M$ \emph{conjuncted} with $\phi$. For instance,
the system $M_1$ in Figure~\ref{fig:example_intro}(b) is a refinement of the
system $M$ in part~(a) of the same figure wrt the formula `$x < 10$'.
Intuitively, $M_1$ exhibits exactly those traces in $M$ that do not go through
a concrete transition from a state that does not satisfy $\phi$.

\begin{definition}[Flat system]
A \emph{flat} counter system is one in which no two distinct cycles among
its control states overlap. That is, all cycles among its control states
are simple cycles.
\end{definition}

\begin{definition}[Flattening]
A flat system $N=\langle Q^\prime, C,\Sigma^\prime,
{\phiinit}^\prime,G^\prime,F^\prime\rangle$ 
is said to be a \emph{flattening} of a system 
$M=\langle Q, C,\Sigma,{\phiinit},G,F\rangle$ if,
(a) the two systems use the same set of counters, (b) there exists a 
function $f:Q^\prime\rightarrow Q$ that maps every control state in $N$
to a control-state in $M$, and (c) any
transition in $N$ from a control-state $q_{ij}$ to a control-state $q_{kl}$ 
such that $f(q_{ij}) = q_i$ and $f(q_{kl}) = q_k$ has the same guard and
action as some transition from $q_i$ to $q_k$ in $M$. 
\end{definition}

Any control state $q'$ of a flattening $N$ of a system $M$ is called a ``copy'' of the control state $f(q')$ of $M$.

As an illustration, parts~(c)-(e) in Figure~\ref{fig:example_intro} show three different flattenings of the system $M_1$ in part~(b) of the same figure. The flattenings in parts~(c) and~(d) each contain a single ``copy'' of the control state $q_0$, while the flattening in part~(e) contains two copies of $q_0$ (i.e., the function $f$ maps both $q_{01}$ and $q_{02}$ to $q_0$ of $M$).

We extend the definition of the function $f$ mentioned above to map states of a flattening $N$ to states of the original counter system $M$, as follows. If $\vec{s^\prime}$ is a state of $N$, then $f(\vec{s^\prime}) = \vec{s}$, where $\vec{s}$ is a state of $M$,
if and only if
$f(\vec{s^\prime}[0]) = \vec{s}[0]$ and
$\forall (1 \leq i \leq m): \vec{s}[i] = \vec{s^\prime}[i]$. We say that any state $\vec{s'}$ of a flattening $N$ of a system $M$ is a ``copy'' of the state $f(\vec{s'})$ of $M$. Note that for each state $\vec{s}$ of $M$, if $\vec{s}[0]$ is a control-state $q$, then $\vec{s}$ has a separate copy in $N$ corresponding to each control state $q'$ of $N$ such that $f(q')=q$. 

Further, we can extend the definition above naturally to map a sequence of states of $N$ to a sequence of states of $M$. That is, for each position $i$, if the $i$th state of the first sequence is $\vec{s^\prime}$, then the $i$th state in the second sequence would be $f(\vec{s^\prime})$.
Finally, we can extend the function $f$ further to work on sets of sequences of states. That is, if $T$ is any set of sequences of states of $N$, then $f(T)$ is the set $\{f(t)| t \in T\}$.

An important property of flattenings is that for any flattening
$N$ of $M$, and for any trace $t^\prime$ in $N$, $f(t^\prime)$ is a trace in $M$. That is, every trace in $N$ is mapped by $f$ to a trace in $M$.  That is, intuitively, $N$ does not exhibit any traces that are not present in $M$. 
Note that in general there could exist traces in $M$ to which no trace in $N$ is mapped by the function $f$; these traces are ``missing'' in $N$, so to say.

For an illustration, let us re-visit Figure~\ref{fig:example_intro}. The flattenings in parts~(c)-(e) all have missing traces when compared to the original system $M_1$ in part~(b) of the figure.  For instance, the flattening in part~(c) (resp.~(d)) misses all traces that go through transition $t_1$ (resp.~$t_0$) even once, while the flattening in part~(e) misses traces that go through any one of the two transitions of $M_1$ more than once consecutively. 

\begin{definition}[Trace flattening]
We say that a flattening $N$ of a counter system $M$ is a trace flattening of $M$ with respect to a set of states $\phi$ of $M$ if and only if for every state $\vec{s}$ in $\phi$, there exists a state $\vec{s'}$ of $N$ such that

\begin{enumerate}
\item $f(\vec{s'})=\vec{s}$ (i.e., $\vec{s'}$ is a copy of $\vec{s}$), and
\item there don't exist two states $\vec{s'_1}$ and $\vec{s'_2}$ of $N$ such that
\begin{enumerate}
\item $\vec{s'_1}$ is reachable from $\vec{s'}$ in $N$ along a trace consisting of zero or more concrete transitions, and
\item there is no concrete transition $\vec{s'_1} \rightarrow \vec{s'_2}$ in $N$ but there is a concrete transition $f(\vec{s'_1}) \rightarrow f(\vec{s'_2})$ in $M$.
\end{enumerate}
\end{enumerate}
\end{definition}

Throughout this paper, wherever we say that all traces from a state $\vec{s}$ or a set of states $\phi$ in a system $M$ are ``preserved'' in a flattening $N$ of $M$, what we mean is that $N$ is a trace flattening of $M$ wrt $\vec{s}$ or $\phi$. Note that if $N$ is a trace flattening of $M$ wrt to any set of states $\phi$, $N$ is also a trace flattening of $M$ wrt to each individual state in $\phi$ and wrt to each subset of $\phi$.

Returning to the illustration in Figure~\ref{fig:example_intro}, if we consider the flattening in part~(c) of the figure as $N$ and the system in part~(b) of the figure as the given system $M_1$, then $N$ is a trace flattening of $M_1$ wrt the formula $x \geq 5$. Note that traces from $x=0$ to $x=4$ are \emph{missing} in $N$. The flattening in part~(d) of the figure preserves no traces at all. Finally, let $N$ be the flattening in part~(e) of the figure. $N$ is a trace flattening of $M_1$ wrt to only the state $x=99$. Furthermore, note that this state has two copies in $N$, one corresponding to each of the two control states of $N$. However, it is only the copy corresponding to control-state $q_{01}$ from which the trace $(x=99) \rightarrow (x=100)$ is preserved. Therefore, we call this copy the ``trace preserving copy'' of the state $x=99$ of $M_1$. Note that for traces from a state $\vec{s}$ of $M$ to be preserved in $N$, at least one of the copies of $\vec{s}$ needs to be a trace preserving copy.

We now consider a variant of the flattening in Figure~\ref{fig:example_intro}(c), wherein we add another control-state $q_{01}$, and a copy of the transition $t_0$ from $q_0$ to $q_{01}$. This is a flattening of the system $M_1$ in Figure~\ref{fig:example_intro}(b). However, unlike with the original flattening in Figure~\ref{fig:example_intro}(c), this extended flattening $N$ is a  trace flattening of $M_1$ only wrt the state $x=99$. This is because, for any other state of $M_1$, from either copy of this state in $N$, one is able to reach a state corresponding to $q_{02}$ in zero or one transitions and then become ``stuck'' (i.e., encounter a missing transition). 

On a related note,  all three flattenings mentioned above actually preserve all traces starting from states $x < 0$, because all such traces in $M_1$ are trivial traces (of length 1).

\begin{definition}[Trace flattable]
We say a counter system $M$ is trace flattable wrt a formula $\phi$ if there exists a flattening
$N$ of $M$ such that $N$ is a trace flattening of $M$ wrt $\phi$.
\end{definition}

It is notable that the system in Figure~\ref{fig:example_intro}(b) is \emph{trace flattable} wrt the set of states $x \geq 5$, but \emph{not} trace flattable wrt any set of states that includes $x=0$ to $x=4$.

\subsection{Reachability and temporal properties}

Given a counter system $M$ and Presburger formula $\phi$, we use the formula $\pre{M}{\phi}$ (which is also a Presburger formula in the counter variables and in $q$) to represent the
set of all states  that have a successor that satisfies $\phi$. 
For the counter system $M$ shown in Figure~\ref{fig:example_intro}(a),  $\pre{M}{(x \leq 2)} \equiv (x \geq 0) \wedge (x \leq 3)$.

An extension of the above definition is the formula $\prek{M}{\phi}$. This represents the set of all states from which some state that satisfies $\phi$ can be reached in exactly $k$ steps (i.e., $k$ concrete transitions). Note that $k$ is an extra free variable in the formula $\prek{M}{\phi}$. For our example system $M$ in Figure(a), 
$\prek{M}{x = 4}$ $\equiv$ $(x \leq 4) \wedge (x \geq (4-k)) \wedge (\mathit{even}(k) \Rightarrow \mathit{even}(x)) \wedge (\mathit{odd}(k) \Rightarrow \mathit{odd}(x))$. 


The backward reachability set for a set of states $\phi$, namely
$\preStar{M}{\phi}$, represents the set of all states from which a state in $\phi$ can be reached in zero or more steps. For our example system $M$, 
$\preStar{M}{x \leq 4}$ $\equiv$ $x \leq 4 \wedge x \geq 0$. 

Similarly, we use the formula $\post{M}{\phi}$  to represent the
set of all states  that have a predecessor that satisfies $\phi$, and the
formula $\postStar{M}{\phi}$ to represent the set of all states 
that can be reached from a state in $\phi$ in zero or more steps.

Temporal properties are used to formally specify properties on sequences of states (i.e, paths) in a transition system. 
$\mathit{CTL}$ properties allow us to quantify over paths in a transition system. Following the idea of Demri et al.\cite{dem2006}, 
we extend the $\mathit{CTL}$ grammar to let Presburger formulas of the form $\phi$ be the basic
propositions in the temporal properties.
Temporal properties in the \emph{existential
normal form} (ENF) of CTL can be defined by the
following grammar
\begin{equation} 
\psi\ \equiv\ \phi\, |\, \neg\psi\, |\, \psi \vee \psi\,
  |\, \enext{\psi}\, |\,
\euntil{\psi}{\psi}\, |\, \eglobally{\psi}
\label{eqn:enf}
\end{equation}
Any $\mathit{CTL}$ property involving universal path quantifiers has an
equivalent $\mathit{CTL}$ property in ENF. Therefore, we
restrict ourselves to CTL properties in ENF. Throughout this paper we use $\psi, \psi_i$, etc., to denote CTL properties in ENF.  Note that in the interest of clarity, throughout this paper, whenever we say ``formula'' we mean a Presburger formula, while whenever we say ``property'' we mean a CTL property.

A state $\vec{s_0}$ satisfies a CTL property $\psi$ in a system $M$, written as
$\vec{s_0} \models \psi$, as per the following definition
\begin{itemize}
 \item $\vec{s_0} \models \phi \iff$ $\vec{s_0}$ is a solution to $\phi$.
  \item $\vec{s_0} \models \neg \psi \iff \vec{s_0} \nvDash \psi$ 
  \item $\vec{s_0} \models \psi_1\vee\psi_2 \iff \vec{s_0}\models
\psi_1$ or
$\vec{s_0}\models\psi_2$
  \item $\vec{s_0} \models \enext{\psi_1} \iff $ there exists a state
    $\vec{s_1}$ such that $\vec{s_1} \models \psi_1$ and $\vec{s_1}$ is
    a successor state of $\vec{s_0}$ in $M$.
  \item $\vec{s_0} \models \euntil{\psi_1}{\psi_2} \iff$ there exists a
trace $\vec{s_0},\vec{s_1},\dots,\vec{s_k}$ in the system $M$ such
that $\vec{s_k} \models \psi_2$ and $\forall i, 0\leq i<k.\, \vec{s_i}
\models \psi_1$. 
\item $\vec{s_0} \models \eglobally \psi \iff$ there exists an infinite trace
$\vec{s_0},\vec{s_1},\dots$ in the system $M$ such that
  $\forall i\geq 0.\,\vec{s_i}\models \psi$.
\end{itemize}

This paper addresses the problem of ``global'' model-checking. That is, given a system $M$ and a CTL property $\psi$, the problem is to find the set of states $\phi$
such that any state $\vec{s}$ is in set $\phi$ iff $\vec{s}$ satisfies property $\psi$ in $M$. 

\section{Until Properties}
\label{sec:until}

\begin{figure}
\begin{center}
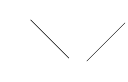
\label{fig:Lattice}
\caption{The lattice $\latticeApprox$}
\end{center}
\end{figure}

\sloppypar In this section we describe our approach to solve for ``until'' properties of the form $\euntil{\phi_1}{\phi_2}$. Our approach is presented as a 
routine $\computeUntil{M}{\phi_1}{\phi_2}{\enumlabel}$, where $M$ is a counter system, $\phi_1$ and $\phi_2$ are Presburger formulas, and $\enumlabel$ is a member of the enumeration $\latticeApprox$ $\equiv$ $\{\precise,\under,\overapprox\}$. The routine either returns a Presburger formula $\phi$ and an enumerator $\approximation$ from $\latticeApprox$, or goes into non-termination. There is a partial ordering on the elements of $\latticeApprox$, as shown in Figure~\ref{fig:Lattice}; in other words, the ordering is $\precise \sqsubseteq \under$, $\precise \sqsubseteq \overapprox$. Whenever the routine terminates, it is guaranteed that the returned enumerator $\approximation$ is such that it is \emph{dominated} by the given enumerator $\enumlabel$; i.e., $\approximation \sqsubseteq \enumlabel$. Furthermore, if $\approximation$ is $\precise$, then, it is guaranteed that $\phi$ represents  precisely the set of states of $M$ that satisfy $\euntil{\phi_1}{\phi_2}$. On the other hand, if $\approximation$ is $\overapprox$ (resp. $\under$), then $\phi$ is guaranteed to be either precise or be an over- (resp. under-) approximation of the set of states of $M$ that satisfy $\euntil{\phi_1}{\phi_2}$. 


\subsection{Approach}
\label{ssec:until:approach}

Note that a state $\vec{s}$ satisfies the property $\euntil{\phi_1}{\phi_2}$ if $\vec{s}$ satisfies $\phi_2$, or $\vec{s}$ satisfies $\phi_1$ and there is a trace in $M$ starting from $\vec{s}$ along which some state $\vec{s'}$ satisfies $\phi_2$ and along which all states prior to $\vec{s'}$ satisfy $\phi_1$. Our approach to finding such states $\vec{s}$, intuitively, is to first conjunct the guard of each transition in $M$ with $\phi_1$; we refer to this step as ``refining'' the system $M$ with the guard $\phi_1$ to obtain a refined system $M'$, and denote this step as $M' = \refineSystem{M}{\phi_1}$. As a result of the refinement, each concrete state of $M'$ that has a successor must satisfy $\phi_1$. Therefore, it is easy to see that a trace (of length $\geq 1$) is present in $M_1$ from a state $\vec{s}$ to some state that satisfies $\phi_2$ iff $\vec{s}$ satisfies the property
$\euntil{\phi_1}{\phi_2}$ in $M$. Hence our problem reduces to finding the set of states from which
we can reach $\phi_2$ in the counter system $M_1$. This problem can be solved using existing reachability analysis tools for counter systems, e.g.,~\cite{Com1998,Ibarra02,finkel02,trex_fast_accelerations05,iteratingOctagons}.

\begin{algorithm}
\caption{$\computeUntil{M}{\phi_1}{\phi_2}{\enumlabel}$}
 \begin{algorithmic}[1]
  \REQUIRE{A counter system $M$, formulas $\phi_1$ and $\phi_2$, and an enumerator $\enumlabel$, which indicates whether the 
  formula $\euntil{\phi_1}{\phi_2}$ has to be computed precisely or over-approximated or under-approximated.}
  \ENSURE{A formula $\phi$ and an enumerator $\approximation$. The enumerator indicates whether $\phi$ is precise or is over-approximation or
under-approximation of the set of states of $M$ that satisfy $\euntil{\phi_1}{\phi_2}$.}
  \STATE $M_1\gets\refineSystem{M}{\phi_1}$\label{line:untilRefine}
  \STATE $(\phi,\approximation)\gets\computePreStar{M_1}{\phi_2}{\enumlabel}$\label{line:untilPreStar}
  \RETURN $(\phi,\approximation)$
 \end{algorithmic}
\label{algo:until}
\end{algorithm}

The approach mentioned above is depicted as Algorithm~\ref{algo:until}. 
The routine $\computePreStarName$, which is invoked in 
line~\ref{line:untilPreStar} as a black-box, is expected to be provided by
a reachability analysis tool. The invocation
$\computePreStar{M_1}{\phi_2}{\enumlabel}$ is expected to either return a
formula $\phi$ and an enumerator $\approximation$ (from the enumeration
$\latticeApprox$), or to go into non-termination.  Whenever the routine
terminates, it is assumed that the returned enumerator $\approximation$ is
such that it is \emph{dominated} by the given enumerator
$\enumlabel$. Furthermore, if $\approximation$ is $\precise$, then, it is
assumed that $\phi$ is equal to $\preStar{M_1}{\phi_2}$, which is the set
of states of $M_1$ from which states that satisfy $\phi_2$ are reachable.  On the other hand, if $\approximation$ is $\overapprox$ (resp. $\under$), then $\phi$ is assumed to be either precisely equal to or be an over- (resp. under-) approximation of the set of states $\preStar{M_1}{\phi_2}$. 

Reachability analysis for Presburger counter systems is in general an
undecidable problem. The reachability analysis tools mentioned above target
different sub-classes of counter systems. If a tool is given a system that is
outside its targeted subclass, the tool may either be inapplicable, or could
potentially go into non-termination. Note that in both these cases, if
$\enumlabel$ is not $\precise$, then the following default action can be taken
in routine $\computeUntilName$ instead of using the results from the invocation
$\computePreStar{M_1}{\phi_2}{\enumlabel}$:  if $\enumlabel$ is $\overapprox$
return ($\phi_1 \vee \phi_2, \overapprox$), else return ($\phi_2, \under$).

Some of the reachability analysis tools that are liable to go into non-termination are capable of providing an approximated solution in a specific direction upon forced premature termination. For instance, the tool Fast~\cite{fast} works by exploring \emph{flattenings} of the given system $M_1$ one by one, uses \emph{acceleration} to compute reachability on each flattening, and unions these reachability results across the flattenings. In other words, it computes a growing under-approximation of the set of states from which states in $\phi_2$ are reachable. If it terminates, then it would have obtained a precise solution. On the other hand, if it is forcibly stopped at any point, then it yields an under-approximate solution. Routine $\computeUntilName$ could make use of this feature whenever $\enumlabel$ is in the supported direction of approximation, by setting a timeout on the invocation to $\computePreStarName$.

\subsection{Correctness of the routine $\computeUntilName$}
In this section we formally state the correctness properties of the routine
$\computeUntilName$. We give the intuition behind proofs; detailed proofs are
provided in Appendix~\ref{app:proofs}

\begin{theorem}
\label{thm:untilPrecise}
Given a counter system $M=\langle Q, C,\Sigma,
\phiinit,G,F\rangle$ and sets of states $\phi_1$ and $\phi_2$, routine $\computeUntilName$ in Algorithm~\ref{algo:until} returns  precisely 
the set of states that satisfy $\euntil{\phi_1}{\phi_2}$ whenever the invocation $\computePreStar{M_1}{\phi_2}{\enumlabel}$ in the routine returns the precise formula $\preStar{M_1}{\phi_2}$, where $M_1 \equiv \refineSystem{M}{\phi_1}$. 
\end{theorem}

We give the intuition behind the proof of this theorem.
Any state $\vec{s}$ that satisfies $\euntil{\phi_1}{\phi_2}$ should either satisfy
$\phi_2$ or, there should be a trace from $\vec{s}$ such that the last state in
that trace satisfies $\phi_2$ and all states along the trace should satisfy
$\phi_1$. Therefore the state $\vec{s}$ should be in $\preStarName$ of $\phi_2$
in the refined system. The refinement ensures that all states from $\vec{s}$ to
the state that satisfies $\phi_2$ will definitely satisfy $\phi_1$.

\begin{theorem}
 \label{thm:untilApprox}
Given a counter system $M=\langle Q, C,\Sigma,
\phiinit,G,F\rangle$ and sets of states $\phi_1$ and $\phi_2$, routine $\computeUntilName$ in Algorithm~\ref{algo:until} returns  an over- (resp. under-) approximation of 
the set of states that satisfy $\euntil{\phi_1}{\phi_2}$ whenever the invocation $\computePreStar{M_1}{\phi_2}{\enumlabel}$ in the routine returns an over- (resp. under-) approximation of the set of states that satisfy the formula  $\preStar{M_1}{\phi_2}$, where $M_1 \equiv \refineSystem{M}{\phi_1}$. 
\end{theorem}

The proof for Theorem~\ref{thm:untilApprox} follows directly from proof of
Theorem~\ref{thm:untilPrecise} in cases where the routine $\preStarName$
provides an approximate result. It is straightforward to see that in cases
where the routine $\preStarName$ approximates, it returns $\phi_1\vee \phi_2$
when the $\enumlabel$ is $\overapprox$ and $\phi_2$ when $\enumlabel$ is
$\under$. These states are clearly an over- (resp. under-) approximation of the
set of states of the states that satisfy $\euntil{\phi_1}{\phi_2}$. Hence the
routine $\computeUntilName$ also approximates the result in the same direction
as that of $\preStarName$.

\section{Global Properties}
\label{sec:global}
In this section we describe our approach to solve for ``global'' properties of the form $\eglobally{\phi}$. Our approach is presented as a 
routine
$\computeGlobal{M}{\phi}{\enumlabel}$, where $M$ is a counter system, $\phi$ is a Presburger formula, and $\enumlabel$ is a member of the enumeration $\latticeApprox$ $\equiv$ $\{\precise,\under,\overapprox\}$ that was depicted in Figure~\ref{fig:Lattice}. This routine either returns a Presburger formula $\phi_1$ and an enumerator $\approximation$ from the enumeration $\latticeApprox$, or goes into non-termination. Whenever the routine terminates, it is guaranteed that the returned enumerator $\approximation$ is such that it is \emph{dominated} by the given enumerator $\enumlabel$; i.e., $\approximation \sqsubseteq \enumlabel$. Furthermore, if $\approximation$ is $\precise$, then, it is guaranteed that $\phi_1$ represents  precisely the set of states of $M$ that satisfy $\eglobally{\phi}$. On the other hand, if $\approximation$ is $\overapprox$ (resp. $\under$), then $\phi_1$ is guaranteed to be either precise or be an over- (resp. under-) approximation of the set of states of $M$ that satisfy $\eglobally{\phi}$. 

\subsection{Overview of the approach}

\begin{algorithm}[t]
\caption{$\computeGlobal{M}{\phi}{\enumlabel}$}
\label{algo:compute-global}
 \begin{algorithmic}[1]
 \REQUIRE A counter system $M$, a formula $\phi$, and an enumerator $\enumlabel\in\latticeApprox$.
 \ENSURE Returns a formula $\phi_1$ and an enumerator $\approximation \in \latticeApprox$. The enumerator $\approximation$ indicates whether $\phi_1$ is precise or is over-approximation or
under-approximation of the set of states of $M$ that satisfy $\eglobally{\phi}$.

  \STATE $M_1\gets\refineSystem{M}{\phi}$\label{line:global-refine}
  \STATE $X=\emptyset, Y= \neg (g_1 \vee g_2 \vee \dots \vee g_n) \, \vee \neg \phi$.\label{line:init-xy}
  \STATE \COMMENT{$X$: growing under-approx. of $\eglobally{\phi}$. $Y$: growing under-approx. of $\neg \eglobally{\phi}$.}
  \WHILE{not forcefully terminated}\label{line:outer-loop-beg}
  \STATE $\mathit{Flat} \gets$ $k_1$ fresh flattenings of $M_1$. \COMMENT{$k_1 \geq 1$ is a parameter to the algorithm.}\label{line:gen-flattenings}
  \FORALL{$N \in \mathit{Flat}$} \label{line:inner-loop-beg}
  \STATE $X\gets X \vee \forall k \geq 0. \prek{N}{\phi}$ \label{line:growX}
  \STATE $(\phi',\mathit{flag})\gets\checkflattening{M_1}{N}{\phi-X-Y}$\label{line:checkfl} 
  \IF {$\mathit{flag}$}\label{line:term-Under}
  \RETURN $(X,\precise)$\label{line:return-Under}
  \ENDIF
   \STATE $Y \gets Y \vee \phi'$\label{line:growYb}
 \ENDFOR\label{line:inner-loop-end}
 \STATE $Y \gets  \computeGlobalOverNew{M_1}{\phi}{Y}$\label{line:growY}.

 \IF{return value from subroutine call above is equal to $Y$ given as argument}\label{line:term-Over}
  	\RETURN $(\neg Y,\precise)$\label{line:return-Over}
  \ENDIF\label{line:endY}
  \ENDWHILE\label{line:outer-loop-end}


  \IF{$\enumlabel=\under$} 
  \RETURN $(X,\under)$\label{line:return-Under-force}
    \ELSE
    \RETURN$(\neg Y,\overapprox)$\label{line:return-Over-force}
     \ENDIF

%
 \end{algorithmic}

\end{algorithm}

Algorithm~\ref{algo:compute-global} depicts our routine $\computeGlobalName$. The overall structure of the algorithm is as follows. In line~\ref{line:global-refine} we refine the system $M$ wrt $\phi$ to obtain a system $M_1$. Note that the set of states that satisfy $\eglobally{\phi}$ in $M$ and $M_1$ are identical. In the remainder of the algorithm, $M_1$ is used instead of $M$ for all purposes. The algorithm is iterative in nature, and maintains two key data structures: a growing under-approximation $X$ of the set of states of $M_1$ that satisfy $\eglobally{\phi}$, and a growing under-approximation $Y$ of the set of states of $M_1$ that \emph{do not} satisfy $\eglobally{\phi}$. Both these sets are encoded as Presburger formulas. The main outer loop of the algorithm is in lines~\ref{line:outer-loop-beg}-\ref{line:outer-loop-end}. In each iteration of this loop the set $X$ and $Y$ are grown. Moreover, the set $Y$ plays a role in determining the states that get added to $X$, while conversely, the set $X$ also plays a role in determining the states that get added to $Y$. The algorithm terminates whenever either set $X$ or set $Y$ is determined to have grown to the fullest extent possible (see the ``return' statements in lines~\ref{line:return-Under} and~\ref{line:return-Over}). If $X$ has grown fully, then it is returned as the precise solution, whereas if $Y$ has grown fully, then $\neg Y$ is returned as the precise solution. Termination of the algorithm in this way is guaranteed on a class of Presburger counter systems. We introduce this class later in this section.

The algorithm can be stopped forcefully at any point, upon which $X$ can be returned as an  under-approximation (see line~\ref{line:return-Under-force}) or $\neg Y$ can be returned as an over-approximation (see line~\ref{line:return-Over-force}).

\subsection{Computing $X$}
\label{ssec:under}
\subsubsection{Using refinement and reachability.}
A state $\vec{s}$ satisfies $\eglobally{\phi}$ in the refined system $M_1$ (and hence in the original system $M$) iff
there is at least one infinite trace from this state in $M_1$; this is because \emph{every} concrete transition in $M_1$ starts from a state that satisfies $\phi$. Our objective therefore is to find a Presburger formula, using reachability analysis, that represents the set of states in $M_1$ that have an infinite trace starting from them. Two key insights that make this possible are: (a) In a finite-branching system, as per K\"{o}enig's Lemma, there is an infinite trace from a state iff there are traces starting from it of \emph{all possible} lengths $k$, for $k \geq 0$. (b) A state has a trace of length $k$ from it iff it satisfies the formula $\prek{M_1}{\phi}$, which can be computed by existing reachability analysis techniques. Therefore, with this formula in hand, one only needs to eliminate $k$ as a free variable from it using universal quantification, as in   $\forall k\geq 0.\ \prek{M_1}{\phi}$, to obtain the precise set of states that satisfy $\eglobally{\phi}$ in $M$. 

\subsubsection{Computing $\forall k\geq 0.\ \prek{M_1}{\phi}$.} 
A key limitation of  existing reachability techniques~\cite{Com1998,finkel02,trex_fast_accelerations05,iteratingOctagons} is that although they can compute the formula $\preStar{M_1}{\phi}$ for interesting subclasses of systems, on the more difficult problem of computing the formula $\prek{M_1}{\phi}$ their applicability is restricted to the narrow class of flat systems. Whereas, most practical systems, such as those provided by the Fast toolkit~\cite{fast} are not flat.
A way out of this quandary is to obtain any flattening $N$ of $M_1$, and to compute the formula $\prek{N}{\phi}$. The presence of an infinite trace in $N$ from any state $\vec{s}$ implies the presence of the same trace in $M_1$. Therefore,
the set of states that satisfy $\eglobally{\phi}$ in $N$ (as represented by the formula
$\forall k\geq 0.\ \prek{N}{\phi}$) is guaranteed to be a subset (i.e., an under-approximation) of the set of states that satisfy $\eglobally{\phi}$ in $M_1$. Furthermore, by systematically enumerating various flattenings of $M_1$, and by accumulating the set of states that satisfy $\eglobally{\phi}$ in these flattenings, we can generate iteratively a non-decreasing under-approximation of the set of states that satisfy $\eglobally{\phi}$ in $M_1$, which would  basically be the set $X$ that was mentioned earlier.

We now discuss in detail how the approach mentioned above is carried out by Algorithm~\ref{algo:compute-global}. The set $X$ is initialized to $\emptyset$ (see line~\ref{line:init-xy}). In each iteration of the outer loop of the algorithm, some $k_1$ \emph{fresh} (i.e., hitherto un-enumerated)  flattenings of $M_1$ are generated, and saved in the set \emph{Flat} (see line~\ref{line:gen-flattenings}), where $k_1 \geq 1$ is a user parameter. In order to be systematic, we enumerate flattenings of $M_1$ in increasing order of \emph{length}, where the length of a flattening is the number of transitions it possesses. The flattenings in \emph{Flat} are then processed one by one in the inner loop in lines~\ref{line:inner-loop-beg}-\ref{line:inner-loop-end}. The set of states $\forall k\geq 0.\ \prek{N}{\phi}$ is computed for each of these flattenings (see line~\ref{line:growX}), and added to $X$. Line~\ref{line:term-Under} is a termination condition -- we will discuss this below. Lines~\ref{line:growY}-\ref{line:endY} pertaining to expanding the set $Y$, which we will discuss later in this section. After this point the algorithm goes onto the next iteration of the inner loop, wherein more flattenings of $M_1$ are enumerated and processed. 

\subsubsection{Termination condition based on $X$.}
\label{sssec:under-termination}

We now discuss the termination condition of the algorithm in line~\ref{line:term-Under}, which is based on the current contents of the set $X$. (We will discuss the other termination condition in the algorithm, in line~\ref{line:term-Over}, later. This condition is  dependent on the subroutine call in line~\ref{line:growY}, which also we discuss later.)
A challenge in devising a suitable termination condition based on the contents of set $X$ is that 
even if no new states get added to $X$ due to flattenings that are enumerated in a given iteration of the outer loop, there is no guarantee that the same will happen when considering more flattenings in the next iteration.
Therefore, the termination condition that we use is as follows: we first invoke
a subroutine $\checkflattening{M_1}{N}{\phi-X-Y}$ in line~\ref{line:checkfl}.
This subroutine returns $\tru$ iff $\traces{M_1}{\phi-X-Y} =
\traces{N}{\phi-X-Y}$. Else it returns a set of states $\phi^\prime$ such that
N is a trace flattening of $M_1$ with respect to $\phi^\prime$. We discuss the
implementation of this check later in an appendix; it suffices to say here that
this check is known to be decidable as per the existing
literature~\cite{dem2006}. $\computeGlobalName$ then returns $(X,\precise)$ if
the subroutine above returned $\tru$.

The correctness of the termination condition can be argued as follows. $Y$ contains states that definitely do not satisfy $\eglobally{\phi}$ in $M_1$ (this will become clear later in this section). Therefore, the only concrete states that are currently not in $X$ and that will possibly get added to $X$ in future iterations of the outer loop of the algorithm are ones that are in $\phi - X - Y$.  Now, no state in $\phi - X - Y$ satisfies $\eglobally{\phi}$ in $N$ (otherwise such a state would have gotten added to $X$ in line~\ref{line:growX}). Therefore, if $\traces{M_1}{\phi-X-Y} = \traces{N}{\phi-X-Y}$, it is clear that these states do not satisfy $\eglobally{\phi}$ in $M_1$ either. It therefore follows that $X$ already contains all states that satisfy $\eglobally{\phi}$ in $M_1$. Therefore, $(X,\precise)$ can be returned. 

Whenever $\checkflattening{M_1}{N}{\phi-X-Y}$ returns $\fls$, it also returns the subset $\phi'$ of the set $\phi-X-Y$ such that traces from states in this subset happen to be identical in $M_1$ and $N$. Since these states do not satisfy $\eglobally{\phi}$ in $N$, it follows that they do not satisfy $\eglobally{\phi}$ in $M_1$ either. Therefore, they are added to the set $Y$ (see line~\ref{line:growYb} in the algorithm). 

\subsection{A variant of our algorithm -- \textbf{Variant X}}
\label{ssec:global:variantX}

At this point we would like to introduce a variant of Algorithm~\ref{algo:compute-global}, which we call~\textbf{Variant X}. This variant is the same as the full algorithm, but \emph{excluding} the logic in lines~\ref{line:growY}-\ref{line:endY} of the algorithm (we have not yet discussed the working of these lines in detail). 

A few noteworthy points about this variant are as follows:

\begin{itemize}
\item This variant is basically very similar to the ``under-approximation component'' that was alluded to earlier in Section~\ref{ssec:intro:globally}. This under-approximation component was originally presented as a standalone algorithm in our previous conference paper~\cite{kvasantafme2014}. The improvement is that the $\checkflatteningName$ subroutine now returns a set of states that can be added to $Y$. The version in our previous conference submission returned only the \emph{flag}. This change enables Variant~X to terminate more quickly than the old under-approximation component on certain systems (although the termination class per se is not wider).

\item This variant shares certain basic properties as the full algorithm. In particular, the set $X$ remains an under-approximation of $\eglobally{\phi}$ at all times during the run of the algorithm, and is definitely equal to $\eglobally{\phi}$ whenever the condition in line~\ref{line:term-Under} is true.

\item This variant terminates on a strict subset of systems as the full algorithm  (due to the missing termination check in line~\ref{line:term-Over}).

\item We introduce this variant in this paper even though it has no practical advantages over the full algorithm primarily because we have a way of characterizing a class of systems on which this variant terminates. A similar exercise for the full algorithm remains an open problem.

\item Analogously, there exists another variant of the full algorithm (\textbf{Variant~Y}), which we introduce later, in Section~\ref{ssec:over}. This variant corresponds to the ``over-approximation component'' that was alluded to in Section~\ref{sec:intro}. 

\item The full Algorithm~\ref{algo:compute-global} basically embodies the integration of the two components mentioned above. As was mentioned in Section~\ref{ssec:intro:globally}, this integration is a contribution of this paper over our previous conference version. 
\end{itemize}

\subsection{An illustration}
\label{sssec:under-illus}

We now illustrate Variant~X of our algorithm.
The example that we consider for illustration is  system $M$ in Figure~\ref{fig:example_intro}(a), with the property being $\eglobally{(x < 10)}$. The corresponding refined system $M_1$ is depicted in Figure~\ref{fig:example_intro}(b). Note that $M_1$ is not a flat system. The precise solution to the property above is `$(x \geq  0) \wedge (x < 5)$'. Intuitively, from these states, one can follow infinitely long traces by alternating between transitions $t_0$ and $t_1$. 

At first, 
$Y$ gets initialized in line~\ref{line:init-xy} of the algorithm to the formula `$x < 0 \vee x \geq 10$'. 

\begin{enumerate}
\item Say the very first flattening of $M_1$ that is considered in line~\ref{line:inner-loop-beg} of the algorithm is the flattening depicted in Figure~\ref{fig:example_intro}(c). 
None of the states satisfy $\eglobally{(x < 10)}$ in this flattening. This is because each trace exhibited by this flattening stops when $x$ reaches the value 100. Therefore, $X$ remains the empty set, while $Y$ grows in line~\ref{line:growYb} to `$x < 0 \vee x \geq 5$' (this growth happens because traces from states that satisfy $x\geq 5$ are identical in $N$ and $M_1$).

\item Say the next flattening considered is the one shown in Figure~\ref{fig:example_intro}(d).  This flattening happens to add no new states to $X$ or to $Y$.

\item Say the next flattening that gets generated is the flattening in Figure~\ref{fig:example_intro}(e). In this flattening infinitely long traces are present from all states that satisfy the formula `$(x \geq  0) \wedge (x < 5)$'. Therefore, $X$ becomes equal to this formula (which is  in fact the final precise result). Also, $\phi - X - Y$ is the empty set of states (recall that $\phi$ is $x < 10$). Therefore, the $\checkflatteningName$ check in line~\ref{line:checkfl} returns $\tru$, which causes the algorithm to terminate.
\end{enumerate}



\subsection{Computing $Y$}
\label{ssec:over}

Recall that the set $Y$ is meant to be a growing under-approximation of the
set of concrete states that \emph{do not} satisfy $\eglobally{\phi}$ in
$M_1$. This set is initialized, in line~\ref{line:init-xy} of 
Algorithm~\ref{algo:compute-global}, to contain ``stuck'' states (i.e., states that do not satisfy
the guards of any of the transitions of $M_1$) as well as states that do
not satisfy $\phi$. These states clearly do not satisfy $\eglobally{\phi}$ in $M_1$. As was mentioned earlier, states are added to $Y$ (in
line~\ref{line:growYb}) during the computation of $X$. In this sub-section we focus on the other mechanism in the algorithm to add states to $Y$, namely, the invocation of  subroutine $\computeGlobalOverNew{M_1}{\phi}{Y}$ in line~\ref{line:growY}.


\subsubsection{The working of the subroutine $\computeGlobalOverName$.}
\label{sssec:over-working}

This subroutine takes as inputs a counter system $M_1$, a set of states $\phi$, and a set of states $\Yinit$. It is assumed that $\Yinit$ is some under-approximation of  the states that do not satisfy $\eglobally{\phi}$ in $M_1$ (e.g., $\Yinit$ is also allowed to be $\emptyset$). The subroutine also uses a user-given parameter $k_2 \geq 1$.  If $\Yinit$ does not contain all states that do not satisfy $\eglobally{\phi}$ in $M_1$, then the subroutine returns a strict superset $Y$ of $\Yinit$ such that $Y$ is also an under-approximation of the set of states that do not satisfy $\eglobally{\phi}$ in $M_1$; else, the subroutine returns $\Yinit$ itself. 

The property mentioned above is made use of in the  second termination condition in line~\ref{line:term-Over} of Algorithm~\ref{algo:compute-global} (recall that the first termination condition for that algorithm was in line~\ref{line:term-Under}). That is, if the set returned by $\computeGlobalOverName$ is equal to the set $Y$ given to it, then $Y$ must already be equal to $\neg \eglobally{\phi}$; therefore Algorithm~\ref{algo:compute-global} returns `$\neg Y$' as the precise answer (see line~\ref{line:return-Over}). 

\begin{algorithm}
 \begin{algorithmic}[1]
	 \REQUIRE {A system $M_1$, a set of states $\phi$, and a set of states $\Yinit$ that is an under-approximation of the set of states that do not satisfy $\eglobally{\phi}$ in $M_1$. It is assumed that the transitions of $M_1$ are refined wrt $\phi$.}
\ENSURE {Returns a set of states $Y$. If $\Yinit$ is a strict under-approximation of $\eglobally{\phi}$, then $Y \supset \Yinit$ and $Y$ under-approximates $\eglobally{\phi}$, else $Y = \Yinit$.}
	\STATE $Y = \Yinit$, count = 1.\label{line:addYinit}
	\WHILE {($(\growone \vee \growtwo)$ is satisfiable) $\wedge$ count $\neq k_2$}\label{line:over-loop-beg}
		\STATE  $Y = Y \vee (\growone \vee \growtwo)$\label{line:over-growY}
		\STATE count = count+1
	\ENDWHILE\label{line:over-loop-end}
	\RETURN $Y$
\end{algorithmic}
\caption {Algorithm $\computeGlobalOverName$}
\label{algo:over-approx}

\end{algorithm} 

Subroutine $\computeGlobalOverName$ is depicted as Algorithm~\ref{algo:over-approx}. This subroutine basically resembles the classical approach
for solving $\eg$ properties for finite-state systems~\cite{book_clarke},
but uses reachability analysis as a black-box to accelerate the process of
adding states to $Y$. A state does not satisfy $\eglobally{\phi}$ in $M_1$ iff all traces starting from it are finite.
Therefore, the subroutine starts by initializing the set $Y$ to the given set $\Yinit$ of states that are known not to satisfy $\eglobally{\phi}$ (see line~\ref{line:addYinit} in Algorithm~\ref{algo:over-approx}).
Subsequently, in each iteration of its loop (lines~\ref{line:over-loop-beg}-\ref{line:over-loop-end}), 
the subroutine identifies states that do not
satisfy $\eglobally{\phi}$ in $M_1$, using two different conditions, namely $\growone$ and $\growtwo$,  and adds them to $Y$. These two conditions are discussed in detail below. The loop iterates at most $k_2$ times. However, the loop may also terminate earlier; in particular, if in some iteration neither one of the two conditions mentioned above adds any states to $Y$, it then follows that $Y$ is already equal to $\neg \eglobally{\phi}$, thus allowing for intermediate termination.


\emph{Condition} $\growone$:
If all successors of a state $\vec{s}$ are in $Y$ then $\vec{s}$ can be
added to
$Y$. The states that satisfy this property can be identified using the 
following Presburger formula:

\rule[1.5em]{0in}{0in}\rule[-1em]{0in}{0in}%
$\growone$ \ $\equiv$ \ $(\forall s^\prime.\, ((g_1 \wedge f_1) \vee (g_2 \wedge f_2) \vee \dots (g_n \wedge f_n) \implies Y[s^\prime/ s])) - Y$

Recall that $g_i, f_i$ are the guard and action of
transition $i$, respectively. Note that $Y$ as well as the $g_i$'s use the counter names as free variables, while the $f_i$'s use both the unprimed as well as the primed versions of the counter names as free variables. Let $s$ denote the (vector of) names of the counter variables, and $s'$ denote the (vector of) names of the primed versions of the counter variables. Therefore, $Y[s^\prime/s]$ denotes the variant of $Y$
wherein each counter names is replaced with its corresponding primed version. Due to the quantification over $s'$, the formula above involves only the unprimed counter names as free variables. 

Intuitively, the formula above is saying that a state $\vec{s}$ satisfies $\growone$ iff (a) for each state $\vec{s'}$, if $\vec{s}$ satisfies any $g_i$ and if ($\vec{s}, \vec{s'})$ satisfies the corresponding $f_i$ (that is, $\vec{s'}$ is a successor state of $\vec{s}$), then $s'$ already satisfies $Y$, and (b) $\vec{s}$ does not already satisfy $Y$.

Note that the condition $\growone$ is identical to the one used in the classical finite-state algorithm.

\begin{figure}
\centering
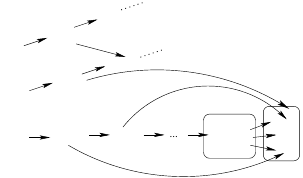
		\caption{Illustration of formula $\growtwo$. The shaded region represents states that satisfy $\growtwo$.}
		\label{over_approx:single_trace2}
\end{figure}

\emph{Condition} $\growtwo$: Ignoring all concrete transitions whose target state is
already in $Y$, if a state $\vec{s}$ is such that (a) there is only one
trace $t$ in $M_1$ starting from $\vec{s}$ (not counting prefixes of this
trace $t$), and (b) $t$ reaches a state that satisfies $\growone$ after a
finite number of steps, then $\vec{s}$ can be added to $Y$. This condition differs from the one used in the classical finite-state algorithm, and uses accelerations to grow more quickly the set of states that do not satisfy $\eglobally{\phi}$. 

In the
illustration in Figure~\ref{over_approx:single_trace2}, states $\vec{s},
\vec{s_1}, \vec{s_2}$, etc., (which are within the shaded region) satisfy both sub-conditions~(a) and~(b) mentioned
above; 
state
$\vec{s_{11}}$ satisfies only sub-condition~(a), while state $\vec{s_{21}}$
  satisfies neither of the two sub-conditions.

The states that satisfy sub-condition~(a) can be identified using the
following Presburger formula:

$$\growtwoa \equiv \neg (\preStar{M_1}{ \neg \asoY})$$

where $\asoY$ represents the states that have at most one successor state
that is not already in $Y$. Therefore, $\neg \asoY$ represents states that
have two
or more successors  outside $Y$. Therefore, the transitive predecessors of these
states are the ones that \emph{don't} belong to $\growtwoa$.


The formula for $\asoY$ is as follows:

\begin{gather*}
    \asoY \equiv\\ 
      \forall i, j \in \Sigma. \, ((g_i \wedge f_i) \wedge (g_j \wedge
    f_j[s^{\prime\prime}/ s^\prime]) \implies\\ 
    (s^\prime = s ^{\prime\prime}) \vee (Y[s^\prime/s] \wedge
    Y[s^{\prime\prime} / s]) \vee \\
    (Y[s^\prime/ s] \wedge \neg Y[s^{\prime\prime}/ s]) \vee  
    (\neg Y[s^\prime/ s] \wedge Y[s^{\prime\prime}/ s]))
\end{gather*}

Intuitively, the part before the `$\implies$' identifies pairs of
successor states $(\vec{s^\prime},\vec{s^{\prime\prime}})$ of the state
$\vec{s}$ under consideration, while the part after the
`$\implies$' requires that $\vec{s'}$ and $\vec{s''}$ be the same state, or that at least
one of them be already in $Y$.

Now, sub-condition~(b) above is captured by the following formula:

$$\growtwob \equiv \preStar{M_1}{\growone}$$

Therefore, the states to be in
added to $Y$ by \emph{Condition 2} are described by the formula $\growtwo
\equiv \growtwoa \wedge \growtwob$.

\subsubsection{Accommodating approximated reachability anaylsis.}

Note that the $\preStarName$ computations referred to above in the context of the formulas $\growtwoa$ and $\growtwob$ need to be performed by the black-box routine $\computePreStarName$, which was introduced in Section~\ref{ssec:until:approach}, that is expected to be provided by reachability analysis tools. Reachability analysis tools that provide this black box need not always terminate with precise results. If forcibly stopped during their execution, some of them may be capable of returning approximate results. Our approach mentioned above can be modified to work with approximate reachability results, while still ensuring that the subroutine $\computeGlobalOverName$ fulfills its contract as specified at the beginning of Section~\ref{ssec:over}, as follows. 
In the context of the formula $\growtwoa$, to compute $\preStar{M_1}{\neg \asoY}$, we would need to invoke $\computePreStar{M_1}{\neg \asoY}{\overapprox}$, and use its return value (whether it is precise or over-approximated). On the other hand, in the context of the formula $\growtwob$, to compute $\preStar{M_1}{\growone}$, we would need to invoke $\computePreStar{M_1}{\growone}{\under}$, and use its return value (whether it is precise or under-approximated).

In case we are using a black-box that is not capable of providing approximate results upon forcible termination, then if the black-box does not terminate in a given iteration of the loop in $\computeGlobalOverName$ (in the context of the computation of either $\growtwoa$ or $\growtwob$), then in that iteration we can fall back upon using only $\growone$ (which needs no reachability analysis). 

\subsection{Another variant of our algorithm -- \textbf{Variant Y}}
\label{ssec:global:variantY}

Analogous to Section~\ref{ssec:global:variantX}, we now introduce a variant of Algorithm~\ref{algo:compute-global} which corresponds to the ``over-approximation component'' that was introduced in Section~\ref{ssec:intro:globally}. This variant, which we call Variant~Y, incorporates several modifications to the full Algorithm~\ref{algo:compute-global}: (1) The main loop in Algorithm~\ref{algo:compute-global} executes for only one iteration. (2) Lines~\ref{line:outer-loop-beg}-~\ref{line:inner-loop-end} are skipped. (The current variant is called Variant~Y because the lines mentioned above are responsible for growing the set $X$.)  (3)
Therefore, the value for $Y$ that is passed to subroutine $\computeGlobalOverName$ in line~\ref{line:growY} is the value to which $Y$ gets initialized in line~\ref{line:init-xy}. (4) The check `count $\neq k_2$' in subroutine $\computeGlobalOverName$ is omitted. Therefore, the loop in this subroutine iterates until $Y$ stops growing. (5) The ``return'' statement in line~\ref{line:return-Over} is executed unconditionally.

\begin{itemize}
\item This variant is basically identical to the ``over-approximation component'' that was introduced earlier in Section~\ref{ssec:intro:globally}. This over-approximation component was originally presented as a standalone algorithm in our previous conference paper~\cite{kvasantafme2014}.

\item This variant shares certain basic properties as the full algorithm. In particular, the set $Y$ remains at all times during the run an under-approximation of the set of states that do not satisfy $\eglobally{\phi}$.  Also, whenever the condition in line~\ref{line:term-Over} is true in this variant, $Y$ would still contain exactly the states that do not satisfy $\eglobally{\phi}$ in $M$. 

\item This variant terminates on a strict subset of systems as the full algorithm. We will discuss the reason for this later in Section~\ref{sec:termination}. 

\item As with Variant~X, we introduce this variant in this paper even though it has no practical advantages over the full algorithm primarily because we have a way of characterizing a class of systems on which this variant terminates. (This class is incomparable with the class of systems on which Variant~X terminates. )
\end{itemize}

\subsection{Illustration of Variant~Y, and illustration of the full algorithm}
\label{sssec:over-illus}

We first illustrate Variant~Y of our algorithm. 
We use the example system $M_1$ in Figure~\ref{fig:example_intro}(b) for the illustration. $Y$ gets initialized in line~\ref{line:init-xy} of the algorithm to the formula `$x < 0 \vee x \geq 10$'. Control then enters subroutine $\computeGlobalOverName$. Here,  $\growone$ gets set to the formula $x = 9$, as this is the only state such that all its successors are in $Y$. $\growtwo$ gets set to $x \geq 5 \wedge x \leq 9$. These are the states from which $x = 9$ is reachable, and from which no state with two successors outside $Y$ are reachable. (Note that each state in the set $x \geq 0 \wedge x < 5$ is such that a state with two successors outside $Y$ is reachable from it. Hence these states do not get included in $\growtwo$.) Therefore, $Y$ grows to $x < 0 \vee x \geq 5$. 
Now, none of the remaining states, i.e., $x \geq 0 \wedge x < 5$, satisfy $\growone$. Therefore, the loop in subroutine $\computeGlobalOverName$ 
terminates, and control returns to Algorithm~\ref{algo:compute-global}. 
Therefore, the final result of $\neg Y$, i.e., $x \geq 0 \wedge x < 5$, is returned from line~\ref{line:term-Over} in Algorithm~\ref{algo:compute-global}. Recall that the same result was returned in the illustration in Section~\ref{sssec:under-illus}.

We now illustrate our full Algorithm~\ref{algo:compute-global}.  Say the parameter $k_1$ (see line~\ref{line:outer-loop-beg} of the algorithm) is set to 1.  $Y$ attains its final value (i.e., $x < 0 \vee x \geq 5$) in the first iteration of the inner loop in lines~\ref{line:inner-loop-beg}-\ref{line:inner-loop-end}.  This happens, as was discussed in the illustration in Section~\ref{ssec:under}, due to the flattening depicted in Figure~\ref{fig:example_intro}(c)). Therefore, when line~\ref{line:growY} is later reached in the first iteration of the outer loop, and control enters subroutine~$\computeGlobalOverName$, the loop in this subroutine (see Algorithm~\ref{algo:over-approx}) is not entered at all. Therefore, control returns to Algorithm~\ref{algo:compute-global}, and execution  terminates at line~\ref{line:return-Over}. It is notable that on this example, with $k_1$ set to 1, the full algorithm explored one flattening in the inner loop in  lines~\ref{line:inner-loop-beg}-\ref{line:inner-loop-end}, whereas Variant~X explored three flattenings (see Section~\ref{sssec:under-illus}). Also, the full algorithm did not enter the loop in subroutine $\computeGlobalOverName$ at all, whereas Variant~Y entered this loop once (see discussion above). In other words, on certain system-property pairs the full algorithm has the potential to do somewhat less work than either variant separately.

\section{Correctness of routine $\computeGlobalName$}
\label{sec:global-correctness}

We formally state and prove the correctness of the routine
$\computeGlobal{M}{\phi}{\enumlabel}$ (i.e., Algorithm~\ref{algo:compute-global}) in this section, via a  lemma and a series of theorems. These results are all stated implicitly in the context of the system
$M_1 \equiv \refineSystem{M}{\phi}$, and characterize properties of this system. The correctness of the algorithm on the given system $M$ follows readily from these results.



\begin{lemma}
\label{lem:grow}
The following properties hold at any point during the execution of Algorithm~\ref{algo:compute-global}:

\begin{enumerate}
\item the set $Y$ is $\postStarName$-\emph{closed}; i.e., every successor of every state in $Y$ is also in $Y$.

\item\label{it:lem:grow:1} A state $\vec{s}$ satisfies formula $\growone$ in the loop in subroutine $\computeGlobalOverName$ iff the state is not in $Y$ and all successors of this state are already in $Y$. 

\item\label{it:lem:grow:2} A state $\vec{s}$ satisfies formula $\growtwo$ in the loop in subroutine $\computeGlobalOverName$ iff:

\begin{enumerate}
\item there is a unique trace $t$ that starts from $\vec{s}$ and ends at a
state that satisfies $\growone$, and
\item no state in this trace is currently in $Y$, and
\item\label{it:outside-t-in-Y}
  every state $\vec{s'}$ that is outside $t$ and that is a successor of 
any state in $t$ is already in $Y$.
\end{enumerate}
\end{enumerate}
\end{lemma}
We give the intuition behind the proof of this theorem. The proof is an
induction on the number of times states are added to $Y$. Initially the set of
states added to $Y$ are stuck states. Then, in each step, the formula $\asoY$
along with $\growone$, ensures that Algorithm~\ref{algo:compute-global} only adds states that along all paths end
up in a state in $Y$. Hence $Y$ remains $\postStarName$ closed.

\begin{theorem}
	\label{thm:computeGlobal-invariant}
	The two sets of states $X$ and $Y$ that are maintained by Algorithm~\ref{algo:compute-global} are such that (a) $X$ is a growing under-approximation of the set of states that satisfy the property $\eglobally{\phi}$ in $M_1$, and (b) $Y$ is a growing under-approximation of the set of states that satisfy $\neg \eglobally{\phi}$ in $M_1$.
\end{theorem}

The proof of the Theorem~\ref{thm:computeGlobal-invariant} is an induction
on the steps of Algorithm~\ref{algo:compute-global}. The states added to $X$ in
Line~\ref{line:growX}, by K\"{o}nig's Lemma, have an infinite path. Note, that
$M_1$ is a refinement of $M$ with respect to $\phi$. Therefore, states with an
infinite path in $M_1$ will satisfy $\eglobally{\phi}$. Therefore any state
added to $X$ will definitely satisfy $\eglobally{\phi}$. Hence $X$ is a growing
under-approximation of the set of states that satisfy $\eglobally{\phi}$.
Addition of states to $Y$ involves a call to subroutine
$\computeGlobalOverName$. If a state $\vec{s}$ satisfies $\growone$ then all successors
of $\vec{s}$ are in $Y$. If $\vec{s}$ satisfies $\growtwo$ then by
Lemma~\ref{lem:grow}, $\vec{s}$ has all traces ending at a state in $Y$.
Therefore in either of the two cases, $Y$ is an under apporximation of the set
of states that satisfy $\neg \eglobally{\phi}$. Detailed proof is provided in
Appendix~\ref{app:proofs}.

\begin{theorem}
\label{thm:combined-correctness}
If Algorithm~\ref{algo:compute-global} terminates on its own and returns a set of states, then this return value is precisely the set of states that satisfy $\eglobally{\phi}$ in the counter system $M_1$.
\end{theorem}
Note that Algorithm~\ref{algo:compute-global} terminates in two cases.
Therefore, we need to prove that, in each case, the returned set of states
precisely represents the set of states that satisfy the property
$\eglobally{\phi}$. We provide an outline of the proof for the purposes of
brevity. Detailed proof is provided in Appendix~\ref{app:proofs}.

Suppose Algorithm~\ref{algo:compute-global} terminates due to condition in
Line~\ref{line:term-Under}. Then $X$ is an under-approximation of the set of
states that satisfies $\eglobally{\phi}$. The termination condition ensures
that from any state that satisfies $\phi$ but not in $X$ or $Y$, the traces in
the system $M_1$ and the flattening $N$ are same. Therefore none of the states
in $\phi-X-Y$ satisfy $\eglobally{\phi}$. Therefore $X$ precisely represents
the set of states that satisfy $\eglobally{\phi}$ in $M_1$.

The other case is when Algorithm~\ref{algo:compute-global} terminates due to
condition in Line~\ref{line:term-Over}. This termination condition becomes true
only when none of states satisfy the formula $\growone$ or $\growtwo$. We show
in Appendix~\ref{app:proofs} that when none of these formula's hold, by
contradiction, $Y$ precisely represents the set of states that do not satisfy
the property $\eglobally{\phi}$. Therfore $\phi - Y$ precisely represents the
set of states that satisfy the property $\eglobally{\phi}$

\begin{theorem}
\label{thm:combined-approximation}
If Algorithm~\ref{algo:compute-global} is forcibly terminated, and returns a pair ($\phi', \enumlabel\in\latticeApprox$), then $\phi'$ is an approximation of $\eglobally{\phi}$ in the direction indicated by $\enumlabel$. 
\end{theorem}
\begin{proof}
    The algorithm returns a result upon forcible termination either at
    line~\ref{line:return-Under-force} or at line~\ref{line:return-Over-force}
    of Algorithm~\ref{algo:compute-global}. In the first scenario, the desired
    result follows immediately, because as per
    Theorem~\ref{thm:computeGlobal-invariant}, $X$ is an under-approximation of
    $\eglobally{\phi}$. In the second scenario, since $\phi$ is an
    over-approximation of $\eglobally{\phi}$, and since by
    Theorem~\ref{thm:computeGlobal-invariant} $Y$ is an under-approximation
    $\neg \eglobally{\phi}$, it follows that $\phi - Y$ is an
    over-approximation of $\eglobally{\phi}$.
\end{proof}
Note that the set of states that satisfy $\eglobally{\phi}$ is the same in $M_1$ as in $M$ (this was discussed at the beginning of this section). Therefore, even though Theorems~\ref{thm:computeGlobal-invariant},~\ref{thm:combined-correctness}, and~\ref{thm:combined-approximation} were proved in the context of the system $M_1$, they apply to the given system $M$ as well.



\section{Termination of the routine $\computeGlobalName$}
\label{sec:termination}

\sloppypar In this section we describe the termination characteristics of
the routine $\computeGlobalName$ (i.e.,  Algorithm~\ref{algo:compute-global}). Defining a single logically coherent termination class for this algorithm is challenging. Therefore, in Section~\ref{ssec:over-term} below, we
consider only counter systems such that (a) the action of each transition is deterministic, and (b) all $\mathit{pre}^*$
queries (i.e., reachability queries) on the refined system $M_1$ terminate.
We show the existence of a class of systems that satisfy the two restrictions above on which  Algorithm~\ref{algo:compute-global}
necessarily terminates.
Later, in Section~\ref{ssec:under-term}, we show that the algorithm also terminates on a class of systems  that may have potentially non-deterministic actions (but with finite branching) and on which
reachability queries may not terminate. It is noteworthy that determinism of actions is a stronger property than finite branching.

\subsection{Termination on systems $M_1$ that have deterministic actions and on which $\mathit{pre}^*$ queries
terminate} 
\label{ssec:over-term}

In the analysis that we employ in this section, we use Variant~Y of our algorithm, which we had introduced in Section~\ref{ssec:global:variantY}.  Recall that this variant consists simply of invoking subroutine $\computeGlobalOverName$ once, with no apriori upper bound on the number of iterations executed by the loop that is within this subroutine.
For the purposes of this proof we do assume one more change to the algorithm on top of Variant~Y, whose intent is to enable a more meaningful termination characterization. This change is that in each iteration of the loop in subroutine $\computeGlobalOverName$, in addition to the states added to $Y$ in line~\ref{line:over-growY}, some additional non-deterministically chosen subset of states that do not satisfy $\eglobally{\phi}$ are added to $Y$. This is change is meant to simulate the potential addition of states to $Y$ in line~\ref{line:growYb} in the (full) Algorithm~\ref{algo:compute-global}.

Since our plan in this section is to show a termination class for Variant~Y (with the change mentioned above) and then claim that this termination class is also valid for the full algorithm, we first need to show that if Variant~Y terminates on an input instance (i.e., a system $M$ and property $\phi$), then the full algorithm would also terminate on the same instance.
The argument for this claim is as follows:

\begin{itemize}
\item The steps performed in each iteration of the outer loop in algorithm   (excluding the call to the subroutine $\computeGlobalOverName$) are   always terminating.

In particular, the two non-trivial steps in the outer   loop are in lines~\ref{line:growX}   and~\ref{line:checkfl}.

Line~\ref{line:growX} will terminate because of   an assumption we make that $\prekName$ queries on flat systems always   terminate. Note that  with reachability approaches such as   Fast~\cite{finkel02}, termination of $\preStarName$ queries on non-flat systems is predicated on  termination of $\prekName$ queries on flat systems. 

Line~\ref{line:checkfl} will terminate because of another assumption we make, which is that any call to $\checkflatteningName$ always terminates. Again, this assumption holds on the class of counter systems addressed by Fast.

\item If the algorithm can be shown to terminate when states are added to $Y$ non-deterministically within the loop in subroutine $\computeGlobalOverName$, clearly it will also terminate if states are added to $Y$ deterministically in line~\ref{line:growYb} in between successive calls to subroutine $\computeGlobalOverName$. 
\end{itemize}

We present a few definitions before we proceed with our proofs. 

\begin{definition}
Given a flat counter system $N$, 
we  define a
function $\mathit{order_{N}}$ thats maps every control state $q \in Q$ in
$N$ to a natural number, as per the following rules:
	\begin{description}
		\item{Rule 1.}\label{order:no-succs}
          $\order{{N}}{(q)} = 0$, if there is no outgoing transition
          from $q$.

        \item{Rule 2.}\label{order:succ-0} $\order{{N}}{(q)} = 1$, if $q$
          is not part of any cycle, and $q$ has a single control-state
          successor of order 0.

        \item{Rule 3.}\label{order:terminal-cycle} $\order{{N}}{(q)} =
          1$, if $q$ is in a cycle $c$ and no control state in cycle $c$
          has a successor outside cycle $c$.  (Note that every control
          state in this cycle will have order 1.)

        \item{Rule 4.}\label{order:single-succ} $\order{{N}}{(q)} = n$,
          if $q$ is not part of any cycle, $q$ has only one successor
          control state $q_1$, $\order{{N}}{(q_1)} = n$, and $n > 0$.

        \item{Rule 5.}\label{order:mult-succ} $\order{{N}}{(q)} = n + 1$,
          if $q$ is not part of any cycle, $q$ has multiple successor
          nodes, the maximum value of the $\mathit{order}$ of any successor
          control-state of $q$ is $n$.

        \item{Rule 6.}\label{order:interm-cycle} $\order{{N}}{(q)} = n +
          1$, if $q$ is part of a cycle $c$, and the maximum value of the
          \emph{order} of any node $q_1$ that is not part of the cycle but
          has a predecessor in cycle $c$ is $n$.  (Note that every control
          state in the cycle will have same value for $\mathit{order}$.)
	\end{description}
\end{definition}

It is easy to see that for any flat system $N$ 
there exists a unique integer $m$ and a unique assignment of orders from the range $[1,m]$ to the control-states in $N$ such that all the rules mentioned above are satisfied. 

We say that  concrete state $\vec{s}$ \emph{corresponds} to a control-state $q$ if the first element of the vector that represents $\vec{s}$ (i.e., the control-state of $\vec{s}$) is $q$.

In the remainder of this section we state and prove two theorems, each of which provides a termination class for our approach. The first class is the class of flat systems.
The second class  is a much broader class, which has a more complex characterization.
We structure our arguments this way because it allows the proofs to be natural and intuitive.  

In Theorem~\ref{thm:flat_systems}, we prove the termination
of Algorithm~\ref{algo:compute-global}.
\begin{theorem}
	\label{thm:flat_systems}
	Given a flat counter system $M$ and any property $\eglobally{\phi}$, Algorithm~\ref{algo:compute-global} will always terminate, provided: (a) All $\preStarName$ queries on $M_1$ terminate and all $\prekName$ queries on flattenings of $M_1$ terminate, (b) The action $f_b$ of each transition $b$ of $M$ is \emph{deterministic}, in the sense that it maps each state $\vec{s}$ to a unique state $\vec{s'}$, and (c) the check $\checkflattening{M_1}{N}{\phi-X-Y}$ always terminates. 
\end{theorem}
The proof of this theorem is given in Appendix~\ref{app:proofs}.

The following proves the termination of Algorithm~\ref{algo:compute-global} on
trace flattable systems.
\begin{theorem}
	\label{thm:over_approx_characterization}
	Given any property $\eglobally{\phi}$, Algorithm~\ref{algo:compute-global} will terminate on a system $M$ if:

    \begin{enumerate}
	\item all $\mathit{pre}^*$ queries on the system $M_1$ terminate, where
$M_1$ is  a refinement of $M$ with respect to $\phi$

    \item the action $f_b$ of each transition $b$ of $M$ is \emph{deterministic}, and

    \item there exists an integer bound $m$, and a (finite or infinite) set
of flattenings $L$ of $M_1$, such that (a) $\order{N}{(q)}$ is at most $m$,  for each control state $q$ of each flattening $N$ in $L$, and (b) for each state $\vec{s}$ that does not satisfy $\eglobally{\phi}$ in $M_1$, there exists a flattening $N$ in $L$ such that all traces from $\vec{s}$ in $M_1$ are preserved in $N$. 
    \end{enumerate}
    \end{theorem}

\begin{proof}
Let $\vec{s}$ be any state that does not satisfy $\eglobally{\phi}$ in $M_1$. Let $N$ be the flattening in $L$ such that all traces from $\vec{s}$ in $M_1$ are preserved in $N$. That is, there exists a state $\vec{s'}$ of $N$ that is a ``trace preserving copy'' of $\vec{s}$ in $M_1$ (see Section~\ref{ssec:prel:traces-flattenings}). Clearly, $\vec{s'}$ does not satisfy $\eglobally{\phi}$ in system $N$, either. As mentioned at the beginning of Section~\ref{ssec:over-term}, we will analyze the termination of Algorithm~\ref{algo:compute-global} by analyzing only the termination of subroutine $\computeGlobalOverName$. Consider an application of subroutine~$\computeGlobalOverName$ on the system $N$. In what follows we will prove that $\vec{s}$ would get added to the set $Y$ (of states that do not satisfy $\eglobally{\phi}$) when the subroutine is applied on the system $M_1$ within the same number of iterations as is needed to add  state $\vec{s'}$ to $Y$ when the subroutine is applied on $N$. Since the control states in the flattenings in $L$ have ``order'' at most $m$ (as per the assumption of this theorem), and since the subroutine terminates on any flat system $N$ within at most as many iterations as the ``order'' of the control state that has the highest order (as per Theorem~\ref{thm:flat_systems}), it would follow that the subroutine terminates on $M_1$ within at most $m$ iterations overall.

We now prove the following property by mathematical induction.\\

	$\textbf{P(n)}:$ Let $\vec{s}$ be any state that does not satisfy $\eglobally{\phi}$ in $M_1$. Let $N$ be the flattening in $L$, and let state $\vec{s'}$ in $N$ be a trace preserving copy of $\vec{s}$.  If state $\vec{s'}$ gets added to $Y$ in the $n$th iteration when the subroutine $\computeGlobalOverName$ is applied on the flat system $N$, then $\vec{s}$ would get added to $Y$ within $n$ iterations when the subroutine is applied on the system $M_1$. \\
    
	\textbf{Base Case(n=0):} This happens when $\vec{s'}$ is a stuck state in $N$. Since all traces from $\vec{s}$ in $M_1$ are preserved from $\vec{s'}$ in $N$, it follows that $\vec{s}$ is a stuck state in $M_1$ also. Therefore, the subroutine would add $\vec{s}$ to $Y$ before the first iteration when it is applied on $M_1$. \\

	\textbf{Inductive case:} There are two scenarios here. The first scenario is that $\vec{s'}$ is added to $Y$ in the $n$th iteration by formula $\growone$ when the subroutine is applied on the flattening $N$. This can happen only if all successors of $\vec{s'}$ are added to $Y$ in fewer than $n$ iterations in the same application of the subroutine. Since $\vec{s'}$ is a trace preserving copy of $\vec{s}$, it follows that each successor of $\vec{s}$ in $M_1$ has a trace-preserving copy in $N$ that is a successor of $\vec{s'}$.  Therefore, by the inductive hypothesis, all successor states of $\vec{s}$ would have been added to $Y$ in fewer than $n$ iterations when the subroutine is applied on $M_1$ also. Therefore, in the $n$th iteration of this application, formula $\growone$ would add $\vec{s}$ to $Y$ (if $\vec{s}$ is not already in $Y$). 

The second scenario is that  $\vec{s'}$ is added to $Y$ in the $n$th iteration by formula $\growtwo$ when the subroutine is applied on the flattening $N$. Therefore, from Lemma~\ref{lem:grow}, it follows that ignoring states that are reachable from $\vec{s'}$ that were added to $Y$ in fewer than $n$ iterations, there is a single outgoing trace $t'$ from $\vec{s'}$ in $N$ which ends in a state $\vec{s'_k}$ that satisfies $\growone$ in the $n$th iteration. Furthermore, it follows from the definition of $\growone$ that all successors of the state $\vec{s'_k}$ would have been added to $Y$ in fewer than $n$ iterations. Let \emph{inY} be the set of states of $N$ that are outside $t'$ but are immediate successors of states in $t'$.

(1) From the observations above, it follows that all states in \emph{inY} would have been added to $Y$ in fewer than $n$ iterations when the subroutine is applied on $N$.

(2) Let $t=f(t')$ be the trace in $M$ of which $t'$ is a copy. Because $\vec{s'}$ is a trace-preserving copy of $\vec{s}$, it follows that no state $\vec{s_1}$ in $t$ can have a successor state $\vec{s_2}$ outside $t$ such that the copy $\vec{s'_1}$ of $\vec{s_1}$ in $t'$ is missing a concrete transition to a copy $\vec{s'_2}$ of $\vec{s_2}$.

(3) In other words, every state of $M$ that is outside $t$ and that is a successor of some state in $t$ has a copy in $N$ that is in the set \emph{inY}. Therefore, due to Point~(1) above and due to the inductive hypothesis, it follows that all these successor states outside $t$ would be added to $Y$ in fewer than $n$ iterations when subroutine $\computeGlobalOverName$ is applied on $M_1$.

(4) From the observation above, and from Lemma~\ref{lem:grow}, it follows that all states in $t$ (including $\vec{s}$) would get added to $Y$ within $n$ iterations of the loop in this subroutine when this subroutine is applied on $M_1$.  \\
\end{proof}


The previous approach of Demri et al.~\cite{dem2006} is able to solve the global model checking problem for a fragment of CTL* properties. The approach addresses the class of counter systems such that (a) there exists a \emph{trace flattening} of each system wrt the initial set of states $\phiinit$, (b) all states that are not reachable from states in $\phiinit$ are stuck states, (c) $\preStarName$ as well as $\prekName$ queries terminate on all flattenings, and (d) the check $\checkflatteningName$ is decidable. The following corollary of Theorem~\ref{thm:over_approx_characterization} above establishes a relationship between the termination class of our algorithm $\computeGlobalOverName$ and the termination class of the approach of Demri et al.

\begin{corollary}
  If the refined system $M_1$ is in the class (mentioned above) that is addressed by the approach of Demri et al., \emph{and} if the action of each transition of $M_1$ is deterministic, then Algorithm~\ref{algo:compute-global} would terminate on $M_1$.
\end{corollary}

Our approach actually terminates on many natural counter systems that are outside the class addressed by Demri et al. This is discussed further in Sections~\ref{ssec:term:discussion} and~\ref{ssec:implementation}.  Also noteworthy is that the systems addressed by Fast have deterministic transitions only. 

\subsection{Termination on systems on which $\mathit{pre}^*$ queries may
not terminate and whose transitions may have non-deterministic actions (with finite branching)}
\label{ssec:under-term}

In the analysis in this section, we use Variant~X of our algorithm, which we had introduced in Section~\ref{ssec:global:variantX}. Since this variant  does \emph{not} invoke $\preStarName$ queries, it could terminate on some systems on which $\preStarName$ queries do not terminate.

For the purposes of this proof we do assume one more change to the algorithm on top of Variant~X, whose intent is to enable a more meaningful termination characterization.
This change is that in each iteration of outer loop in Algorithm~\ref{algo:compute-global}, in place of lines~\ref{line:growY}-\ref{line:endY} (which are anyway not included in Variant~X),  some  non-deterministically chosen subset of states that do not satisfy $\eglobally{\phi}$ are added to $Y$. This is to simulate the potential addition of states to $Y$ in line~\ref{line:growY} in the (full) Algorithm~\ref{algo:compute-global}.

Since our plan in this section is to show a termination class for Variant~X (with the change mentioned above) and then claim that this termination class is also valid for the full algorithm, we first need to show that if Variant~X terminates on an input instance, then the full algorithm would also terminate on the same instance. This result is easy to see provided the full algorithm is restructured slightly, as follows:

\begin{itemize}
\item The first time line~\ref{line:growY} in Algorithm~\ref{algo:compute-global} is visited, subroutine~$\computeGlobalOverName$ is forked off and run in a separate thread, with the main thread transferring immediately   to line~\ref{line:outer-loop-beg} to continue exploring flattenings of $M_1$. This way, any non-termination in the $\preStarName$ operations in subroutine~$\computeGlobalOverName$ does not prevent the outer loop from continuously exploring flattenings  of $M_1$.

\item During any subsequent visit to line~\ref{line:growY}, the main thread   checks whether the most recent child thread that was forked during a   previous visit to this location has completed. If yes, the main thread   forks off the next invocation of subroutine~$\computeGlobalOverName$ as a   separate thread, adds the returned set of states from the completed   previous invocation of subroutine~$\computeGlobalOverName$ to $Y$   (instead of just copying these states over to $Y$), and then continues to   line~\ref{line:term-Over}. If not, the main thread transfers back   immediately to line~\ref{line:outer-loop-beg}.
\end{itemize}

With the changes mentioned above, it is easy to see that if Variant~X is able to reach the ``return'' statement in line~\ref{line:return-Under} on any input instance, then the full algorithm would be able to reach this statement also on the same input instance. 


\begin{theorem}
  \label{thm:under_approx_characterization}
    Let $M$ be the given counter system, $\eglobally{\phi}$ be the given
property to check, and  $M_1$ be the refinement of $M$ with respect to 
$\phi$.	Algorithm~\ref{algo:compute-global} is guaranteed to terminate if:

\begin{enumerate}
	\item\label{it:under:prek} $\mathit{pre}^k$ queries terminate on all
flattenings of $M_1$ that the algorithm generates, 

    \item There exists a (finite) set $L$ of flattenings of $M_1$ such that:

      \begin{enumerate}
      \item\label{it:under:X} For each state $\vec{s_1}$ that satisfies $\eglobally{\phi}$ in $M_1$, there exists a flattening $N$ in $L$ and a state $\vec{s'_1}$ in $N$ that is a copy of $\vec{s_1}$ such that  there is at least one infinitely long trace from $\vec{s'_1}$ in $N$. 

      \item\label{it:under:Y} For each state $\vec{s_2}$ that does not satisfy $\eglobally{\phi}$ in $M_1$, there exists a flattening $N$ in $L$ and a state $\vec{s'_2}$ in $N$ that is a copy of $\vec{s_2}$ such that $\vec{s'_2}$ is a trace-preserving copy of $\vec{s_2}$.
      \end{enumerate}

    \item\label{it:under:fl} Calls to subroutine $\checkflatteningName$ in the algorithm always terminate.

\end{enumerate}
\end{theorem}

As mentioned earlier in this section, we will prove the result above (which describes a property of the full Algorithm~\ref{algo:compute-global}) by actually analyzing Variant~X (with the change mentioned earlier in this section about non-deterministically adding states to $Y$).

\begin{proof}
  
The algorithm will eventually generate all the flattenings in the set $L$. This is because the algorithm  enumerates all
flattenings of $M_1$ in a systematic manner (i.e., in increasing order of
the \emph{length} of the flattenings). 

The algorithm does not go into non-termination during the processing of any flattening that it enumerates.  This is because (a)
the $\mathit{pre}^k$ queries (in line~\ref{line:growX} of the algorithm)
have been assumed to be terminating, and (b) the
$\checkflattening{M_1}{N}{\phi-X-Y}$ operation in line~\ref{line:checkfl}
is assumed to be always terminating.

When a  flattening $N$ of $M_1$ is generated, for each state $\vec{s'_1}$ that satisfies $\eglobally{\phi}$ in $N$, the state $f(\vec{s'_1})$ of $M_1$ gets added to the set $X$  (see line~\ref{line:growX} in Algorithm~\ref{algo:compute-global}). Also,  each state $\vec{s_2}$ of $M_1$, such that there exists a state $\vec{s'_2}$ in $N$ that is a copy of $\vec{s_2}$ and that does not satisfy $\eglobally{\phi}$ in $N$, gets added to the set $Y$ (see line~\ref{line:growYb} in Algorithm~\ref{algo:compute-global}).  Therefore, by the time all flattenings in $L$ have been enumerated and processed,  by the assumption on $L$ as stated in the theorem's statement, it follows that $X$ contains all states that satisfy $\eglobally{\phi}$ in $M_1$, and $Y$ contains all other states. Note that the non-deterministic addition of states that do not satisfy $\eglobally{\phi}$ in $M_1$ to $X$ (as discussed in the beginning of this section) does not affect the observation made above. 
Therefore, after all flattenings in $L$ have been processed, in the following iteration of the loop in lines~\ref{line:inner-loop-beg}-\ref{line:inner-loop-end}, $X$ would not grow at all, while call in line~\ref{line:checkfl} will return $\tru$ (because $\phi -X -Y$ would be the empty set). Therefore, the algorithm would terminate.$\Box$
\\

\end{proof}

As discussed in Section~\ref{ssec:over-term}, the CTL* model checking approach of Demri et al.~\cite{dem2006} addresses the class of counter systems such that (a) there exists a \emph{trace flattening} of each system wrt the initial set of states $\phiinit$, (b) all states that are not reachable from states in $\phiinit$ are stuck states, (c) $\preStarName$ as well as $\prekName$ queries terminate on all flattenings, and (d) the check $\checkflatteningName$ is decidable. The following corollary of Theorem~\ref{thm:under_approx_characterization} above establishes a relationship between the termination class of the variant of Algorithm~\ref{algo:compute-global} introduced above and the termination class of the approach of Demri et al.

\begin{corollary}
  If the refined system $M_1$ is in the class (mentioned above) that is addressed by the approach of Demri et al., \emph{and} if the system $M_1$ exhibits finite branching,  then the variant of Algorithm~\ref{algo:compute-global} introduced above would terminate on $M_1$ with the precise result.
\end{corollary}

\subsection{Comparative discussion on termination}
\label{ssec:term:discussion}


In this section we compare and contrast the termination characteristics of
the various variants of our algorithm, namely, Variant~X, Variant~Y, as
well as the ``full routine'' $\computeGlobalName$
(Algorithm~\ref{algo:compute-global}), using illustrative examples.

\subsubsection{Systems where Variant X does better.}

Variant~X is only applicable on systems that exhibit finite branching (it
can incorrectly produce over-approximated solutions upon termination on
systems with infinite branching). However, within the class of systems that
satisfy this restriction, there are systems with non-deterministic actions
on which Variant~X terminates while Variant~Y does not.

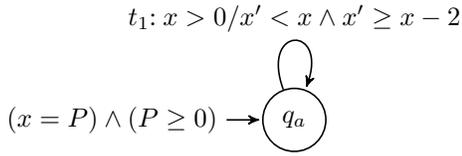
\begin{figure}
    \begin{tikzpicture}[->,>=stealth',shorten >=1pt,auto,
        semithick,/tikz/initial text=, transform shape]
        \tikzstyle{every state}=[fill=white,draw=black,text=black]
        \node[state,initial, initial text=${(x = P) \wedge (P \geq 0)}$] (A)  {$q_a$};
        \path (A) edge [out =110, in=70,loop] node {$t_1$: $x>0 / x^\prime < x \wedge x^\prime \geq x -2$} (A);
    \end{tikzpicture}
    \caption{Example where Variant X alone terminates}
    \label{fig:under-alone-terminates}
\end{figure}

For instance, consider the system shown in
Figure~\ref{fig:under-alone-terminates}. $P$ is a \emph{parameter} to the
system; what we mean by a parameter is a counter that is never modified by
the system. 
The action of the sole transition in this system is
non-deterministic. Let us consider the property $\eglobally{(x >
  0)}$. The traces from any state $x=k$ in this system are at most $k$
steps long. In other words, none of the states satisfy the given
property. Since the system is flat, Variant~X terminates on it within one
iteration itself. Variant~Y, on the other than, does not terminate at all
on this example. At the beginning, the stuck states, i.e., $x \leq 0$, are
added to $Y$. In the first iteration of the loop $x=1$ satisfies
$\growone$, while $x=2$ satisfies $\growtwo$. $x=3$ and other states do not
satisfy $\growtwo$, because each from these states two states that are not
in $Y$ are reachable in one step. In the second iteration, $x=3$ and $x=4$
get added to $Y$. And so on. Therefore, all states $x \geq 0$ will not get
added to $Y$ in a finite number of iterations.

In other words, there exist systems with non-deterministic transitions on
which Variant~X terminates but Variant~Y does not.

\subsubsection{Systems where Variant Y does better.}

We begin this section with a note that Variant~Y is correct (i.e., produces
precise solutions upon termination) even on systems with infinite
branching. However, since Variant~X as well as the full routine
$\computeGlobalName$ are correct only on systems with finite branching, for
convenience throughout this paper we assume that any given system $M$ has
finite branching.

We now make a useful observation about Variant~Y: When
attention is restricted to systems that have deterministic transitions and
on which all $\preStarName$ queries terminate, Variant~Y terminates on
\emph{each} system-property pair that Variant~X terminates on. The argument for this is
easy: The conditions stated in the statement of
Theorem~\ref{thm:under_approx_characterization} together with the two
restrictions stated above clearly imply all the conditions that appear in
the statement of Theorem~\ref{thm:over_approx_characterization}.

\begin{figure}
\begin{tikzpicture}[->,>=stealth',shorten >=1pt,auto,node distance=2.8cm,
                    semithick,/tikz/initial text=]
  \tikzstyle{every state}=[fill=white,draw=black,text=black]

  \node[state,initial, initial text=${(y = -1) \wedge (z = 0) \wedge
  (R \geq 0)}$] (A)                    {$q_b$};

  \path (A) edge [out=40,in=20,loop] node {$t_2$} (A);
  \path (A) edge [out=-40,in=-60,loop] node {$t_3$} (A);
  \path (A) edge [out=-100,in=-120,loop] node {$t_4$} (A); 

\end{tikzpicture}

$t_2$ : $(y=-1)  / y^\prime = 0, z^\prime=z$\\
$t_3$ : $(y \neq -1) \wedge (y<z)  / y^\prime = y + 1, z^\prime=z$\\
$t_4$ : $(y \neq -1) \wedge (y=z)  \wedge (z < R) / $ $y^\prime = -1, z^\prime = z + 1$ \\
  \caption{Example where Variant Y alone terminates}
  \label{fig:over-alone-terminates}
\end{figure}

There do exist systems on which Variant~Y terminates, but Variant~X does
not. The example system $M$ in Figure~\ref{fig:over-alone-terminates}, with
property $\eglobally{(y \geq 1)}$, is one such example.  In this example,
the system exhibits only a single trace of finite length, whose length is
determined by the value of the parameter $R$. For $R=0$, the trace is
$t_2.t_3^0$ ($t_3^0$ stands for the transition $t_3$ being taken zero
times), for $R=1$ it is $t_2.t_3^0.t_4.t_2.t_3^1$, for $R=2$ it is
$t_2.t_3^0.t_4.t_2.t_3^1.t_4.t_2.t_3^2$, and so on. Therefore, none of the
states satisfy $\eglobally{(y \geq 1)}$. An infinite number of flattenings
of the system shown above are required to preserve all traces exhibited by
the system; in particular, for each trace $t$ exhibited by the system, a
linear (i.e., acyclic and branching-free) flattening that contains
sequentially the transitions taken by $t$ (in the same order), is able to
exhibit $t$. Variant~X does not terminate, because this is not a finite set
of flattenings. Variant~Y terminates on this example in a single iteration;
note that the maximum ``order'' of any control state in any of these
flattenings is 1.

In our empirical studies with real benchmarks that we describe in
Section~\ref{ssec:implementation}, there were many system-property pairs on
which Variant~Y terminated in a few hundred milliseconds, but on which
Variant~X did not terminate even after 1 hour. On the other hand, there was
no system-property pair on which Variant~X terminated but Variant~Y did not
within one hour.

\subsubsection{Systems where our full algorithm does better than both
variants.} 
\label{sssec:combined-does-better}

We have already shown that the full algorithm is guaranteed to
terminate on each system-property pair on which \emph{either} Variant~X or
Variant~Y terminates, provided the algorithm is restructured using
multi-threading as was discussed in Section~\ref{ssec:under-term}. 

\label{sec:term-sys}
There exists a family of system-property pairs on which neither Variant~X
nor Variant~Y terminates, but the full routine $\computeGlobalName$
terminates. In fact, given any system-property pair
$M_x, \eglobally{\phi_1}$
such that Variant~X terminates on this pair but Variant~Y does not terminate, and given any other system-property pair $M_y,\eglobally{\phi_2}$ such that Variant~Y terminates on this pair but Variant~X does not, it is possible to construct a new
``combined'' system $M$ from $M_x$ and $M_y$ such that neither Variant~X
nor Variant~Y terminates on this combined system wrt the property $\eglobally{(\phi_1
\vee \phi_2)}$, but the full routine $\computeGlobalName$
terminates. In the interest of brevity we omit a formal description of
this construction. However, we illustrate it using an example.

Recall that Variant~Y does not terminate on the counter system shown in
Figure~\ref{fig:under-alone-terminates} for the property $\eglobally{(x >
0)}$. In the rest of this section let us call this system $M_x$.
Similarly, recall that Variant~X does not terminate on the counter system
shown in Figure~\ref{fig:over-alone-terminates} for the property
$\eglobally{(y\geq 1)}$. Let us call this system $M_y$. In this rest of this section we refer to the formula $x>0$ as $\phi_1$ and to the formula $y\geq 1$ as $\phi_2$.

\begin{figure}
    \begin{tikzpicture}[->,>=stealth',shorten >=1pt,auto,node distance=2.8cm,
    semithick,/tikz/initial text=, transform shape]
    \tikzstyle{every state}=[fill=white,draw=black,text=black]
    \node[state,initial, initial text=
    {\begin{tabular}{c}$((x = P) \wedge (P \geq 0))\vee$ \\
        $(y =-1 \wedge z=0 \wedge R\geq 0)$\end{tabular}}] (S) {$\qstart$};
\node[state] (A)[above right of = S]  {$q_a$};
\node[state] (B)[below right of = S] {$q_b$};

\path (S) edge node {$t_5$} (A);
    \path (S) edge  node {$t_6$} (B);
\path (A) edge [out =110, in=70,loop      ] node {$t_1$}  (A);
    \path (B) edge [out=10,in=-10,loop    ] node {$t_2$} (B);
    \path (B) edge [out=-65,in=-85,loop   ] node {$t_3$} (B);
    \path (B) edge [out=-110,in=-130,loop ] node {$t_4$} (B);

\end{tikzpicture}
$t_1$: $x>0\wedge w=1 / x^\prime < x \wedge x^\prime \geq x-2, w^\prime=1,y^\prime=y,z^\prime=z$\\
$t_2$: $(y=-1) \wedge w=2/ y^\prime = 0, z^\prime=z,w^\prime=2,x^\prime=x$\\
$t_3$: $(y \neq -1) \wedge (y<z) \wedge w=2 / y^\prime = y + 1, z^\prime=z,w^\prime=2,x^\prime=x$\\
$t_4$: $(y \neq -1) \wedge (y=z)  \wedge (z < R) \wedge w=2/ $ $y^\prime = -1, z^\prime = z + 1,w^\prime=2,x^\prime=x$\\
$t_5$: $(x = P) \wedge (P \geq 0)/ x^\prime=x,w^\prime=1,y^\prime=y,z^\prime=z$\\ 
$t_6$: $y =-1 \wedge z=0 \wedge R \geq 0 / y^\prime=y,z^\prime=z,w^\prime=2,x^\prime=x$\\
    \caption{``Combined'' system $M$ on which the routine $\computeGlobalName$ alone terminates}
    \label{fig:combined-alone-terminates}
\end{figure}

The ``combined''
counter system $M$ constructed from the two systems referred to above is
shown in Figure~\ref{fig:combined-alone-terminates}.
Notice that this combined system has all the counters of $M_x$ and
$M_y$, i.e., $x$ and $\{y,z\}$, and one additional counter $w$. $\qstart$
is the starting control-state of system $M$, while its set of initial
states is nothing but the union of the set of initial states of systems
$M_x$ and $M_y$. The portion of $M$ ``above'' $\qstart$ is nearly
identical to $M_x$, with only the following additional aspects: the new counter $w$ is set to 1 in every transition, and the counters $y,z$ that are inherited from $M_y$ retain their original values continuously. Similarly, the portion of $M$ ``below'' $\qstart$ is nearly
identical to $M_y$, with only the following additional aspects: the new counter $w$ is set to 2 in every transition, and the counter $x$ that is inherited from $M_x$ retains its original value continuously.  Note that the value of counter $w$, which is maintained at either 1 or 2, basically encodes whether the current state is from the ``upper'' part of $M$ or from the ``lower'' part.

Let us consider the property  $\eglobally{(\phi_1 \wedge (w=1)) \vee (\phi_2 \wedge (w=2))}$. Intuitively, Variant~X will not terminate on $M$ wrt this property because it does not terminate on system $M_y$ for property $\eglobally{(\phi_2 \wedge (w=2))}$. Similarly, Variant~Y will not terminate on $M$ wrt this property because it does not terminate on system $M_x$ for property $\eglobally{(\phi_1 \wedge (w=1))}$. Intuitively, the full algorithm is able to terminate for this property because (a) the traces that go via state $q_a$ of $M$, which only impact the sub-property $(\phi_1 \wedge (w=1))$, are handled in finite time by processing the flattenings of $M$ (this follows from the fact that Variant~X is able to terminate on the system $M_x$), while (b) the traces that go via state $q_b$ of $M$, which only impact the sub-property $(\phi_2 \wedge (w=2))$, are handled in finite time by the iterations of the loop in the subroutine $\computeGlobalOverName$ that is invoked by routine $\computeGlobalName$ (this follows from the fact that Variant~Y is able to terminate on the system $M_y$).

We now elaborate upon the intuition above  by making a sequence of observations:

\begin{enumerate}
\item Let us call the traces of the combined system $M$ that go via
the control-state $q_a$ the ``upper'' traces.  (1) It is easy to see that these traces are identical to the traces
exhibited by system $M_x$, provided the value of the counter $w$ as well as
the counters inherited from system $M_y$ (i.e., $y$ and $z$) are ignored. (2) Conversely, for every trace $t$ of $M_x$, and for every possible valuation for counters $y$ and $z$, there is an upper trace $t_1$ of $M$ such that (a) all states of $t_1$ have the just-mentioned valuation for $y$ and $z$, and (b) every pair of corresponding states of $t$ and $t_1$ agree on the value of counter $x$. Note that both observations above hold only when the first transition in each upper trace (the transition from $\qstart$ to $q_a$) is ignored.

\item Let us call the traces of the combined system $M$ that go via the control-state $q_b$ the ``lower'' traces.  (1) It is easy to see that these traces are identical to the traces
exhibited by system $M_y$, provided the value of the counter $w$ as well as
the counter $x$ are ignored. (2) Conversely, for every trace $t$ of $M_y$, and for every possible value for counter $x$, there is an lower trace $t_2$ of $M$ such that (a) all states of $t_2$ have the just-mentioned value for counter $x$, and (b) every pair of corresponding states of $t$ and $t_2$ agree on the value of counters $y$ as well as $z$. Note that both observations above hold only when the first transition in each lower trace (the transition from $\qstart$ to $q_b$) is ignored.

Note that  the upper traces and the lower traces are
non-overlapping; no concrete state belongs to both an upper trace as well as  a lower trace.

\item We claim  that only Variant~X (and not Variant~Y) will terminate on the system $M$ for the property $\eglobally{(\phi_1 \wedge
(w=1))}$. The argument for this follows. This
property is satisfied only by states of $M$ that the upper traces go
through. Therefore, from the trace equivalence mentioned in Point~1 above,
and from the fact that only Variant~X terminates on the system $M_x$ for
property $\eglobally{\phi_1}$, our claim gets proved.

\item By an analogous argument, only Variant~Y terminates on system $M$ for
property $\eglobally{(\phi_2 \wedge
(w=2))}$.

\item We claim that neither Variant~X nor Variant~Y will terminate on system $M$ for the property $\eglobally{(\phi_1 \wedge (w=1)) \vee (\phi_2 \wedge (w=2))}$. Our argument is as follows. 
Since the upper and the lower traces are non-overlapping, either variant will terminate on $M$ for the property $\eglobally{(\phi_1 \wedge (w=1)) \vee (\phi_2 \wedge (w=2))}$ only if it terminates on $M$ separately for both properties $\eglobally{(\phi_1 \wedge
(w=1))}$ and $\eglobally{(\phi_2 \wedge
(w=2))}$. Therefore, from Points~3 and~4 above, our claim is proved. 

The remaining observations go towards showing that the full routine $\computeGlobalName$ terminates on the system $M$ for the property  $\eglobally{(\phi_1 \wedge (w=1)) \vee (\phi_2 \wedge (w=2))}$. 

\item Given a state $\vec{s_x}$ of $M_x$, we define $\extend{\vec{s_x}}$ to be the set of all states of $M$ that are obtained by inheriting the value of counter $x$ from $\vec{s_x}$, letting the counters $y$ and $z$ have arbitrary values, and letting the new counter $w$ have value 1.  If $\vec{S_x}$ is a set of states of $M_x$, then $\extend{\vec{S_x}}$ is the union of $\extend{\vec{s_x}}$ over all $\vec{s_x} \in \vec{S_x}$.  Analogously, $\extend{\vec{s_y}}$ and $\extend{\vec{S_y}}$ are defined for states of the system $M_y$. Informally, we say that $\extend{\vec{s_x}}$ (resp. $\extend{\vec{s_y}}$) is the set of  ``extensions'' of $\vec{s_x}$ (resp. $\vec{s_y}$).

\item Any flattening $N$ of $M$ contains within it a flattening of $M_x$; let us call this flattening $N_x$. $N_x$ consists of the control-states of $N$ that are copies of control-states of $M$ that are inherited from $M_x$, along with the transitions of $N$ that are incident on these control states.

\item We claim that for any flattening $N$ of $M$, if $N$ is analyzed by routine $\computeGlobalName$ wrt any given temporal property $\eglobally{\phi}$, for every state $\vec{s_x}$ of $M_x$ that would have been added to set $X$ (resp. set $Y$) by Variant~X were it to process the flattening $N_x$ wrt to the same property, routine $\computeGlobalName$ would add \emph{every} state of $N$ in the set $\extend{\vec{s_x}}$ to set $X$ (resp. set $Y$). This claim follows easily from the observations in Points~1 and~2 above.

\item We now argue that the full routine $\computeGlobalName$, when applied on system $M$ for property $\eglobally{(\phi_1 \wedge (w=1))}$, will end up adding to set $X$ all states of $M$ that satisfy this property, within a finite number of steps.  We are given that Variant~X terminates on the system $M_x$ for the property $\eglobally{\phi_1}$. That is, there exists a set $\vec{S_x}$ of flattenings of $M_x$, such that once these flattenings have been processed by Variant~X, the set $X$ contains all states of $M_x$ that satisfy $\eglobally{\phi_1}$ in $M_x$ and the set $Y$ contains all states of $M_x$ that satisfy $\phi_1 - X$.  Therefore, it follows that there exists a set $S$ of flattenings of $M$ such that every flattening in $\vec{S_x}$ is contained within some flattening in $S$. Due to the fair strategy of exploring flattenings by length, routine $\computeGlobalName$, when applied on system $M$, would at some point have explored all flattenings in $S$.
Therefore, it follows from Point~8 above that at this point $\computeGlobalName$ would have added to $X$ all states of $M$ that are ``extensions'' of states of $M_x$ that satisfy $\eglobally{\phi_1}$. These are exactly the states of $M$ that satisfy the property $\eglobally{\phi_1 \wedge (w=1)}$ (no states in the ``lower'' part of $M$ can satisfy this property because in these states the counter $w$ has value 2).

\item We now argue that the full routine $\computeGlobalName$, when applied on system $M$ for property $\eglobally{(\phi_1 \wedge (w=1))}$, will end up adding to set $Y$ all states of $M$ that do not satisfy this property, within a finite number of steps.

None of the states in the ``lower'' part of $M$ satisfy the property $\eglobally{(\phi_1 \wedge (w=1))}$. All these states get added to $Y$ at the very beginning of algorithm, because they do not satisfy the condition $w=1$. The states in the ``upper'' part of $M$ that do not satisfy $\eglobally{(\phi_1 \wedge (w=1))}$ get added to $Y$ by the time all the flattenings of $M$ in the set $S$ mentioned above are processed, by the same reasoning as used in the previous point. Note that subroutine $\computeGlobalOverName$ could also add states to $Y$ (in line~\ref{line:growY} in Algorithm~\ref{algo:compute-global}), in addition to states being added to $Y$ due to processing of flattenings (in line~\ref{line:growYb}). However, it is easy to see that if a set is to be added to $Y$ during the processing of a flattening, then this does not get impacted at all by the pre-existing content of $Y$ before this flattening is processed.

\item It follows from Points 9 and 10 above that routine $\computeGlobalName$ terminates on system $M$ for property $\eglobally{(\phi_1 \wedge (w=1))}$. 

\item We now argue that routine $\computeGlobalName$ terminates on system $M$ wrt property $\eglobally{(\phi_2 \wedge (w=2))}$. We argue this by showing that all states of $M$ that do not satisfy the property $\eglobally{(\phi_2 \wedge (w=2))}$ get added to $Y$ in a finite number of steps. Therefore, termination is guaranteed to happen via line~\ref{line:return-Over} in Algorithm~\ref{algo:compute-global}. 

All ``upper'' states of $M$ (all of which do not satisfy the property mentioned above) get added to $Y$ at the very beginning of the algorithm, because they do not satisfy the condition $w=2$. Regarding the ``lower'' states, it is easy to see that for every state $\vec{s_y}$ that Variant~Y would add to set $Y$ within $i$ (cumulative) iterations of the loop in subroutine $\computeGlobalOverName$,  every extension of $\vec{s_y}$ will get added to $Y$ within $i$ (cumulative) iterations of the same loop when $\computeGlobalName$ is applied on system $M$ wrt property $\eglobally{(\phi_2 \wedge (w=2))}$ (certain states could get added even earlier than when Variant~Y would add them because of line~\ref{line:growYb} of Algorithm~\ref{algo:compute-global}). Therefore, since Variant~Y is known to terminate on system $M_x$ wrt property $\eglobally{(\phi_2)}$, it follows that all states of $M$ that do not satisfy property $\eglobally{(\phi_2 \wedge (w=2))}$ will get added to $Y$ by routine $\computeGlobalName$ in a finite number of steps.

\item We now finally argue that routine $\computeGlobalName$ terminates on system $M$ wrt property $\eglobally{((\phi_1 \wedge (w=1))\vee(\phi_2 \wedge (w=2)))}$.  Consider three runs of the routine $\computeGlobalName$ on the system $M$, (a) wrt property $\eglobally{(\phi_1 \wedge (w=1))}$, (b) wrt property $\eglobally{(\phi_2 \wedge (w=2))}$, and (c) $\eglobally{((\phi_1 \wedge (w=1))\vee(\phi_2 \wedge (w=2)))}$. Consider any state $\vec{s'}$ that does not satisfy the property $\eglobally{((\phi_1 \wedge (w=1))\vee(\phi_2 \wedge (w=2)))}$. Say this state is in the ``upper'' traces. Clearly, this state does not satisfy the property $\eglobally{(\phi_1 \wedge (w=1))}$ either. Since run~(a) terminated (see Point~11 above), and since any run of routine $\computeGlobalName$ can terminate only when the set $Y$ contains all states that do not satisfy the  property that is being checked by the run, it follows that run~(a) added state $\vec{s'}$ when a certain flattening was processed or within a certain number of cumulative iterations of the loop in subroutine $\computeGlobalOverName$. It is easy to see that $\vec{s'}$ would also be added to $Y$ by run~(c) by the time the same flattening mentioned above is processed or within the loop in subroutine $\computeGlobalOverName$ within the same number of iterations mentioned above. Analogously, using a similar argument involving run~(b), it can be argued that any state in the lower traces that does not satisfy the property $\eglobally{((\phi_1 \wedge (w=1))\vee(\phi_2 \wedge (w=2)))}$ gets added to set $Y$ in run~(c) within a finite number of steps. Putting these two observations together, it follows that at some point in run~(c) set $Y$ contains all states that do not satisfy the property $\eglobally{((\phi_1 \wedge (w=1))\vee(\phi_2 \wedge (w=2)))}$. When this happens, one of the two termination checks in Algorithm~\ref{algo:compute-global} is guaranteed to succeed. In other words, run~(c) is guaranteed to terminate.

That is, the full routine $\computeGlobalName$ terminates on the combined system $M$ wrt the property $\eglobally{((\phi_1 \wedge (w=1))\vee(\phi_2 \wedge (w=2)))}$.

\end{enumerate}


\subsubsection{An example where non-termination occurs.}

For the sake of completeness, we now provide an example on which neither
variant of our algorithm terminates.  This is a system with a
single control state $q$, a single
counter $x$, and a two transitions $t_1$ and $t_2$, with both these
transitions going from
$q$ to $q$. The guards and actions of these two transitions are as follows: 

$t_1$: Guard: $x > 0$, Action: $x' = x - 1$

$t_2$: Guard: $x > 0$, Action: $x' = x - 2$

Notice that this system is equivalent in terms of traces exhibited to the
system that was presented earlier on which Variant~X terminated but
Variant~Y did not. On this current system Variant~Y does not terminate for
the same reason as it did not terminate on the previous system. However,
Variant~X also fails to terminate on this system. This is because
no set of flattenings of this system preserves all traces exhibited by the
system; this is because the traces in this system are able to follow arbitrary
interleavings of the two transitions. The full algorithm also fails to
terminate on this example.

\subsubsection{The approach of Demri et al.}
The approach of Demri et al.~\cite{dem2006}, which we had introduced in
Section~\ref{ssec:over-term}, is the closest related work to our work. As
we had mentioned at that point, this approach is applicable only to systems
that possess a trace flattening wrt the initial set of states
$\phiinit$. This restriction \emph{is not} obeyed by our running example in
Figure~\ref{fig:example_intro}(a), nor by the example that we mentioned
earlier on which Variant~Y (but not Variant~X) terminates, nor by many of
the real benchmarks that come with the Fast toolkit~\cite{fast}, on which our approach
terminates (we present more details about this in
Section~\ref{ssec:implementation}).

In the converse direction, consider a system with a single control-state
$q$, a single counter $x$, and a single transition with guard $x>0$ and
action $x' < x$. Let the property of interest be $\eglobally{(\tru)}$. This
system exhibits infinite branching; hence our algorithm (or Variant~X) are
not applicable on it. Variant~Y is applicable on it, but does not terminate
(it adds $x \leq 0$ to $Y$ initially, $x=1$ and $x=2$ to $Y$ in the first
iteration, $x=3$ and $x=4$ to $Y$ in the second iteration, and so
on). Demri's approach is applicable on this example because it is a flat
system (and hence possesses a trace flattening).




\section{Algorithm for full CTL}
\label{sec:algo}
In this section we describe an approach to answer $\ctl$ properties on counter systems. The inputs to the approach are a counter system 
$M = \langle Q, C,\Sigma,\phiinit,G,F\rangle$, a temporal property  $\psi$ in existential normal form, and an enumerator $\enumlabel$  from the enumeration $\latticeApprox$ $\equiv$ $\{\precise,\under,\overapprox\}$. 
If $\enumlabel$ is $\precise$, then the approach either terminates and returns the precise solution, namely, a formula that precisely encodes the set of concrete states of $M$ that satisfy $\psi$, or goes into non-termination. If the label is $\under$ (resp. $\overapprox$), then the approach either terminates on its own and returns the precise solution, or can be terminated forcefully anytime during its working, in which case it returns a formula that is in general an under-approximation (resp. over-approximation) of the precise solution. The algorithm is \emph{inductive}, in that recursively invokes itself to solve sup-properties of the given property $\psi$ in order to eventually solve for $\psi$. Also, it makes use of the routines $\computeUntilName$ and $\computeGlobalName$, which we discussed in preceding sections of this paper. We discuss our approach in detail in the rest of this section. 



\subsection{The inductive approach}
\renewcommand{\algorithmicrequire}{\textbf{Input:}}
\renewcommand{\algorithmicensure}{\textbf{Output:}}
\begin{algorithm}[H]
\caption{$\sat{M}{\psi}{\enumlabel}$}
\label{algo:sat}
\begin{algorithmic}[1]
 \REQUIRE A counter system $M=\langle Q, C,\Sigma,\phiinit,G,F\rangle$, a
  $\ctl$ temporal property $\psi$ in existential normal form, and an enumerator $\enumlabel$, which is an element of the enumeration $\latticeApprox$ $\equiv$ $\{\precise,\under,\approx\}$.  

\ENSURE A pair ($\phi, \approximation$), where $\phi$ is a Presburger formula representing a set of concrete states, and $\approximation$ is an element of $\latticeApprox$. Basically, $\approximation$ indicates whether $\phi$ represents precisely the set of concrete states of $M$ that satisfy $\psi$, or whether it represents an under- or over-approximation of this precise set. $\approximation$ is guaranteed to be $\sqsubseteq$ $\enumlabel$, as per the partial order relation depicted in Figure~\ref{fig:Lattice}. 

  \IF {$\psi=\phi_i$   \COMMENT{i.e., $\psi$ is a basic proposition}} \label{line:phi}
       \RETURN $(\phi_i,\precise)$
   \ELSIF {$\psi=\neg \psi_1$} \label{line:neg}
      \STATE $(\phi_1,\approximation)\gets \sat{M}{\psi_1}{\neg \enumlabel}$\label{line:neg-call}
       \RETURN $(\neg \phi_1,\neg \approximation)$
   \ELSIF {$\psi=\psi_1\vee\psi_2$} \label{line:or}
	\STATE $(\phi_1,\approximation_1)\gets\sat{M}{\psi_1}{\enumlabel}$
	\STATE
    $(\phi_2,\approximation_2)\gets\sat{M}{\psi_2}{\enumlabel}$
	\RETURN $(\phi_1\vee\phi_2,\approximation_1 \join \approximation_2)$
    \ELSIF{$\psi=\enext{\psi_1}$} \label{line:next}
	\STATE $(\phi_1,\approximation)\gets\sat{M}{\psi_1}{\enumlabel}$
	\RETURN $(\pre{M}{\phi_1},\approximation)$\label{line:predecessor}
    
     \ELSIF{$\psi=\euntil{\psi_1}{\psi_2}$} \label{line:until}
	\STATE $(\phi_1,\approximation_1)\gets\sat{M}{\psi_1}{\enumlabel}$
	\STATE $(\phi_2,\approximation_2)\gets\sat{M}{\psi_2}{\enumlabel}$
	\STATE $(\phi,\approximation_3)\gets\computeUntil{M}{\phi_1}{\phi_2}{\enumlabel}$\label{line:computeUntil}
	\RETURN $(\phi,\approximation_1 \join \approximation_2 \join \approximation_3)$

    \ELSIF {$\psi=\eglobally{\psi_1}$} \label{line:global}
	\STATE $(\phi_1,\approximation_1)\gets\sat{M}{\psi_1}{\enumlabel}$
	\STATE $(\phi,\approximation_2)\gets\computeGlobal{M}{\phi_1}{\enumlabel}$\label{line:computeGlobal}
	\RETURN $(\phi,\approximation_1 \join \approximation_2)$
  \ENDIF
 \end{algorithmic}
\end{algorithm}

Algorithm~\ref{algo:sat} contains a procedure named \emph{SAT}, which embodies the inductive approach. The procedure has a case-based structure, dependent on the root operator of
$\psi$. If $\psi$ is a basic proposition $\phi_i$ (i.e., a Presburger formula), then $\phi_i$ itself is the solution (see line~\ref{line:phi}).
If $\psi$ is of the form $\neg \psi_1$, then the procedure \emph{SAT} is first invoked recursively on the sub-property $\psi_1$ with an inverted approximation label. This inversion operation on the enumeration $\latticeApprox$ is defined as follows:

$$\neg
\enumlabel \equiv (\enumlabel=\under)\ ?\ \overapprox\ : $$
$$ (\enumlabel=\overapprox)\ ?\ \under\ :\ \precise$$

The basic intuition behind this is that if we need to approximate the solution to $\neg \psi_1$ in a certain direction, then this can be done by computing an approximate solution to $\psi_1$ in the opposite direction.

The case where the root operator is an `$\vee$' (lines~\ref{line:or} onward) involves solving for the two operands of this operator in the same approximation direction as desired for the root, and then taking the disjunction of the two formulas obtained. A simple approach to returning the direction of approximation of the result would have been simply to return the given enumerator $\enumlabel$. However, it is possible to do better than this, because, even if $\enumlabel$ is $\under$ or $\overapprox$, it is possible in certain cases that the solutions returned for both the operands is $\precise$; in this case, $\precise$ would be the ideal enumerator to return. The algorithm  applies this logic by  computing 
the approximation direction of the returned solution  as the \emph{least upper bound} (i.e., `$\sqcup$') of the approximation directions of the solutions of the two operands as per the partial ordering depicted in Figure~\ref{fig:Lattice}. 

To find the set of states that satisfy $\enext{\psi_1}$ (lines~\ref{line:next} onward), we first find the set of
states $\phi_1$ that satisfy $\psi$ (in the same direction of approximation as the root operator). $\pre{M}{\phi_1}$ then gives the set of states that satisfies $\enext{\psi}$.
$\pre{M}{\phi_1}$ can be computed easily because the transitions in the counter system give the \emph{pre-post} relation between the concrete states of the counter
system.

For the ``until'' property $\euntil{\psi_1}{\psi_2}$ (lines~\ref{line:until} onward), the procedure first recursively finds the set of states $\phi_1$ and 
$\phi_2$ that satisfy $\psi_1$ and $\psi_2$ respectively, in the approximation direction indicated by $\enumlabel$. Then, the routine $\computeUntilName$, which we discussed in Section~\ref{sec:until}, is invoked to find the solution for $\psi$. It is easy to see that if $\phi_1$ and $\phi_2$ are both under- (over-) approximations of the respective precise solutions for $\psi_1$ and $\psi_2$, then the solution  returned by $\computeUntilName$ would also be an under- (over-) approximation of the precise solution for $\psi$. Therefore, $\enumlabel$ would have been a sound approximation direction to return. However, in certain cases, it is possible that $\phi_1$ and $\phi_2$ both are precise, and $\computeUntilName$ also returns the enumerator $\precise$, even though $\enumlabel$ is $\under$ or $\overapprox$. Therefore, we can actually return a more precise approximation direction, namely,  $\approximation_1 \sqcup \approximation_2 \sqcup \approximation_3$. For ``global'' properties (lines~\ref{line:global} onward) the idea is very similar to that for ``until'' properties, except that routine $\computeGlobalName$ (which was discussed in Section~\ref{sec:global})  is used instead of routine $\computeUntilName$.

 \subsection{Correctness}

\begin{theorem}
 \label{thm:labeling-correctness}
Given a temporal property $\psi$ and a counter system $M$, say the procedure $\sat{M}{\psi}{\enumlabel}$ returns the enumerator 
$\approximation$ along with a formula $\phi$. Then, (a) $\approximation \sqsubseteq \enumlabel$, and (b) $\phi$ approximates the set of states of $M$ that actually satisfy 
$\psi$ in the direction given by the returned enumerator $\approximation$. 
\end{theorem}

\begin{proof}
The temporal property $\psi$ can be thought of as an expression-tree. We prove this theorem  by induction on the \emph{height} of this tree. 


\emph{Base case:} For $\psi$ to be of height 1, it must be a  basic proposition of the form $\phi_i$. In this case $\sat{M}{\psi}{\enumlabel}$
returns $(\phi_i,\precise)$. Since $\phi_i$ represents precisely the set of states that satisfy $\phi_i$, the theorem statement holds.

\emph{Inductive step:} Let $\psi$ be a formula of height $k+1$, $k > 0$. The induction hypothesis is that the theorem holds for all formulas  of height at most $k$. We prove the theorem case by case, based on the root operator of $\psi$.

\begin{itemize}
\item Let $\psi \equiv \neg\psi_1$ and $(\phi_1,\approximation)=\sat{M}{\psi_1}{\neg\enumlabel}$. It is easy to see that the set of states 
that satisfy $\psi$ is given by the formula $\neg\phi_1$. As per the induction hypothesis,
$\approximation\sqsubseteq\neg \enumlabel$. It now follows from the definition of the  negation operator that
$\neg\approximation \sqsubseteq \enumlabel$. Therefore the theorem holds. 

\item Let $\psi \equiv \psi_1 \vee \psi_2$. The argument is similar to the one for the previous case. By the inductive hypothesis, $\phi_1$ and $\phi_2$ are approximations of the precise solutions for $\psi_1$ and $\psi_2$, respectively, in the direction indicated by $\enumlabel$. Therefore, $\phi_1 \vee \phi_2$ is an approximation of the precise solution for $\psi$ in the direction indicated by $\enumlabel$. Also, by the inductive hypothesis, $\approximation_1 \sqsubseteq \enumlabel$ and $\approximation_2 \sqsubseteq \enumlabel$. Therefore, it follows that $\approximation_1 \sqcup \approximation_2$, which is the returned enumerator, is dominated by $\enumlabel$ in the partial ordering. 

\item Let $\psi\equiv\enext{\psi_1}$. As per the inductive hypothesis, the formula $\phi_1$ that is returned by the recursive call to \emph{SAT} approximates the precise solution to $\psi_1$ in the direction indicated by $\enumlabel$. Therefore, following the definition of $\enext{\psi_1}$, it is easy to see that the formula returned, namely, $\pre{M}{\phi_1}$, is an approximation of the precise solution to $\enext{\psi_1}$ in the direction indicated by $\enumlabel$. Also, it follows from the inductive hypothesis that the enumerator returned, namely $\approximation$, is dominated by $\enumlabel$ in the partial ordering. 

\item\sloppypar Let $\psi \equiv \euntil{\psi_1}{\psi_2}$, $(\phi_1,\approximation_1)=\sat{M}{\psi_1}{\enumlabel}$ and 
$(\phi_2,\approximation_2)=\sat{M}{\psi_2}{\enumlabel}$.
Then, by the induction hypothesis, $\phi_1$ approximates the precise solution to $\psi_1$ in the direction of $\enumlabel$, and $\phi_2$ approximates the precise solution to $\psi_2$ in the direction of $\enumlabel$. Therefore, from the  definition of  $\euntil{\psi_1}{\psi_2}$, it follows that $\euntil{\phi_1}{\phi_2}$ approximates $\euntil{\psi_1}{\psi_2}$ in the direction of $\enumlabel$. Furthermore, we had proved in Section~\ref{sec:until} that $\computeUntil{M}{\phi_1}{\phi_2}{\enumlabel}$ returns a formula $\phi$ that approximates $\euntil{\phi_1}{\phi_2}$ in the direction indicated by $\enumlabel$. Therefore, the two observations above imply that the formula $\phi$ (which is returned by \emph{SAT}) approximates $\euntil{\psi_1}{\psi_2}$ in the direction indicated by $\enumlabel$. 

From the inductive hypothesis it also follows that $\approximation_1 \sqsubseteq \enumlabel$ and $\approximation_2 \sqsubseteq \enumlabel$. Also, as was proved in Section~\ref{sec:until}, the enumerator $\approximation_3$ returned by $\computeUntilName$ is guaranteed to be dominated by $\enumlabel$ in the partial ordering. Therefore, it follows that the enumerator returned by \emph{SAT}, namely,  $(\approximation_1\join\approximation_2\join\approximation_3)$ is guaranteed to be dominated by $\enumlabel$ in the partial ordering.

\item Let $\psi \equiv \eglobally{\psi_1}$, and $(\phi_1,\approximation_1)=\sat{M}{\psi_1}{\enumlabel}$.
By the induction hypothesis, $\phi_1$ approximates the precise solution to $\psi_1$ in the direction of $\enumlabel$. Therefore, from the  definition of  $\eglobally{\psi_1}$, it follows that $\eglobally{\phi_1}$ approximates $\eglobally{\psi_1}$ in the direction of $\enumlabel$. Furthermore, we had proved in Section~\ref{sec:global} that $\computeGlobal{M}{\phi_1}{\enumlabel}$ returns a formula $\phi$ that approximates $\eglobally{\phi_1}$ in the direction indicated by $\enumlabel$. Therefore, the two observations above imply that the formula $\phi$ (which is returned by \emph{SAT}) approximates $\eglobally{\psi_1}$ in the direction indicated by $\enumlabel$. 

From the inductive hypothesis it also follows that $\approximation_1 \sqsubseteq \enumlabel$. Also, as was proved in Section~\ref{sec:global}, the enumerator $\approximation_2$ returned by $\computeGlobalName$ is guaranteed to be dominated by $\enumlabel$ in the partial ordering. Therefore, it follows that the enumerator returned by \emph{SAT}, namely,  $(\approximation_1\join\approximation_2)$ is guaranteed to be dominated by $\enumlabel$ in the partial ordering.

\end{itemize}
\end{proof}

\subsection{Termination}

The procedure \emph{SAT} is recursive, with invocation depth bounded by the depth of the given formula $\psi$. The only actions that it performs other than recursively calling itself are to invoke the `\emph{pre}' operator (in the ``next'' case), which always terminates, and to invoke routines $\computeUntilName$ and $\computeGlobalName$. Therefore, a call to \emph{SAT} terminates iff each invocation that it makes to the two routines above terminate. Note that whenever these two routines are passed a $\enumlabel$ whose value is $\under$ or $\overapprox$, they can be forcefully terminated at any point desired and made to return an approximate solution in the indicated direction.

\section{Implementation and Results}
\label{ssec:implementation}
\sloppypar We have implemented our routine $\computeGlobalName$ for model-checking ``global'' properties, which we had described in Algorithm~\ref{algo:compute-global}. We have also implemented its two variants, namely Variant~X (see Section~\ref{ssec:global:variantX}) and Variant~Y (see Section~\ref{ssec:global:variantY}).
(At this point we have not implemented
the full CTL checking algorithm described in Section~\ref{sec:algo}.). Note
that Algorithm~\ref{algo:compute-global} needs an underlying reachability analysis tool, to compute $\prek{N}{\phi}$ (see line~\ref{line:growX} in the algorithm), as well as to compute $\preStarName$ (see formula $\growtwoa$ in Section~\ref{sssec:over-working}, which is used within subroutine~$\computeGlobalOverName$, which is described in Algorithm~\ref{algo:over-approx}). For this purpose
we use the reachability analysis black-box provided by the \emph{Fast} toolkit~\cite{fast}. Fast is applicable on counter systems whose guards and actions satisfy certain constraints~\cite{finkel02}.
Fast provides a routine for computing $\mathit{pre}^*$ formulas on systems, which necessarily terminates on flat systems as well as on systems that have a trace flattening with respect to their respective sets of initial states, but also terminates on many other systems that do not satisfy these restrictions.  Fast also provides (an always terminating) routine to compute
$\mathit{pre}^k$ formulas on simple cycles, which we extended in a straightforward way to work on flat systems.
We have implemented the subroutine $\mathit{isTraceFlattening}$,
which is invoked in line~\ref{line:checkfl} in Algorithm~\ref{algo:compute-global} and which was discussed in Section~\ref{sssec:under-termination},  using the 
formula referred to by Demri et al.~\cite{dem2006} and shared with us by them via private communication.

Recall that as defined in Section~\ref{sec:prelim}, every system $M$ is expected to come with a set of \emph{initial} states $\phiinit$.
We define the set of \emph{reachable} states $\phireach$ of the system $M$ as the states that belong to $\postStar{M}{(\phiinit)}$. 
In our experiments, we wish to find not the set of \emph{all} states of $M$ that satisfy the given property $\eglobally{\phi}$, but only the set of \emph{reachable} states of $M$ that satisfy this property. This is because users are typically interested in knowing about only the properties of reachable states. We achieve this objective by using the property $\eglobally{(\phi\wedge \phireach)}$ instead of the property $\eglobally{(\phi)}$. To determine the formula $\phireach$ we once again use Fast. In the remainder of this section, for the sake of brevity, we denote $\phi\wedge \phireach$ simply as $\phi$.

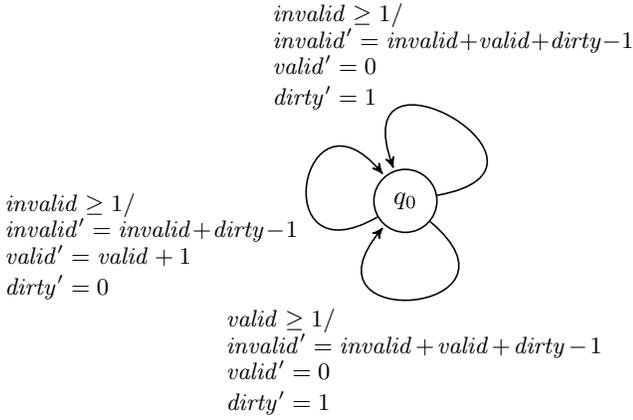
\begin{figure}[t]
\begin{tikzpicture}[->,>=stealth',shorten >=1pt,auto,node distance=2.8cm,
    semithick,/tikz/initial text=, transform shape]
    \tikzstyle{every state}=[fill=white,draw=black,text=black]
    \node[state] (A) at (0,0)  {$q_0$};

    \path (A) edge [out =210, in=130, loop] node [below left,  text width =3.9cm,align=left] {\small$\mathit{invalid} \geq 1/$ \\ $\mathit{invalid}^\prime = \mathit{invalid}+ \mathit{dirty}-1$ \\ $\mathit{valid}^\prime = \mathit{valid}+1$\\ $\mathit{dirty^\prime}=0$}  (A);
    \path (A) edge [out =320, in=230, loop] node [below, text width =5cm,align=left] {\small$\mathit{valid}\geq 1/$ \\ $\mathit{invalid}^\prime = \mathit{invalid}+\mathit{valid}+\mathit{dirty}-1$ \\ $\mathit{valid^\prime}=0$\\ $\mathit{dirty}^\prime=1$}  (A);
    \path (A) edge [out =10, in=110 , loop] node [above, text width =4.8cm,align=left] {\small$\mathit{invalid}\geq 1/$ \\ $\mathit{invalid}^\prime = \mathit{invalid}+\mathit{valid}+\mathit{dirty}-1$ \\ $\mathit{valid^\prime}=0$\\ $\mathit{dirty}^\prime=1$}  (A);
%
\end{tikzpicture}
%
\caption{MSI cache coherence protocol}
\label{fig:msi}
\end{figure}

\subsection{Benchmarks selection}
\label{ssec:benchmarks-selection}
The Fast toolkit comes bundled with about 45 sample counter systems, which model a variety of protocols and mechanisms, such as  cache coherence protocols, client-server interactions,
control systems for lifts and trains, producer-consumer systems, etc. We use these systems for our experimental evaluation.
For instance, the counter system shown in Figure~\ref{fig:msi} is from this bundle, and models the MSI cache coherence protocol for a single cache line.
The counters $\inval$, $\vald$ and $\dirt$ represent the number of
processors in the respective states for the modeled cache line. 
Transition $t_1$ represents a read \emph{miss} by a processor in invalid state.
The action of this transition, which is $((\inval^\prime =  \inval + \dirt - 1) \wedge (\dirt^\prime = 0))$, $\vald^\prime = \vald + 1$, encodes the property  that all processors that previously had a modified copy write to the memory and 
move to the invalid state, while 
the requesting processor gets a copy from memory and moves to shared state from the invalid state. Similarly, transition $t_2$ represents a write by a processor that is in the \emph{valid} state; this processor goes to the \emph{dirty} state as a result of this transition, while all other processors move to the \emph{invalid} state. And $t_3$ represents a write-miss by a processor in the invalid state. The processor goes to
the dirty state and the remaining processors invalidate their copies.




One of the requirements of the MSI protocol is that when a cache line of a processor
is in \emph{shared} state, it remains in this
state until there is a write to this cache line by some processor. This temporal property can be expressed in CTL as follows:

\begin{equation*}
\varphi\equiv\ \ \mathit{\vald\geq1}\implies\auntil{(\vald\geq1)}{(\dirt=1)}.
\end{equation*}

The property above can be represented in existential normal form as follows:

\begin{eqnarray}
    \varphi & \equiv & [\mathit{\vald<1}] \vee
    \neg[(\eglobally{(\dirt\neq1)})\vee \nonumber \\ 
    && \euntil{(\dirt\neq1)}{(\dirt\neq1\wedge \vald<1)}] \nonumber
\end{eqnarray}

Our routine $\computeGlobalName$ returns the following formula as the solution for the sub-property $\eglobally{(\dirt\neq 1)}$ of $\varphi$:

$$\phi_1 \equiv \fls$$

Intuitively this is because any trace in the system along which \emph{dirty} is continuously not equal to 1 needs to take transition $t_1$ repeatedly (the actions of transitions $t_2$ and $t_3$ set $\dirt$ to 1). However,  transition $t_1$ can be taken sequentially at most as many times as the initial value of \emph{invalid}. Therefore, there exists no infinite trace in the system along which $\dirt\neq 1$ holds continuously. Routine $\computeGlobalName$ is able to terminate on this example with the solution $\fls$. 







From the 45 systems that come with the Fast toolkit (version 2.1), we chose, using a simple
sufficient condition that we designed, $14$ systems that are guaranteed to
\emph{not} have a trace-flattening with respect to their respective sets of initial states. We chose such
systems because they lie \emph{outside} the class of systems  addressed by the
previous approach of Demri et al.~\cite{dem2006}, and are amenable to our
approach using Fast as the underlying reachability analysis tool. In other 
words, they are the more challenging systems.

\begin{table*}[!htbp]
\begin{scriptsize}
\caption{Experimental results for all 3 algorithms. Sys: short name of system. \#c: num. counters in system. \#t: num. transitions. RT - running time in milliseconds. 
 FL - max. length of flattenings explored. NFE - number of flattenings
 explored. NI: cumulative num. iterations of loop in Algo~$\computeGlobalOverName$. Unexp? Whether algo's output does not match expected output (Empty cell indicates ``no''.) {\timeout} - timed out. }
\label{tbl:results}
 \centering
\begin{tabular}{|l|c|r|r|r|r|r|r|r|r|r|r|}
\hline
S.  & \multicolumn{3}{c|}{Sys. info}&\multicolumn{3}{c|}{Variant-X}&\multicolumn{2}{c|}{Variant-Y}&\multicolumn{3}{c|}{$\computeGlobalName$}\\
\cline{2-12}
No. & (1) & (2) & (3) & (4) & (5) & (6) & (7) & (8) & (9) & (10) &
(11) \\

 & Sys& \#c & \#t &RT(ms)& FL &NFE &RT (ms)& NI &RT (ms)& NI & Unexp? \\\hline

1 & firefly
    & 4 & 8 &  \timeout &  & 2295$k$ & 24  & 2  & 28  &  2 &  Y \\
2 &    &   &   &  13       &  & 1       & 15  & 2  & 13  &  1 & Y       \\
3 &    &   &   &  \timeout &  & 2094$k$ & 47  & 3  & 54  &  3 & Y \\
4 &    &   &   &  1        &  & 0       & 1   & 0  & 1   &  0 &        \\ \hline

5 & illinois   & 4 & 9 &  52       &  & 13      & 25  & 1  & 40  &  2 & Y \\
6 &    &   &   &   2       &  & 0       & 2   & 0  & 2   &  0 & \\\hline

7 & mesi
    & 4 & 4 &  9        &  & 1       & 10  &  2 & 9   &  1 & \\
8 &    &   &   &  12       &  & 1       & 13  &  2 & 13  &  1 & \\
9 &    &   &   &  \timeout &  & 579$k$  & 12  &  2 & 14  &  2 &  Y  \\
10 &    &   &   &  \timeout &  & 2383$k$ & 25  &  2 & 32  &  2 & Y \\\hline

11 & moesi
     & 5 & 4 &  \timeout &  & 2134$k$ & 34  & 2  & 42  &  2 & Y \\
12 &    &   &   &  10       &  & 1       & 9   & 1  & 10  &  1 & Y \\
13 &    &   &   &  21       &  & 1       & 22  & 2  & 21  &  1 & \\
14 &    &   &   &  17       &  & 1       & 18  & 2  & 18  &  1 & \\\hline

15 & synapse
    & 3 & 3 &  7        &  & 1       & 7   & 2  & 7   &  1 & \\
16 &    &   &   &  8        &  & 1       & 9   & 2  & 9   &  1 & \\
17 &    &   &   &  \timeout &  & 3165$k$ & 21  & 2  & 26  &  2 & \\\hline

18 & centralserver
    & 12& 8 &  409      &  & 53      & 88  & 1  & 96  &  1 & * \\
19 &    &   &   &  587      &  & 38      & 81  & 1  & 84  &  1 & * \\\hline

20 & consistency
    & 11& 8 &  2        &  & 0       & 2   & 0  & 2   &  0 & Y \\
21 &    &   &   &  2        &  & 0       & 2   & 0  & 2   &  0 & Y \\
22 &    &   &   &  2        &  & 0       & 2   & 0  & 2   &  0 & \\
23 &    &   &   &  2        &  & 0       & 2   & 0  & 2   &  0 & \\\hline

24 & futurbus
   & 9 & 9 &  64       &  & 2       & 59  & 1  & 62  &  1 & Y \\
25 &    &   &   &  1        &  & 0       & 1   & 0  & 1   &  0 & \\\hline

26 & lastinfirstserved
    & 7 & 10&  \timeout &  & 219$k$  & 206 & 3  & 235 &  3 & Y \\
27 &    &   &   &  67       &  & 7       & 46  & 1  & 48  &  1 & Y \\
28 &    &   &   &  63       &  & 7       & 43  & 1  & 51  &  1 & Y \\\hline

29 & lift
     & 4 & 5 &  \timeout &  & 1.7$k$  & 19  & 2  & 22  &  2 & \\
30 &    &   &   &  1        &  & 0       & 1   & 0  & 1   &  0 & \\
31 &    &   &   &  \timeout &  & 1.7$k$  & 17  & 2  & 20  &  2 & \\
32 &    &   &   &  1        &  & 0       & 1   & 0  & 1   &  0 & \\\hline

33 & ticket2i
    & 6 & 6 & \timeout  &  & 912$k$  & 28  & 2  & 31  &  2 & \\
34 &    &   &   & \timeout  &  & 919$k$  & 28  & 2  & 31  &  2 & \\
35 &    &   &   & 56        &  & 22      & 20  & 1  & 24  &  1 & \\
36 &    &   &   & 56        &  & 22      & 20  & 1  & 25  &  2 & \\\hline
    
37 & train
    & 3 & 12& 86        &  & 1       & 28  & 2  & 27  &  1 & \\
38 &    &   &   & 21        &  & 1       & 20  & 2  & 20  &  1 & \\
39 &    &   &   & \timeout  &  & 1732$k$ & 187 & 11 & 247 & 11 & \\
40 &    &   &   & 1         &  & 0       & 1   & 0  & 1   &  0 & \\
41 &    &   &   & 59        &  & 5       & 36  & 1  & 42  &  1 & \\
42 &    &   &   & \timeout  &  & 2716$k$ & 159 & 11 & 244 & 10 & \\\hline

43 & ttp2 & 9 & 17 & \timeout &  & 138$k$  & \timeout & 10  & \timeout & 10 & \\

44 & swimmingpool & 9 & 6 & \timeout &  & -  & \timeout & -  & \timeout & - & \\

\hline
\end{tabular}

 \end{scriptsize}
\end{table*}

Table~\ref{tbl:results} provides an overview of our entire experiment. Columns~(1)-(3) in this table provide basic information about the systems that we have selected. Column~``Sys'' is the name of the system (excluding the ``.fst'' file name extension). Column~``\#c'' gives the number of counters in the system, while the
Column~``\#t'' gives the number of transitions in the system. The number of control states is not mentioned in the table; it is 1
for every input system except for train and ttp2, which have four control states each.

The 14 systems that we chosen (and also, the remaining 31 systems that come with the Fast toolkit) have been used so far only for evaluating reachability analysis techniques (Demri et al. do not discuss empirical evaluation of their approach). As such, the Fast toolkit does not
come with any sample temporal properties for these systems.  Therefore, we had to study these 14 systems, and identify natural temporal properties. Since our implementation targets only $\eg$ properties, we took care to identify CTL properties that contained $\eg$ sub-properties, and supplied only these $\eg$ sub-properties to our tool. Across the 14 systems, we identified a total of 44 properties.  We provide more information about the 14 systems as well as the 44 properties in Appendix~\ref{app:sys-prop}.
In the sequel, we refer to each $\eg$ sub-property simply as the ``property''.

\subsection{Overview of results}
In Table~\ref{tbl:results}, each row in Columns~(5) onward depicts information corresponding to a single $\eg$ property of a single benchmark system. Thus, there are 44 rows in the table, corresponding to the 44 $\eg$ properties that we have chosen. We describe the properties themselves in more detail in the Appendix~\ref{app:sys-prop}.

We ran all our three algorithms (i.e., the full routine
$\computeGlobalName$, and its two variants separately) on each of 44 properties. We used a 1-hour timeout for each property for each algorithm.

 Columns~(4)-(6) of the table correspond to results from Variant-X of the
algorithm. Columns~(7)-(8) correspond to results from Variant-Y. Finally,
  Columns~(9)-(11) correspond to the results from the full routine
 $\computeGlobalName$.

The meanings of all the columns in the table are explained in the caption of the table. A note about the columns titled ``NFE'' and ``NI'': Recall that each iteration of the outer loop in the full routine~$\computeGlobalName$ involves generation of $k_1$ flattenings in the first inner loop of the routine (lines~\ref{line:inner-loop-beg}-\ref{line:inner-loop-end} in Algorithm~\ref{algo:compute-global}), followed by
$k_2$ iterations of the loop in sub-routine $\computeGlobalOverName$ (i.e., Algorithm~\ref{algo:over-approx}). In our implementation we choose
both $k_1$ and $k_2$ to be 1. Therefore, even though the full routine~$\computeGlobalName$ enumerates flattenings, we omit an ``NFE'' column for this routine, because with $k_1$ set to 1,  NFE is equal to NI (which occurs in Column~(10)). We have shown an ``FL'' column for Variant~X, but not for the full routine, because the value of this metric in all cases is quite small with the full routine.

The output from our tool for each property, which is a Presburger formula expressed in Fast's syntax, is sometimes quite verbose for systems of larger size such as the ones selected for evaluation in this section. This makes the output difficult to interpret manually. The verbosity arises due to two reasons:

\begin{itemize}
\item Fast's $\preStarName$ and $\postStarName$ operations, which our approach uses as black boxes, often return verbose formulas on larger systems.

\item Since our approach involves multiple applications of the operations mentioned above, the resultant formula's size grows with each application. 
\end{itemize}

Therefore, in order to proceed with the evaluation, for each property, using our understanding of the counter system and/or the underlying protocol being modeled by the system, we identified manually the set of states of the systems that
are expected to satisfy the given property. We call these the ``expected solutions''. We encoded these expected solutions programatically using the Fast APIs for constructing formulas (since these properties were specified manually, they were compact). We then wrote a simple piece of code that compared the actual computed solution for each property with the expected solution, and reported cases where they differed. The results of these comparisons are presented in Column~(11) in Table~\ref{tbl:results}. A `Y' in this column indicates that the actual computed solution \emph{does not} match the expected solution. Note that on two of the properties, namely, the ones represented by Rows~18 and~19 in Table~\ref{tbl:results}, the expected solution was actually too complex for a full programmatic check to be encoded. In these two cases, we selected a few states manually and checked that these states were either present in both the expected and the actual solutions, or not present in either. These two rows are marked with a `*' in Column~(11).

All our runs were done on a desktop computer with an Intel Core i5-2400 processor running at 3.01 GHz, with 8 GB of main memory. Our implementation is  single-threaded, because the data structures used in the reachability analysis part of Fast are not thread safe. 

\subsection{Discussion on termination and running time}

The first thing we would like to note with the results in Table~\ref{tbl:results} is that full routine $\computeGlobalName$ terminated on 42 of the 44 properties within the 1-hour timeout, not taking more than 247 milliseconds  on any of these 42 properties. Considering that these systems are pre-existing models of real protocols, and considering that we have attempted to choose natural temporal properties for each system, we believe that these results are very encouraging.

A related point to note is that the results we report here are significantly better than the preliminary results that we had reported in our earlier conference paper~\cite{kvasantafme2014}. In that paper, Variant~X or Variant~Y had terminated within the 1-hour timeout on only 10 properties out of 19 properties that we had identified at that time (the combined routine $\computeGlobalName$ was not yet developed at that time). Furthermore, on certain properties on which they terminated within the timeout, they took several seconds to a couple of minutes to terminate. There are two primary reasons for the improved performance that we report here:

\begin{itemize}
\item We now restrict the analysis of all our routines to within the reachable state-space of the systems. We had discussed this point earlier in this section.

\item  We remove all \emph{insignificant} transitions from the given counter system $M$. These are transitions such that only stuck states can be reached via this transition (in one step itself). 

Note that for some of the systems, for some temporal properties, all transitions get removed due to this effect. Therefore, our approaches terminate instantaneously. This is the reason why the `NFE' and `NI' columns are zero for some of the properties in Table~\ref{tbl:results}. 
\end{itemize}
 
We now make a set of more detailed observations about our results. The first is that Variant~Y (and the full routine~$\computeGlobalName$) terminated on a superset of the input properties as Variant~X. This is not a surprising outcome, because all example systems in the Fast toolkit have deterministic actions, and because all $\preStarName$ queries that Variant~Y or the full routine issued for all properties other than the last property (swimmingpool) happened to terminate. Recall our observation in  Section~\ref{ssec:term:discussion} that
on systems that have deterministic transitions
and on which all $\preStarName$ queries terminate, Variant~Y (and hence, the full routine also) necessarily terminates on
each property that Variant~X terminates on.  And on most of the properties where both variants terminated, Variant~Y was generally significantly faster whenever Variant~X had to enumerate multiple flattenings, while Variant~X was generally slightly faster whenever it had to enumerate either no flattenings or just one flattening. 

Note on the system swimmingpool, Variant~X did not finish processing even one flattening, while Variant~Y and the full routine did not complete one iteration of the loop in routine $\computeGlobalOverName$. This happens because, on this system, each  $\preStarName$ and $\prekName$ query itself takes more than 1 hour time. 

Coming to the full routine~$\computeGlobalName$, this routine generally needed slightly more running time than Variant~Y. This was because of the need to generate and processing flattenings (as many flattenings as mentioned in the `NI' column.) 
Note that in our experiments there was no input property on which $\computeGlobalName$ terminated but on which Variant~Y did not terminate.

\subsection{Usefulness of our tool}
\label{ssec:usefulness-of-results}


A tool such as ours can be useful to protocol and mechanism modelers in a couple of different ways. We explore these use cases in this section.

The first use of our tool is in verifying properties.  In 26 out of the 44 properties that we evaluated our approaches on, the solution computed by the full routine $\computeGlobalName$ matched exactly the expected solution (we had discussed earlier in Section~\ref{ssec:benchmarks-selection} how we checked these matches programmatically). These are the properties that do not have a `*' or a `Y' in the last column and don't have a ``(TO)'' in Column~(9) in Table~\ref{tbl:results}. In other words, these 26 properties can be considered as being verified.

A notable point is that on several of these properties, initially we found that the computed solution did not match the expected solution. On closer inspection, we came to the conclusion that we had misunderstood the system as well as the expected solution to some extent. We then corrected the expected solution, upon which a full match was detected. In other words, a second use of our tool is to assist users in coming to a better understanding of a system via the route of automated property checking.

On 14 properties, which are marked with a `Y' in the last column in Table~\ref{tbl:results}, the computed solution did not match the expected solution. In these cases also our approach helped us observe surprising facts about the models, and hence to understand them better.  For instance, in the system ``firefly'', all traces of the system turned out to be finite traces. This was surprising, given this system (and indeed, all other systems in the suite) are reactive system. It was not clear to us whether and how the system could be modified to remedy this situation. As another example, we identified an issue with the system `consistency', namely, that it came with an initial states specification (i.e., $\phiinit$) that was smaller than what we expect it to be. It is specified as `idle = 1', whereas we expected it to be `idle $\geq$ 1'.  As a final example, in the system `lastinfirstserved', in which there are processors requesting resources, our expectation that there would be no traces where a processor would be starved upon issuing a request; this expectation did not hold true. However, in this instance, we were not sure whether starvation is naturally a feature of the underlying protocol that the system is modeling.

All told, we feel that our tool enables model designers to understand the properties of their models better, and in some cases to refine their models. 

We provide a more in-depth version of the discussion above in Appendix~\ref{app:sys-prop}.

\subsection{Summary of empirical evaluation}

To summarize, our full routine  terminated on 42 of the 44 properties that we evaluated it on, taking only 247 ms in the worst case to solve for a single property. Also, our approach was able to verify 26 of the given properties as being satisfied, and was able to improve our understanding of the systems substantially in many of the remaining cases. 

In comparison with pre-existing
approaches for checking CTL properties~\cite{dem2006,gaporder12}, we are the first to report empirical
evidence on real systems using an implementation. Furthermore, all the systems that we have used for evaluation are non-trace-flattable, and therefore outside the class of systems handled by the approach of Demri et al~\cite{dem2006}.




\section{Related Work}
\label{sec:related_work}
Research work on model-checking $\ctl$ properties in counter systems has progressed along side the
developments in techniques to answer reachability on these systems.  The approach of Bultan et al.~\cite{sip1997} is an early approach; it does not use accelerations~\cite{Com1998,finkel02,trex_fast_accelerations05,iteratingOctagons} to traverse sequences of concrete transitions at one go, and is subsumed by subsequently proposed approaches~\cite{dem2006,gaporder12} that do use accelerations.
These approaches both build a summary of all possible traces in the given counter system using accelerations. This summary is then checked against the given temporal property. The two key technical differences of our approach over these are: (a) Rather than attempting to summarize \emph{all} the traces in the system, we use refinement and then accelerations to characterize only the traces that satisfy the \emph{given property}. (b) We use \emph{repeated} reachability queries, and not a single phase of applying accelerations. The consequences of these differences are as follows. Due to features~(a) and~(b) above, as discussed  in Sections~\ref{ssec:under-approx-thms} and~\ref{sec:over_approx_correctness}, we target systems beyond trace flattable systems, and terminate with precise results on a wider class of systems than the approach of Demri et al.~\cite{dem2006}. 
The practical importance of this is borne out by our empirical studies. Feature~(a) also enables us to solve arbitrarily nested CTL properties, while feature~(b) enables us to compute approximated solutions in cases where a precise computation may not be possible, which is very useful in practice. The previous approaches do not possess these advantages.  Finally, the previous approaches did not provide empirical results using implementations.

There are a few other noteworthy points about the previous approaches mentioned above.  The approach of Bozelli et al.~\cite{gaporder12} does not have the finite-branching restriction. Also, although neither previous approach addresses arbitrarily nested CTL properties, they address certain operators of CTL* that we do not address. 

Cook et al.~\cite{cook-2011,cook-2013} proposed a technique to model check arbitrarily nested
temporal properties in a restricted class of C programs.  The major
difference is that we address the ``global'' model-checking problem, wherein we return a formula that encodes \emph{all} states that satisfy a property. In their case they check whether a \emph{given} set of states satisfies a property. Also, they do not have capabilities for approximations. Nevertheless, an interesting investigation for future work would be to compare the classes of systems targeted by them and by us.

\rule[1.5em]{0in}{0in}\textbf{Acknowledgments:} We thank A. Finkel and J. Leroux for their suggestions and for their help with the Fast tool.

%
\bibliographystyle{ieeetr}\bibliography{references}

\begin{thebibliography}{10}

\bibitem{fast}
``{FASTer}.''
\newblock http://altarica.labri.fr/forge/projects/faster/wiki/.

\bibitem{Bouajjani97}
A.~Bouajjani, J.~Esparza, and O.~Maler, ``Reachability analysis of pushdown
  automata: Application to model-checking,'' in {\em Proc. Int. Conf. on
  Concurrency Theory}, CONCUR '97, pp.~135--150, 1997.

\bibitem{esparza1998decidability}
J.~Esparza, ``Decidability and complexity of petri net problems—an
  introduction,'' in {\em Lectures on Petri Nets I: Basic Models},
  pp.~374--428, Springer, 1998.

\bibitem{Com1998}
H.~Comon and Y.~Jurski, ``Multiple counters automata, safety analysis and
  presburger arithmetic,'' in {\em Proc. Comp. Aided Verification (CAV)},
  pp.~268--279, 1998.

\bibitem{Ibarra02}
O.~H. Ibarra, J.~Su, Z.~Dang, T.~Bultan, and R.~A. Kemmerer, ``Counter machines
  and verification problems,'' {\em Theoret. Comp. Sc.}, vol.~289, no.~1,
  pp.~165--189, 2002.

\bibitem{finkel02}
A.~Finkel and J.~Leroux, ``How to compose presburger-accelerations:
  Applications to broadcast protocols,'' {\em Technical Report, Labor. Specif.
  et Verif. (LSV)}, 2002.

\bibitem{trex_fast_accelerations05}
C.~Darlot, A.~Finkel, and L.~{Van Begin}, ``About fast and trex
  accelerations,'' {\em Electronic Notes in Theoretical Computer Science},
  vol.~128, pp.~87--103, May 2005.

\bibitem{iteratingOctagons}
M.~Bozga, C.~G\^{\i}rlea, and R.~Iosif, ``Iterating octagons,'' in {\em Proc.
  Tools and Algorithms for the Constr. and Analysis of Systems (TACAS)},
  pp.~337--351, 2009.

\bibitem{book_clarke}
J.~E.~M. Clarke, O.~Grumberg, and D.~A. Peled, {\em Model checking}.
\newblock MIT Press, 1999.

\bibitem{dem2006}
S.~Demri, A.~Finkel, V.~Goranko, and G.~V. Drimmelen, ``Towards a model-checker
  for counter systems,'' in {\em Proc. Automated Tech. for Verif. and Analysis
  (ATVA)}, pp.~493--507, 2006.

\bibitem{gaporder12}
L.~Bozzelli and S.~Pinchinat, ``Verification of gap-order constraint
  abstractions of counter systems,'' in {\em Proc. of Verif., Model Checking,
  and Abs. Interp. (VMCAI)}, pp.~88--103, 2012.

\bibitem{alpernLiveness1984}
B.~Alpern and F.~B. Schneider, ``{Defining liveness},'' {\em Information
  Processing Letters}, vol.~21, no.~4, pp.~181--185, 1985.

\bibitem{sip1997}
T.~Bultan, R.~Gerber, and W.~Pugh, ``Symbolic model checking of infinite state
  programs using presburger arithmetic,'' in {\em Proc. Comp. Aided Verif.
  (CAV)}, pp.~400--411, 1996.

\bibitem{cook-2011}
B.~Cook, E.~Koskinen, and M.~Vardi, ``Temporal property verification as a
  program analysis task,'' in {\em Proc. Conf. Comp. Aided verification (CAV)},
  pp.~333--348, 2011.

\bibitem{cook-2013}
B.~Cook and E.~Koskinen, ``Reasoning about nondeterminism in programs,'' in
  {\em Proc. Conf. on Progr. Lang. Design and Impl. (PLDI)}, pp.~219--230,
  2013.

\end{thebibliography}
\newpage
\appendices
\section{Proofs of Theorems}
\label{app:proofs}
\setcounter{theorem}{0}
\begin{theorem}
    Given a counter system $M=\langle Q, C,\Sigma,
    \phiinit,G,F\rangle$ and sets of states $\phi_1$ and $\phi_2$, routine
    $\computeUntilName$ in Algorithm~\ref{algo:until} returns  precisely
    the set of states that satisfy $\euntil{\phi_1}{\phi_2}$ whenever the
    invocation $\computePreStar{M_1}{\phi_2}{\enumlabel}$ in the routine returns
    the precise formula $\preStar{M_1}{\phi_2}$, where $M_1 \equiv
    \refineSystem{M}{\phi_1}$.
\end{theorem}

\begin{proof}

    \textbf{Claim A:} If a state $\vec{s}\models\euntil{\phi_1}{\phi_2}$ in the
    counter system $M$, then
    $\vec{s}\in \preStar{M_1}{\phi_2}$.

    \begin{flalign}
        \label{eqn:untilForward}
        &\vec{s}\models\euntil{\phi_1}{\phi_2}&\nonumber\\
        \implies &\exists k\geq 0.\, \exists \vec{s_0},\dots,\vec{s_k}.&\nonumber\\
                 & \ \ \vec{s_k}\models\phi_2 \wedge \vec{s}=\vec{s_0} \wedge
        (\forall i, 0 \leq i < k.\, (\vec{s_i} \models \phi_1) \wedge & \nonumber \\
                                                                      &(\exists
        b\in\Sigma.\, \transition{\vec{s_i}}{M}{b}{\vec{s_{i+1}}}))&\nonumber\\
        \implies &\exists k\geq 0.\, \exists \vec{s_0},\dots,\vec{s_k}.&\nonumber\\
                 & \ \ \vec{s_k}\models\phi_2 \wedge
        \vec{s}=\vec{s_0} \wedge & \nonumber \\
                                 &(\forall i, 0 \leq i < k.\,  \exists b\in\Sigma.\, (\vec{s_i} \models (g_b
        \wedge \phi_1)) \wedge
        \transition{\vec{s_i}}{M}{b}{\vec{s_{i+1}}}),&\nonumber\\
                                                     &\mbox{where $g_b$ is the guard of transition $b$ in $M$}&\nonumber\\
        \implies &\exists k\geq 0.\, \exists \vec{s_0},\dots,\vec{s_k}.&\nonumber\\
                 & \ \ \vec{s_k}\models\phi_2 \wedge 
        \vec{s}=\vec{s_0} \wedge& \nonumber \\
                                &(\forall i, 0 \leq i < k.\,  \exists b\in\Sigma.\, (\vec{s_i} \models g_{b_1})
        \wedge \transition{\vec{s_i}}{M_1}{b}{\vec{s_{i+1}}}),&\nonumber\\
                                                              &\mbox{where $g_{b_1}$ is the guard of transition $b$ in $M_1$}&\nonumber\\
        \implies &\vec{s}\models\preStar{M_1}{\phi_2}&
    \end{flalign}

    Proof of the converse is as follows:\\
    \textbf{Claim B:} If $\vec{s} \in \preStar{M_1}{\phi_2}$, then
    $\vec{s}\models\euntil{\phi_1}{\phi_2}$ in the
    counter system $M$.

    The steps in this proof are simply the reverse of the steps in the proof of
    Claim A.

    The proof of Theorem~\ref{thm:untilPrecise} follows directly from Claims~A
    and~B above.
\end{proof}

\begin{theorem}
    Given a counter system $M=\langle Q, C,\Sigma,
    \phiinit,G,F\rangle$ and sets of states $\phi_1$ and $\phi_2$, routine
    $\computeUntilName$ in Algorithm~\ref{algo:until} returns  an over- (resp.
    under-) approximation of
    the set of states that satisfy $\euntil{\phi_1}{\phi_2}$ whenever the
    invocation $\computePreStar{M_1}{\phi_2}{\enumlabel}$ in the routine
    returns an over- (resp. under-) approximation of the set of states that
    satisfy the formula  $\preStar{M_1}{\phi_2}$, where $M_1 \equiv
    \refineSystem{M}{\phi_1}$.
\end{theorem}

\begin{proof}
    Note that the formula that is returned by $\computeUntilName$ is nothing
    but the formula $\phi$ that is returned by  the invocation
    $\computePreStar{M_1}{\phi_2}{\enumlabel}$. Now, as per Claim~A (resp.
    Claim~B) in the proof of Theorem~\ref{thm:untilPrecise},
    $\preStar{M_1}{\phi_2}$ over- (resp. under-) approximates the set of
    states that satisfy  $\euntil{\phi_1}{\phi_2}$ in $M$. Therefore, it
    follows  that if $\phi$ over- (resp. under-) approximates
    $\preStar{M_1}{\phi_2}$, then it also over- (resp. under-) approximates
    the set of states that satisfy $\euntil{\phi_1}{\phi_2}$ in $M$.
\end{proof}

\begin{lemma}
    The following properties hold at any point during the execution of
    Algorithm~\ref{algo:compute-global}:

    \begin{enumerate}
        \item the set $Y$ is $\postStarName$-\emph{closed}; i.e., every
            successor of every state in $Y$ is also in $Y$.

        \item
            $\growone$ in the loop in subroutine $\computeGlobalOverName$ iff
            the state is not in $Y$ and all successors of this state are
            already in $Y$.

        \item
            $\growtwo$ in the loop in subroutine $\computeGlobalOverName$ iff:

            \begin{enumerate}
                \item there is a unique trace $t$ that starts from $\vec{s}$
                    and ends at a
                    state that satisfies $\growone$, and
                \item no state in this trace is currently in $Y$, and
                \item
                    every state $\vec{s'}$ that is outside $t$ and that is a
                    successor of
                    any state in $t$ is already in $Y$.
            \end{enumerate}
    \end{enumerate}
\end{lemma}

\begin{proof}
    Consider Algorithm~\ref{algo:compute-global}. States get added to $Y$ at
    two locations: line~\ref{line:growYb} in this algorithm itself, and in
    line~\ref{line:over-growY} in the subroutine $\computeGlobalOverName$
    (which is described as Algorithm~\ref{algo:over-approx}). We prove this
    lemma by induction on the number of visits to these locations during the
    run of Algorithm~\ref{algo:compute-global}.

    \textbf{Base case:} $Y$ is initialized in line~\ref{line:init-xy} to
    stuck states and states that do not satisfy $\phi$. In the refined
    system $M_1$ these states are clearly $\postStarName$-closed. Also,
    since control is not inside subroutine $\computeGlobalOverName$ at this
    point, Properties~\ref{it:lem:grow:1} and~\ref{it:lem:grow:2} in this
    lemma's statement are true trivially.\\

    \textbf{Inductive case:} The inductive hypothesis is that the two
    locations mentioned above have been reached a total of $k$ times so far,
    and that $Y$ is $\postStarName$-closed after these $k$ visits. We now
    consider the $(k+1)$th visit to any of these two locations. There are
    three possible scenarios here.\\

    \emph{Scenario 1.} In this scenario the $(k+1)$th visit is to
    line~\ref{line:growYb}.

    Recall, from Section~\ref{sssec:under-termination}, that the set of
    states $\phi'$ returned by $\checkflatteningName$ is the set of states
    that do not satisfy $\eglobally{\phi}$ in the flattening $N$ of $M_1$
    and such that traces from these states are the same in $N$ as in $M_1$.
    It follows from this $\phi'$ is $\postStarName$-closed. Therefore, $Y$
    remains $\postStarName$ closed. Furthermore, since
    Properties~\ref{it:lem:grow:1} and~\ref{it:lem:grow:2} of the lemma
    statement are applicable only when control is inside subroutine
    $\computeGlobalOverName$, they are true trivially.\\

    \emph{Scenario 2.} In this scenario, the $(k+1)$th visit is a visit to
    line~\ref{line:over-growY} in subroutine $\computeGlobalOverName$.

    Property~\ref{it:lem:grow:1} of the lemma's statement is easy to see.
    Also, since all successors of any state $\vec{s}$ that gets added to $Y$
    due to formula $\growone$ are already in $Y$, it is easy to see that $Y$
    remains $\postStarName$-closed.

    We now consider Property~\ref{it:lem:grow:2} of the lemma. Formula
    $\growtwo$ is defined as the conjunction of formulas $\growtwoa$ and
    $\growtwob$ (see Section~\ref{sssec:over-working}). $\growtwoa$ is
    defined as $\neg (\preStar{M_1}{ \neg \asoY})$. $\asoY$, as per its
    definition, identifies states that do not have two different successors
    such that both are not currently in $Y$. In other words, $\growtwoa$ is
    precisely the set of states $\vec{s}$ from which no state that has two
    or more successor states that are currently not in $Y$ can be reached.
    $\growtwob$ is nothing but $\preStar{M_1}{\growone}$.

    The ``if'' direction of the property can be argued as follows. Let
    $\vec{s''}$ be the last state in the trace $t$. Therefore, all
    successors of $\vec{s''}$ are already in $Y$. Furthermore, as per the
    statement of Property~\ref{it:lem:grow:2}, all successors of states in
    $t$ that are outside $t$ are already in $Y$. From this, and by the
    inductive assumption that $Y$ is currently $\postStarName$-closed, it
    follows that the only states not in $Y$ that are reachable from
    $\vec{s}$ are the states in $t$. These states form a single trace.
    Therefore, $\vec{s}$ satisfies formula $\growtwoa$. Also, as per
    Property~\ref{it:lem:grow:2}, state $\vec{s''}$ satisfies $\growone$.
    Therefore,  every state in $t$ (including $\vec{s}$) satisfies
    $\growtwob$. Therefore, state $\vec{s}$ (as well as all other states in
    $t$ until $\vec{s''}$ satisfy formula $\growtwo$.

    We now prove the ``only if'' direction. (i) Since state $\vec{s}$
    satisfies $\growtwo$, it satisfies $\growtwoa$ and $\growtwob$. (ii)
    Since $\vec{s}$ satisfies $\growtwob$, there exists a state $\vec{s''}$
    that satisfies $\growone$ that is reachable from $\vec{s}$. Also, since
    $\vec{s''}$ satisfies $\growone$, it follows that it is not in Y. (iii)
    Consider any trace $t$ from $\vec{s}$ to $\vec{s''}$. Since $\vec{s'}$
    is not in $Y$, and since $Y$ is assumed (by the inductive hypothesis) to
    be $\postStarName$-closed, it follows that none of the states in $t$
    (including $\vec{s}$ and $\vec{s''}$) are in $Y$. (iv) Since $\vec{s}$
    satisfies $\growtwoa$, it follows that no state in $t$ has a successor
    outside $t$ such that the successor is not in $Y$. It also follows from
    this that other than $t$ there can be no other trace from $\vec{s}$ to
    $\vec{s''}$. Therefore, we are done showing that
    Property~\ref{it:lem:grow:2} in the lemma statement is satisfied.

    Note that in the scenario above, along with $\vec{s}$, all other states
    in $t$ also get added to $Y$. We need to show this preserves the
    $\postStarName$-closedness of $Y$. This follows from the following
    reasoning: (i) every successor of every state in $t$ is either already
    in $Y$, or is in $t$. (ii) By the inductive hypothesis, $Y$ is
    $\postStarName$-closed at the point before the states in $t$ are added
    to $Y$.
\end{proof}
\begin{theorem}
    The two sets of states $X$ and $Y$ that are maintained
    by Algorithm~\ref{algo:compute-global} are such that (a)
    $X$ is a growing under-approximation of the set of
    states that satisfy the property $\eglobally{\phi}$ in
    $M_1$, and (b) $Y$ is a growing under-approximation of
    the set of states that satisfy $\neg \eglobally{\phi}$
    in $M_1$.
\end{theorem}

\begin{proof}
    The proof is by induction on the number of iterations of the outer
    loop in Algorithm~\ref{algo:compute-global}.\\

    \textbf{Base case:} $X$ is initialized (in
    line~\ref{line:init-xy} of
    Algorithm~\ref{algo:compute-global}) to $\emptyset$, which
    is obviously an under-approximation of $\eglobally{\phi}$.
    $Y$ is initialized to the set of states that do not satisfy
    $\phi$ or that are stuck states. These states clearly
    satisfy $\neg \eglobally{\phi}$.\\

    \textbf{Inductive case:} After $n$ iterations, for any $n >
    0$, assume that $X$ is an under-approximation of the set of
    states that satisfy $\eglobally{\phi}$, and that $Y$ is an
    under-approximation of the set of states that satisfy $\neg
    \eglobally{\phi}$ in $M$. Consider iteration $n+1$. \\

    \emph{Addition to $X$ in line~\ref{line:growX} of
    Algorithm~\ref{algo:compute-global}:} Say a state $\vec{s}$
    gets added to $X$ due to a flattening $N$ of $M_1$. That
    is, $\vec{s} \, \in \, \forall k\geq 0.\ \prek{N}{\phi}$.
    That is, there exist traces of all possible lengths
    starting from $N$. Due to our finite-branching assumption
    (see Section~\ref{sec:prelim}), K\"{o}nig's Lemma is
    applicable, which means there is an infinite length trace
    in $N$ starting from $\vec{s}$. Now, since  $N$ is a
    flattening of $M_1$,
    $\traces{N}{\phi}\subseteq\traces{M_1}{\phi}$. Therefore,
    the same infinite trace is present in $M_1$. Since $M_1$ is
    refined wrt $\phi$,  every state on this infinite trace
    satisfies $\phi$. In other words, $\vec{s}$ satisfies
    $\eglobally{\phi}$ in $M_1$. Therefore, after addition of
    $\vec{s}$ to $X$, $X$ remains an under-approximation of
    $\eglobally{\phi}$. \\

    \emph{Addition to $Y$ in line~\ref{line:growYb} of
    Algorithm~\ref{algo:compute-global}:} The states $\phi'$
    that are added to $Y$ here are the states that do not
    satisfy $\eglobally{\phi}$ in the flattening $N$ and from
    which traces are identical in $M_1$ and $N$. Therefore,
    these states do not satisfy $\eglobally{\phi}$ in $M_1$,
    either. Therefore, $Y$ remains an under-approximation of
    $\neg \eglobally{\phi}$. \\

    \emph{Addition to $Y$ in line~\ref{line:growY} of
    Algorithm~\ref{algo:compute-global}:} This location in the
    algorithm  involves a call to subroutine
    $\computeGlobalOverName$, which was described in
    Algorithm~\ref{algo:over-approx}. This subroutine is itself
    iterative. It suffices for us to prove that in any
    iteration of this subroutine, if $Y$ was an
    under-approximation of $\neg \eglobally{\phi}$ at the
    beginning of the iteration, then $Y$ remains an
    under-approximation of $\neg \eglobally{\phi}$ at the end
    of the iteration. A state $\vec{s}$ could get added to $Y$
    in the iteration due to two reasons, each of which is
    discussed separately below.

    If $\vec{s}$ satisfies $grow_1$, then all successors of
    $\vec{s}$ are in $Y$. Since states in $Y$ are assumed to
    satisfy $\neg \eglobally{\phi}$, the same would be true
    about $\vec{s}$.

    If $\vec{s}$ satisfies $grow_2$, then as per
    Lemma~\ref{lem:grow}, all traces from $\vec{s}$ end at a
    state in $Y$ within a finite number of transitions.
    Therefore, following the same reasoning as in the previous
    paragraph, $\vec{s}$ would not satisfy $\eglobally{\phi}$.

\end{proof}

\begin{theorem}
    If Algorithm~\ref{algo:compute-global} terminates on its own and returns a
    set of states, then this return value is precisely the set of states that
    satisfy $\eglobally{\phi}$ in the counter system $M_1$.
\end{theorem}

\begin{proof}
    Algorithm~\ref{algo:compute-global} terminates on its own in two cases. We
    consider each of these cases separately.\\

    \emph{Case 1 (the condition in line~\ref{line:term-Under} in
    Algorithm~\ref{algo:compute-global} is true):} As per
    Theorem~\ref{thm:computeGlobal-invariant}, $X$ is always an under-approximation
    of $\eglobally{\phi}$. Therefore, all we need to show now is that for state
    $\vec{s}$ that is not in $X$, this state cannot satisfy $\eglobally{\phi}$.

    If the state $\vec{s}$ does not satisfy $\phi$ then it cannot satisfy
    $\eglobally{\phi}$. Hence, we are done.

    If $\vec{s}\in Y$, then by Theorem~\ref{thm:computeGlobal-invariant} $\vec{s}$
    does not satisfy $\eglobally{\phi}$. Hence, we are done.

    The final possibility is that $\vec{s}$ is not in $Y$. Now, $\vec{s}\notin X$
    implies that
    $\vec{s}\notin\forall k\geq 0.\ \prek{N}{\phi}$.
    That is, there is no infinite length path in $N$
    starting from $\vec{s}$. As per the working of subroutine
    $\checkflatteningName$, which was discussed in
    Section~\ref{sssec:under-termination}, the condition in
    line~\ref{line:term-Under} will be true only  when
    $\traces{M_1}{\phi-X-Y}=\traces{N}{\phi-X-Y}$. Since $\vec{s}$ is present in
    $\phi-X-Y$, it follows that there is no infinite length path in
    $M_1$ from the state $\vec{s}$. Since $M_1$ is refined wrt $\phi$, it follows
    that $\vec{s}$ does not satisfy $\eglobally{\phi}$ in $M_1$. \\

    \emph{Case 2 (the condition in line~\ref{line:term-Over} in
    Algorithm~\ref{algo:compute-global} is true):}
    \sloppypar The condition mentioned above can be true only when subroutine
    $\computeGlobalOverName$ (which is invoked in line~\ref{line:growY}) returns
    the same set $Y$ that is passed to it. As per Algorithm~\ref{algo:over-approx},
    which implements the subroutine $\computeGlobalOverName$, this happens only
    when neither condition $\growone$ nor condition $\growtwo$ is satisfiable. We
    show below that under this scenario $Y$ already represents precisely the set of
    all states that do not satisfy $\eglobally{\phi}$. Note that in this case the
    ``return'' statement in line~\ref{line:return-Over} will return precisely the
    set of states that satisfy $\eglobally{\phi}$.

    Our argument is by contradiction.  Let $\vec{s}$ be a state that does not
    satisfy $\eglobally{\phi}$ and has not been included in $Y$. Our strategy is to
    show by construction the presence of a trace $t$ that starts from $\vec{s}$,
    that visits only states not in $Y$, and that ends in a state that satisfies
    $\growone$. This state would have been added to $Y$ in line~\ref{line:growY}.
    Therefore, the condition in line~\ref{line:term-Over} would have been false,
    which contradicts our assumption.

    The first state in $t$ is $\vec{s}$ itself. We then grow $t$ by extending it
    repeatedly at each step into any successor state that is not in $Y$. This
    extension process cannot happen indefinitely, because that would imply the
    presence of an infinitely long trace from $\vec{s}$, which would in turn imply
    that $\vec{s}$ satisfies $\eglobally{\phi}$. Therefore, say $\vec{s_i}$ is the
    last state to have been identified as part of $t$ by this process. $\vec{s_i}$
    cannot be a stuck state, because stuck states are added to $Y$ by the algorithm
    initially itself. Therefore, $\vec{s_i}$ must have one more successors, and all
    of the successors must be outside $Y$ (otherwise $t$ could be extended
    further). This means that  $\vec{s_i}$ satisfies $\growone$, which means we are
    done.

\end{proof}

\begin{theorem}
    If Algorithm~\ref{algo:compute-global} is forcibly terminated, and returns
    a pair ($\phi', \enumlabel\in\latticeApprox$), then $\phi'$ is an
    approximation of $\eglobally{\phi}$ in the direction indicated by
    $\enumlabel$. 
\end{theorem}

\begin{proof}
    The algorithm returns a result upon forcible termination either at
    line~\ref{line:return-Under-force} or at line~\ref{line:return-Over-force}
    of Algorithm~\ref{algo:compute-global}. In the first scenario, the desired
    result follows immediately, because as per
    Theorem~\ref{thm:computeGlobal-invariant}, $X$ is an under-approximation of
    $\eglobally{\phi}$. In the second scenario, since $\phi$ is an
    over-approximation of $\eglobally{\phi}$, and since by
    Theorem~\ref{thm:computeGlobal-invariant} $Y$ is an under-approximation
    $\neg \eglobally{\phi}$, it follows that $\phi - Y$ is an
    over-approximation of $\eglobally{\phi}$.
\end{proof}

\begin{theorem}
    Given a flat counter system $M$ and any property
    $\eglobally{\phi}$, Algorithm~\ref{algo:compute-global}
    will always terminate, provided: (a) All $\preStarName$
    queries on $M_1$ terminate and all $\prekName$ queries on
    flattenings of $M_1$ terminate, (b) The action $f_b$ of
    each transition $b$ of $M$ is \emph{deterministic}, in the
    sense that it maps each state $\vec{s}$ to a unique state
    $\vec{s'}$, and (c) the check
    $\checkflattening{M_1}{N}{\phi-X-Y}$ always terminates.
\end{theorem}

\begin{proof}
    Let $M_1$ be the refinement of $M$ with respect to
    $\phi$. We actually prove a stronger property that
    Algorithm~\ref{algo:compute-global} will
    necessarily terminate if $M_1$ is flat. Note that
    this is a stronger property: $M_1$ will be flat if
    $M$ is flat, but not necessarily vice versa.\\[1em]

    \textbf{Base Case(n=0):} If a control state $q$ has
    order 0, it must have been assigned this order by
    Rule~\ref{order:no-succs}. It follows that $q$ has
    no control-state successors.  Therefore, all states
    that correspond to $q$ are stuck states.
    Therefore, all these states will be added to $Y$ in
    subroutine $\computeGlobalOverName$ before the loop
    in this subroutine is entered.\\[1em]

    \textbf{Inductive case:} Let the statement $P(k)$
    be true.  We now prove that statement $P(k+1)$ is
    true.  Let $q$ be a control state such that
    $\order{{M_1}}{(q)} = k+1$.  There are four
    different cases under which a state $\vec{s}$ that
    has not been added to $Y$ so far and that
    corresponds to control state $q$ will not satisfy
    $\eglobally{\phi}$.\\

    \emph{Case 1:} Let $q$ be a control state that is
    not part of any cycle, and let $q$ have only one
    successor control-state, namely, $q_1$. Let
    $\order{M_1}{(q_1)}$ be zero and let
    $\order{M_1}{(q)}$ be 1. (This scenario is
    addressed by Rule~\ref{order:succ-0} above.) Since
    the order of $q_1$ is zero, all states that
    correspond to $q_1$ are stuck states, and would
    have been added to $Y$ before the loop in
    subroutine $\computeGlobalOverName$.  Since $q$'s
    only successor is $q_1$, and due to our assumption
    that each transition of $M_1$ is deterministic,
    each state that corresponds to $q$ has single
    successor state, which is a stuck state.
    Therefore, all states that correspond to $q$
    satisfy $\growone$ in the first iteration of the
    loop in subroutine $\computeGlobalOverName$, and
    would have been added to $Y$ in the first
    iteration.\\

    \emph{Case 2:} Let $q$ be a control state such that
    it does not belong to any cycle in $M_1$ and has
    multiple successor control states. For any state
    $\vec{s}$ that corresponds to $q$, if $\vec{s}$
    does not     satisfy $\eglobally{\phi}$, then
    $\vec{s}$ could be a stuck state itself, in which
    case it would have been added to $Y$ before the
    loop in subroutine $\computeGlobalOverName$. Or
    else, all successor states of $\vec{s}$ are such
    they do not satisfy $\eglobally{\phi}$, and
    correspond to control states of order at most $k$
    (as per Rule~\ref{order:mult-succ}).  Hence, these
    states would have been added to $Y$ within $k$
    iterations (due to the inductive hypothesis).
    Therefore, if $\vec{s}$ is not yet in $Y$ by the
    end of $k$ iterations, then it satisfies $\growone$
    and will hence get added to $Y$ in the $(k+1)$th
    iteration.\\

    \emph{Case 3:} Let $q$ be a control state that is
    not part of any cycle, and let $q$ have only one
    successor control-state, namely, $q_2$.  Let
    $\order{M_1}{(q_1)}$ be $k+1$, where $k+1 > 0$.
    (This scenario is addressed by
    Rule~\ref{order:single-succ} above.) In this
    scenario, there must exist a unique maximal
    sequence of control states $(q_1 = q), q_2, \dots,
    q_l$ such that $\forall i. 1 \leq i < l$, (a)
    $\order{{M_1}}{(q_i)} = (k+1)$, (b) $q_i$ does not
    belong to any cycle, (c) $q_{i+1}$ is the only
    successor control state of $q_i$, and (d)
    $\order{{M_1}}{(q_l)} = (k+1)$.  (Note that it is
    possible that that the sequence above is of length
    1, in which case $l=1$.)

    Consider any state $\vec{s}$ that corresponds to
    $q$, that does not satisfy $\eglobally{\phi}$, and
    that has not yet been added to $Y$ at the end of
    $k$ iterations of the loop in subroutine
    $\computeGlobalOverName$.  One of the following
    four scenarios has to hold. 

    \begin{description} 
        \item{Scenario 1.} There exists a single trace $t$ from $\vec{s}$, and
            this trace ends in a stuck state before leaving the sequence of
            control states $q_1, q_2, \dots, q_l$.  In this scenario, $\vec{s}$
            could be a stuck state, in which case it will get added to $Y$
            before the loop in subroutine $\computeGlobalOverName$ is entered.
            Else, as per Lemma~\ref{lem:grow}, the trace mentioned above will
            cause $\vec{s}$ to satisfy $\growone$ or $\growtwo$ in the first
            iteration of the loop in subroutine $\computeGlobalOverName$.
            $\growone$ requires only a \emph{pre} computation, which always
            terminates.  $\growtwo$ requires a $\preStarName$ computation;
            however, we have assumed that all $\preStarName$ queries on the
            system $M_1$ terminate.  Therefore, it follows that $\vec{s}$ will
            be added to $Y$ in the first iteration of the loop mentioned above.
            (In the remainder of this proof, for purposes of brevity, we will
            implicitly assume that any state that satisfies $\growone$ or
            $\growtwo$ in an iteration will get added to $Y$ in that
            iteration.) 

        \item{Scenario 2.} $q_l$ has multiple control-state successors, and all
            control-state successors of $q_l$ are of order at most $k$.  In
            this scenario, it follows that there is a single trace from
            $\vec{s}$ with no branching until a state $\vec{s_m}$ which
            corresponds to $q_l$ is reached, and that $\vec{s_m}$ is not a
            stuck state, and that $\vec{s_m}$ does not satisfy
            $\eglobally{\phi}$.  Let $t$ be the trace mentioned above that
            starts at $\vec{s}$ and ends at $\vec{s_m}$.  If any state in this
            trace is already in $Y$ by the end of $k$ iterations of the loop in
            subroutine $\computeGlobalOverName$, then due to
            Lemma~\ref{lem:grow}, $\vec{s}$ will be added to $Y$ in iteration
            $k+1$.  Otherwise, $s_m$ is not yet in $Y$; however, all its
            successor states are such that they (a) correspond to
            control-states of order at most $k$ (as per the definition of the
            current scenario), (b) do not satisfy $\eglobally{\phi}$, and (c)
            have already been added to $Y$ within $k$ iterations (as per the
            inductive hypothesis).  Therefore, $\vec{s_m}$ satisfies $\growone$
            in iteration $k+1$.  Hence, $\vec{s}$ satisfies $\growtwo$ in the
            same iteration, and will therefore get added to $Y$ in this
            iteration. 

        \item{Scenario 3.} $q_l$ is part of a cycle $c$.  Therefore, there
            exists a sequence of states starting from $\vec{s}$ that reaches a
            state $\vec{s_m}$ that corresponds to $q_l$, such every state in
            this sequence that precedes $\vec{s_m}$ has only the next state in
            the sequence as its successor.  Let us call this sequence (i.e.,
            trace) $t_1$.  Due to determinism of the actions of all transitions
            of $M_1$, there is a unique trace $t_2$ that starts from
            $\vec{s_m}$ that does not leave $c$.  If this trace is infinite,
            then $\vec{s_m}$ (and hence $\vec{s}$ satisfy $\eglobally{\phi}$.
            Therefore, in the rest of this argument we only consider the case
            where $t_2$ ends in a stuck state $\vec{s_n}$ that is inside $c$.
            Let $t_3$ be a trace that represents the concatenation of $t_1$ and
            $t_2$.  Let $E$ be the set of states \emph{outside} $c$ that are
            successors of states in $t_2$.  If any of these states satisfy
            $\eglobally{\phi}$, then $\vec{s}$ would also satisfy
            $\eglobally{\phi}$.  Therefore, in the rest of this argument we
            only consider the case where these states do not satisfy
            $\eglobally{\phi}$.  If the set of states $E$ is empty, then an
            argument similar to the one in Scenario~1 can be used.  That is, as
            per Lemma~\ref{lem:grow}, all states in trace $t_3$ (including
            $\vec{s}$) would satisfy $\growtwo$ in the first iteration of the
            loop in subroutine $\computeGlobalOverName$.  Therefore, all these
            states would be added to $Y$ in the first iteration of this loop.

            If the set of states $E$ is non-empty, then from
            Rule~\ref{order:interm-cycle} it follows that the control states
            that these states correspond to are all of order $k$ or less.
            Therefore, it follows from the inductive hypothesis that all states
            in $E$ would have been added to $Y$ within the first $k$ iterations
            of the loop in subroutine $\computeGlobalOverName$.  Therefore, at
            this point, the trace $t_3$ is such that all states that are
            outside $t_3$ and that are successors of states in $t_3$ are
            already in $Y$.  Therefore, it follows from Lemma~\ref{lem:grow}
            that all states in trace $t_3$ (including $\vec{s}$) would satisfy
            $\growtwo$ in iteration $(k+1)$ of the loop in subroutine
            $\computeGlobalOverName$.  Therefore, all these states would be
            added to $Y$ in this iteration.
    \end{description}

    \emph{Case 4:} Let $q$ be part of a cycle $c$.  We basically argued in
    Case 3, Scenario 3, that any state $\vec{s_m}$ that corresponds to a
    control-state that is in a cycle of order $k+1$ gets added to $Y$ within
    at most $k+1$ iterations of the loop in subroutine
    $\computeGlobalOverName$. The same argument is applicable in the current
    case, and implies that $\vec{s}$ gets added to $Y$ within at most $k+1$
    iterations.

\end{proof}

\begin{theorem} 
    Given any property $\eglobally{\phi}$, Algorithm~\ref{algo:compute-global}
    will terminate on a system $M$ if: 
\begin{enumerate} \item all $\mathit{pre}^*$
            queries on the system $M_1$ terminate, where $M_1$ is  a refinement of
        $M$ with respect to $\phi$ \item the action $f_b$ of each transition $b$ of
        $M$ is \emph{deterministic}, and \item there exists an integer bound $m$,
            and a (finite or infinite) set of flattenings $L$ of $M_1$, such that
            (a) $\order{N}{(q)}$ is at most $m$, for each control state $q$ of each
            flattening $N$ in $L$, and (b) for each state $\vec{s}$ that does not
            satisfy $\eglobally{\phi}$ in $M_1$, there exists a flattening $N$ in
            $L$ such that all traces from $\vec{s}$ in $M_1$ are preserved in $N$.

    \end{enumerate} 
\end{theorem}

\begin{proof}
    Let $\vec{s}$ be any state that does not satisfy $\eglobally{\phi}$ in
    $M_1$. Let $N$ be the flattening in $L$ such that all traces from $\vec{s}$
    in $M_1$ are preserved in $N$. That is, there exists a state $\vec{s'}$ of
    $N$ that is a ``trace preserving copy'' of $\vec{s}$ in $M_1$ (see
    Section~\ref{ssec:prel:traces-flattenings}). Clearly, $\vec{s'}$ does not
    satisfy $\eglobally{\phi}$ in system $N$, either. As mentioned at the
    beginning of Section~\ref{ssec:over-term}, we will analyze the termination
    of Algorithm~\ref{algo:compute-global} by analyzing only the termination of
    subroutine $\computeGlobalOverName$. Consider an application of
    subroutine~$\computeGlobalOverName$ on the system $N$. In what follows we
    will prove that $\vec{s}$ would get added to the set $Y$ (of states that do
    not satisfy $\eglobally{\phi}$) when the subroutine is applied on the
    system $M_1$ within the same number of iterations as is needed to add
    state $\vec{s'}$ to $Y$ when the subroutine is applied on $N$. Since the
    control states in the flattenings in $L$ have ``order'' at most $m$ (as per
    the assumption of this theorem), and since the subroutine terminates on any
    flat system $N$ within at most as many iterations as the ``order'' of the
    control state that has the highest order (as per
    Theorem~\ref{thm:flat_systems}), it would follow that the subroutine
    terminates on $M_1$ within at most $m$ iterations overall.

    We now prove the following property by mathematical induction.\\

    $\textbf{P(n)}:$ Let $\vec{s}$ be any state that does not satisfy
    $\eglobally{\phi}$ in $M_1$. Let $N$ be the flattening in $L$, and let
    state $\vec{s'}$ in $N$ be a trace preserving copy of $\vec{s}$.  If state
    $\vec{s'}$ gets added to $Y$ in the $n$th iteration when the subroutine
    $\computeGlobalOverName$ is applied on the flat system $N$, then $\vec{s}$
    would get added to $Y$ within $n$ iterations when the subroutine is applied
    on the system $M_1$. \\

    \textbf{Base Case(n=0):} This happens when $\vec{s'}$ is a stuck state in
    $N$. Since all traces from $\vec{s}$ in $M_1$ are preserved from $\vec{s'}$
    in $N$, it follows that $\vec{s}$ is a stuck state in $M_1$ also.
    Therefore, the subroutine would add $\vec{s}$ to $Y$ before the first
    iteration when it is applied on $M_1$. \\

    \textbf{Inductive case:} There are two scenarios here. The first scenario
    is that $\vec{s'}$ is added to $Y$ in the $n$th iteration by formula
    $\growone$ when the subroutine is applied on the flattening $N$. This can
    happen only if all successors of $\vec{s'}$ are added to $Y$ in fewer than
    $n$ iterations in the same application of the subroutine. Since $\vec{s'}$
    is a trace preserving copy of $\vec{s}$, it follows that each successor of
    $\vec{s}$ in $M_1$ has a trace-preserving copy in $N$ that is a successor
    of $\vec{s'}$.  Therefore, by the inductive hypothesis, all successor
    states of $\vec{s}$ would have been added to $Y$ in fewer than $n$
    iterations when the subroutine is applied on $M_1$ also. Therefore, in the
    $n$th iteration of this application, formula $\growone$ would add $\vec{s}$
    to $Y$ (if $\vec{s}$ is not already in $Y$).

    The second scenario is that  $\vec{s'}$ is added to $Y$ in the $n$th
    iteration by formula $\growtwo$ when the subroutine is applied on the
    flattening $N$. Therefore, from Lemma~\ref{lem:grow}, it follows that
    ignoring states that are reachable from $\vec{s'}$ that were added to $Y$
    in fewer than $n$ iterations, there is a single outgoing trace $t'$ from
    $\vec{s'}$ in $N$ which ends in a state $\vec{s'_k}$ that satisfies
    $\growone$ in the $n$th iteration. Furthermore, it follows from the
    definition of $\growone$ that all successors of the state $\vec{s'_k}$
    would have been added to $Y$ in fewer than $n$ iterations. Let \emph{inY}
    be the set of states of $N$ that are outside $t'$ but are immediate
    successors of states in $t'$.

    (1) From the observations above, it follows that all states in \emph{inY}
    would have been added to $Y$ in fewer than $n$ iterations when the
    subroutine is applied on $N$.

    (2) Let $t=f(t')$ be the trace in $M$ of which $t'$ is a copy. Because
    $\vec{s'}$ is a trace-preserving copy of $\vec{s}$, it follows that no
    state $\vec{s_1}$ in $t$ can have a successor state $\vec{s_2}$ outside $t$
    such that the copy $\vec{s'_1}$ of $\vec{s_1}$ in $t'$ is missing a
    concrete transition to a copy $\vec{s'_2}$ of $\vec{s_2}$.

    (3) In other words, every state of $M$ that is outside $t$ and that is a
    successor of some state in $t$ has a copy in $N$ that is in the set
    \emph{inY}. Therefore, due to Point~(1) above and due to the inductive
    hypothesis, it follows that all these successor states outside $t$ would be
    added to $Y$ in fewer than $n$ iterations when subroutine
    $\computeGlobalOverName$ is applied on $M_1$.

    (4) From the observation above, and from Lemma~\ref{lem:grow}, it follows
    that all states in $t$ (including $\vec{s}$) would get added to $Y$ within
    $n$ iterations of the loop in this subroutine when this subroutine is
    applied on $M_1$.
\end{proof}

\begin{theorem}
    Let $M$ be the given counter system, $\eglobally{\phi}$ be the given
    property to check, and  $M_1$ be the refinement of $M$ with respect to
    $\phi$. Algorithm~\ref{algo:compute-global} is guaranteed to terminate if:
    
    \begin{enumerate}
        \item
            $\mathit{pre}^k$ queries terminate on all
            flattenings of $M_1$ that the algorithm generates,
        \item There exists a (finite) set $L$ of flattenings of $M_1$ such
            that:

            \begin{enumerate}
                \item
                    For each state $\vec{s_1}$ that
                    satisfies $\eglobally{\phi}$ in $M_1$, there exists a
                    flattening $N$ in $L$ and a state $\vec{s'_1}$ in $N$ that
                    is a copy of $\vec{s_1}$ such that  there is at least one
                    infinitely long trace from $\vec{s'_1}$ in $N$. 

                \item
                    For each state $\vec{s_2}$ that does
                    not satisfy $\eglobally{\phi}$ in $M_1$, there exists a
                    flattening $N$ in $L$ and a state $\vec{s'_2}$ in $N$ that
                    is a copy of $\vec{s_2}$ such that $\vec{s'_2}$ is a
                    trace-preserving copy of $\vec{s_2}$.
            \end{enumerate}

        \item
            Calls to subroutine $\checkflatteningName$ in
            the algorithm always terminate.

    \end{enumerate}
\end{theorem}
\begin{proof}

    The algorithm will eventually generate all the flattenings in the set $L$.
    This is because the algorithm  enumerates all
    flattenings of $M_1$ in a systematic manner (i.e., in increasing order of
    the \emph{length} of the flattenings).

    The algorithm does not go into non-termination during the processing of any
    flattening that it enumerates.  This is because (a)
    the $\mathit{pre}^k$ queries (in line~\ref{line:growX} of the algorithm)
    have been assumed to be terminating, and (b) the
    $\checkflattening{M_1}{N}{\phi-X-Y}$ operation in line~\ref{line:checkfl}
    is assumed to be always terminating.

    When a  flattening $N$ of $M_1$ is generated, for each state $\vec{s'_1}$
    that satisfies $\eglobally{\phi}$ in $N$, the state $f(\vec{s'_1})$ of
    $M_1$ gets added to the set $X$  (see line~\ref{line:growX} in
    Algorithm~\ref{algo:compute-global}). Also,  each state $\vec{s_2}$ of
    $M_1$, such that there exists a state $\vec{s'_2}$ in $N$ that is a copy of
    $\vec{s_2}$ and that does not satisfy $\eglobally{\phi}$ in $N$, gets added
    to the set $Y$ (see line~\ref{line:growYb} in
    Algorithm~\ref{algo:compute-global}).  Therefore, by the time all
    flattenings in $L$ have been enumerated and processed,  by the assumption
    on $L$ as stated in the theorem's statement, it follows that $X$ contains
    all states that satisfy $\eglobally{\phi}$ in $M_1$, and $Y$ contains all
    other states. Note that the non-deterministic addition of states that do
    not satisfy $\eglobally{\phi}$ in $M_1$ to $X$ (as discussed in the
    beginning of this section) does not affect the observation made above.
    Therefore, after all flattenings in $L$ have been processed, in the
    following iteration of the loop in
    lines~\ref{line:inner-loop-beg}-\ref{line:inner-loop-end}, $X$ would not
    grow at all, while call in line~\ref{line:checkfl} will return $\tru$
    (because $\phi -X -Y$ would be the empty set). Therefore, the algorithm
    would terminate.
\end{proof}

\begin{theorem}
     Given a temporal property $\psi$ and a counter system $M$, say the
     procedure $\sat{M}{\psi}{\enumlabel}$ returns the enumerator
     $\approximation$ along with a formula $\phi$. Then, (a) $\approximation
     \sqsubseteq \enumlabel$, and (b) $\phi$ approximates the set of states of
     $M$ that actually satisfy $\psi$ in the direction given by the returned
     enumerator $\approximation$.
\end{theorem}

\begin{proof}
    The temporal property $\psi$ can be thought of as an expression-tree. We
    prove this theorem  by induction on the \emph{height} of this tree.

    \emph{Base case:} For $\psi$ to be of height 1, it must be a  basic
    proposition of the form $\phi_i$. In this case $\sat{M}{\psi}{\enumlabel}$
    returns $(\phi_i,\precise)$. Since $\phi_i$ represents precisely the set of
    states that satisfy $\phi_i$, the theorem statement holds.

    \emph{Inductive step:} Let $\psi$ be a formula of height $k+1$, $k > 0$.
    The induction hypothesis is that the theorem holds for all formulas  of
    height at most $k$. We prove the theorem case by case, based on the root
    operator of $\psi$.

    \begin{itemize}
        \item Let $\psi \equiv \neg\psi_1$ and
            $(\phi_1,\approximation)=\sat{M}{\psi_1}{\neg\enumlabel}$. It is
            easy to see that the set of states
            that satisfy $\psi$ is given by the formula $\neg\phi_1$. As per
            the induction hypothesis,
            $\approximation\sqsubseteq\neg \enumlabel$. It now follows from the
            definition of the  negation operator that
            $\neg\approximation \sqsubseteq \enumlabel$. Therefore the theorem
            holds.

        \item Let $\psi \equiv \psi_1 \vee \psi_2$. The argument is similar to
            the one for the previous case. By the inductive hypothesis,
            $\phi_1$ and $\phi_2$ are approximations of the precise solutions
            for $\psi_1$ and $\psi_2$, respectively, in the direction indicated
            by $\enumlabel$. Therefore, $\phi_1 \vee \phi_2$ is an
            approximation of the precise solution for $\psi$ in the direction
            indicated by $\enumlabel$. Also, by the inductive hypothesis,
            $\approximation_1 \sqsubseteq \enumlabel$ and $\approximation_2
            \sqsubseteq \enumlabel$. Therefore, it follows that
            $\approximation_1 \sqcup \approximation_2$, which is the returned
            enumerator, is dominated by $\enumlabel$ in the partial ordering.

        \item Let $\psi\equiv\enext{\psi_1}$. As per the inductive hypothesis,
            the formula $\phi_1$ that is returned by the recursive call to
            \emph{SAT} approximates the precise solution to $\psi_1$ in the
            direction indicated by $\enumlabel$. Therefore, following the
            definition of $\enext{\psi_1}$, it is easy to see that the formula
            returned, namely, $\pre{M}{\phi_1}$, is an approximation of the
            precise solution to $\enext{\psi_1}$ in the direction indicated by
            $\enumlabel$. Also, it follows from the inductive hypothesis that
            the enumerator returned, namely $\approximation$, is dominated by
            $\enumlabel$ in the partial ordering.

        \item\sloppypar Let $\psi \equiv \euntil{\psi_1}{\psi_2}$,
            $(\phi_1,\approximation_1)=\sat{M}{\psi_1}{\enumlabel}$ and
            $(\phi_2,\approximation_2)=\sat{M}{\psi_2}{\enumlabel}$.
            Then, by the induction hypothesis, $\phi_1$ approximates the
            precise solution to $\psi_1$ in the direction of $\enumlabel$, and
            $\phi_2$ approximates the precise solution to $\psi_2$ in the
            direction of $\enumlabel$. Therefore, from the  definition of
            $\euntil{\psi_1}{\psi_2}$, it follows that
            $\euntil{\phi_1}{\phi_2}$ approximates $\euntil{\psi_1}{\psi_2}$ in
            the direction of $\enumlabel$. Furthermore, we had proved in
            Section~\ref{sec:until} that
            $\computeUntil{M}{\phi_1}{\phi_2}{\enumlabel}$ returns a formula
            $\phi$ that approximates $\euntil{\phi_1}{\phi_2}$ in the direction
            indicated by $\enumlabel$. Therefore, the two observations above
            imply that the formula $\phi$ (which is returned by \emph{SAT})
            approximates $\euntil{\psi_1}{\psi_2}$ in the direction indicated
            by $\enumlabel$.

            From the inductive hypothesis it also follows that
            $\approximation_1 \sqsubseteq \enumlabel$ and $\approximation_2
            \sqsubseteq \enumlabel$. Also, as was proved in
            Section~\ref{sec:until}, the enumerator $\approximation_3$ returned
            by $\computeUntilName$ is guaranteed to be dominated by
            $\enumlabel$ in the partial ordering. Therefore, it follows that
            the enumerator returned by \emph{SAT}, namely,
            $(\approximation_1\join\approximation_2\join\approximation_3)$ is
            guaranteed to be dominated by $\enumlabel$ in the partial ordering.

        \item Let $\psi \equiv \eglobally{\psi_1}$, and
            $(\phi_1,\approximation_1)=\sat{M}{\psi_1}{\enumlabel}$.
            By the induction hypothesis, $\phi_1$ approximates the precise
            solution to $\psi_1$ in the direction of $\enumlabel$. Therefore,
            from the  definition of  $\eglobally{\psi_1}$, it follows that
            $\eglobally{\phi_1}$ approximates $\eglobally{\psi_1}$ in the
            direction of $\enumlabel$. Furthermore, we had proved in
            Section~\ref{sec:global} that
            $\computeGlobal{M}{\phi_1}{\enumlabel}$ returns a formula $\phi$
            that approximates $\eglobally{\phi_1}$ in the direction indicated
            by $\enumlabel$. Therefore, the two observations above imply that
            the formula $\phi$ (which is returned by \emph{SAT}) approximates
            $\eglobally{\psi_1}$ in the direction indicated by $\enumlabel$.

            From the inductive hypothesis it also follows that
            $\approximation_1 \sqsubseteq \enumlabel$. Also, as was proved in
            Section~\ref{sec:global}, the enumerator $\approximation_2$
            returned by $\computeGlobalName$ is guaranteed to be dominated by
            $\enumlabel$ in the partial ordering. Therefore, it follows that
            the enumerator returned by \emph{SAT}, namely,
            $(\approximation_1\join\approximation_2)$ is guaranteed to be
            dominated by $\enumlabel$ in the partial ordering.

    \end{itemize}
\end{proof}

\section{Systems and properties that were used for our experiments}
\label{app:sys-prop}

\begin{table}
  \caption{Descriptions of $\egloballyName$ properties}
  \label{tbl:sys-prop-tbl}
  \centering
 \begin{tabular}{|l|l|l|c|c|}
 \hline
& System & Property & Expected soln. & \\\hline

1 & firefly & $\eglobally{(\mathit{dirty}=0)}$ & Non-empty & Y \\\hline

2 & firefly & $\eglobally{(\mathit{shared}\geq 1)}$ & Non-empty & Y \\\hline

3 & firefly & $\eglobally{(\mathit{exclusive}\geq 0)}$ & Non-empty & Y
\\\hline

4 & firefly & $\eglobally{(\mathit{exclusive}= 1)}$ & $\emptyset$ &  \\\hline

5 & illinois &  $\eglobally{(\mathit{dirty}=0)}$ & $\emptyset$ & Y \\\hline

6 & illinois & $\eglobally{(\mathit{exclusive}\geq 1)}$ & $\emptyset$ & 
\\\hline

7 & mesi & $\eglobally{(\mathit{shared}\geq 1)}$ & $\emptyset$ & \\\hline

8 & mesi & $\eglobally{(\mathit{exclusive}= 0)}$ & $\emptyset$ & 
\\\hline

9 & mesi &  $\eglobally{(\mathit{invalid}=0)}$ & Non-empty & Y \\\hline

10 & mesi &  $\eglobally{(\mathit{modified}=0)}$ & $\emptyset$  & Y \\\hline

11 & moesi & $\eglobally{(\mathit{modified}=0)}$ & $\emptyset$  & Y \\\hline

12 & moesi & $\eglobally{(\mathit{exclusive}\geq 1)}$ & $\emptyset$   & Y
\\\hline

13 & moesi & $\eglobally{(\mathit{exclusive}= 0)}$ & $\emptyset$ & 
\\\hline

14 & moesi & $\eglobally{(\mathit{shared}\geq 1)}$ & $\emptyset$ & \\\hline

15 & synapse & $\eglobally{(\mathit{dirty}=0)}$ & $\emptyset$ & \\\hline

16 & synapse &  $\eglobally{(\mathit{valid}\geq 1)}$ & $\emptyset$ & \\\hline

17 & synapse &  $\eglobally{(\mathit{invalid}\geq 1)}$ & Non-empty &
\\\hline

18 & centra$\ldots$ & $\eglobally{(\mathit{BusyD} =
  0)}$ & \parbox{0.75in}{No overlap with  \mbox{\emph{post}(\emph{WaitD}
    $\geq 1$)}} & * \\\hline

19 & centra$\ldots$ & $\eglobally{(\mathit{BusyC} =
  0)}$ & \parbox{0.75in}{No overlap with  \mbox{\emph{post}(\emph{WaitC}
    $\geq 1$)}} & * \\\hline

20 & consis$\ldots$ &
  $\begin{array}{c}
  \egloballyName (\mathit{ServeS} \geq 1\, \wedge \\
  \mathit{GrantS} =0)
  \end{array}$

  & Non-empty & Y \\\hline

21 & consis$\ldots$ & 
  $\begin{array}{c}
  \egloballyName (\mathit{ServeE} \geq 1\, \wedge \\
  \mathit{GrantE} =0)
  \end{array}$
  
& Non-empty& Y \\\hline

22 & consis$\ldots$ &

$\begin{array}{c}
  \egloballyName (\mathit{ServeS} \geq 1\, \wedge  \\
  \mathit{GrantS} = 0 \wedge\, \\
  \mathit{GrantE} =0)
\end{array}
$

& $\emptyset$ & \\\hline

23 & consis$\ldots$ &
$\begin{array}{c}
  \egloballyName (\mathit{ServeE} \geq 1\, \wedge \\
  \mathit{GrantS} = 0\, \wedge \\
  \mathit{GrantE} =0)
\end{array}
$
& $\emptyset$ & \\\hline

24 & futurbus & $\eglobally{(\mathit{pendingR} \geq 1)}$ & $\emptyset$ & Y 
\\\hline 

25 & futurbus & $\eglobally{(\mathit{pendingW} \geq 1)}$ & $\emptyset$ &
\\\hline 

26 & lastin$\ldots$ & $\eglobally{(I \geq 1)}$ & $\emptyset$ & Y \\\hline

27 & lastin$\ldots$ & $\eglobally{(\mathit{Ea} \geq 1)}$ &
  $\emptyset$ & Y \\\hline

28 & lastin$\ldots$ & $\eglobally{(\mathit{Eb} \geq 1)}$ &
  $\emptyset$ & Y \\\hline

29 & lift & $\eglobally{(c \neq g)}$ &  $\emptyset$ & \\\hline

30 & lift & $\eglobally{(a = 2)}$ &  $\emptyset$ & \\\hline

31 & lift & $\eglobally{(a = 1)}$ &  $\emptyset$ & \\\hline

32 & lift & $\eglobally{(a = 0)}$ &  $\emptyset$ & \\\hline

33 & ticket2i & $\eglobally{(\mathit{p1} = 1)}$ & $\emptyset$ & \\\hline

34 & ticket2i & $\eglobally{(\mathit{p2} = 1)}$ & $\emptyset$ & \\\hline

35 & ticket2i & $\eglobally{(\mathit{p1} = 0)}$ &
$\begin{array}{c}(\mathit{p1} = 0) 
\end{array}
$ & \\\hline

36 & ticket2i & $\eglobally{(\mathit{p2} = 0)}$ &
$
\begin{array}{c}
(\mathit{p2} = 0)
\end{array}
$ & \\\hline

37 & train & $\eglobally{(\mathit{state} = \mathit{stopped})}$ &
$\emptyset$ & \\\hline

38 & train & $\eglobally{(\mathit{state} = \mathit{late})}$ &
$\emptyset$ & \\\hline

39 & train & $\eglobally{(b > s+9)}$ & $\emptyset$ & \\\hline

40 & train & $\eglobally{(b < s-9)}$ & $\emptyset$ & \\\hline

41 & train & $\eglobally{(\mathit{state} = \mathit{ontime})}$ &

$
\begin{array}{{c}}
  b \geq s - 9\, \wedge \\
  b \leq s + 9\, \wedge \\
  \mathit{state} = 
  \mathit{ontime}
\end{array}
$
&
\\\hline

42 & train & $\eglobally{(\mathit{state} = \mathit{onbrake})}$ &
$\emptyset$ & \\\hline

43 & ttp2 & $\eglobally{(df = cf)}$ & $\emptyset$ & \\\hline

44 & swim$\ldots$ & $\eglobally{(\mathit{x7} < \mathit{p1})}$ & & \\\hline

 \end{tabular}

\end{table}

In this appendix we introduce the 14 systems and the 44 ``EG''
properties that we use for our experiments (see
Section~\ref{ssec:implementation}). 

Table~\ref{tbl:sys-prop-tbl} summarizes the information that is discussed
in this appendix.  The column titled ``Property'' provides the actual
properties that we have used, while the column titled ``Expected soln.''
defines our expected solutions to the properties. The final column in
Table~\ref{tbl:sys-prop-tbl} is a copy of the final column in
Table~\ref{tbl:results}. That is, a ``Y'' in a row indicates that for the
corresponding property, the actual solution differs from the expected
solution.

We divide the discussion in this appendix into three sections. The first
section below describes the cache coherence systems and related properties
that we have selected for our evaluation. These are represented in
Rows~1-17 in Tables~\ref{tbl:results} and~\ref{tbl:sys-prop-tbl}. The
second section below covers the remaining systems and properties. In
Section~\ref{ssec:usefulness-of-results} we had briefly discussed how for
certain properties of certain systems the actual solutions differ from the
expected solutions. In the third section below expand upon this discussion,
and point out specific issues in the benchmark systems that came to our
notice during the process of comparing the actual solutions with the
expected solutions.

\subsection{Cache coherence systems and properties}

The systems firefly, illinois, mesi, moesi, and synapse (see Rows~1-17 in
Tables~\ref{tbl:results} and~\ref{tbl:sys-prop-tbl}) are models of
similarly named cache coherence protocols. (``synapse'' is also known as
the ``MSI'' protocol.) All these systems share the following common design
principles: Each system models the state-evolution of a \emph{single} cache
over an execution trace. The number of processors is unbounded. The cache
line is never evicted from any cache (because there is only one line being
modeled). However, a cache line in a processor can become ``invalid'' due
to a write into the line by another processor. There is a set of counters
in the system, with each counter corresponding to a distinct ``state'' that
the cache line can be in as per the underlying protocol being modeled. The
value of each counter at each point of time indicates the number of
processors whose cache lines are in the corresponding state at that point
of time in the execution. Each transition in the system corresponds to a
read-hit, read-miss, write-hit, or a write-miss, and changes the values of
the counters accordingly.

There is a large degree of overlap among the different protocols in terms
of the states they employ. Moreover, any state that is used in multiple
protocols has the same meaning in all the protocols.  Therefore, we provide
a combined discussion below of all the states in all these
protocols:\\

\emph{(M)odified} or \emph{Dirty}:
A processor in this state has
written its copy of the line, and this updated value has not yet been sent to
memory or to the other processors.

\emph{(S)hared}: A processor in this state has a valid copy of the
cache that is coherent with the value in main memory as well as with any
copies in the caches of other processors. 

\emph{(I)nvalid}: A processor in this state does not have an updated
copy of this line. Any read by the processor will result in a read
``miss''. 

\emph{(E)xclusive}: A processor in this state possesses a valid copy
of this line that is coherent with memory, and no other processor possesses
a valid copy of this line. All the protocols that we consider except
synapse employ this state. 

\emph{(O)wned}: A processor in this state possesses a valid copy of
this line. The line may not be coherent with memory, but is coherent with
other copies of this line in other processors.\\[1em]

All the protocols that we experiment with employ the
states \emph{modified/dirty}, \emph{shared}, and \emph{invalid}. All
protocols other than synapse make use of the \emph{exclusive} state, while
the MOESI protocol makes use of the \emph{owned} state.  

As a more detailed illustration a cache coherence system, we refer the
 reader to the discussion of the synapse system in
 Section~\ref{ssec:benchmarks-selection}.

We now discuss the ``EG'' properties that we have devised for the cache
coherence systems.  Many of these properties are based on the following
observation about these systems.  Since only a single cache line is being
modeled, and since the sole cache line being modeled can never be evicted
from any cache, we expect each system to keep performing writes (from some
processor) indefinitely often along every execution trace. Viewed in
another way, if writes stop happening across all processors, then the
system has essentially stopped doing anything.

Our first property is $\eglobally{(\mathit{modified}=0)}$ or
$\eglobally{(\mathit{dirty}=0)}$. We have tried this property with all five
systems.  Due to the expectation of writes occurring indefinitely often in
all traces, in four of the systems (i.e., all except ``firefly''), we
except no state to satisfy the property mentioned above. ``firefly'' is an
exception. As per available descriptions of this protocol, a write by a
processor in the \emph{shared} state may preserve the states of all
processors, because the written value is propagated to all processors
immediately. Therefore, we expect certain states to satisfy this property.

For the same reason as above, we expect no state to satisfy the property
$\eglobally{(\mathit{shared} \geq 1)}$, except with system firefly. And for
the same reason, in all systems we expect no states to satisfy the
properties $\eglobally{(\mathit{exclusive} = 1)}$,
$\eglobally{(\mathit{exclusive} \geq 1)}$, and
$\eglobally{(\mathit{valid} \geq 1)}$.

We tried the property $\eglobally{(\mathit{exclusive} = 0)}$ with the
systems mesi and moesi. In both these systems, a processor is supposed to
transition to \emph{exclusive} state first before doing a write
operation. Because writes are to happen indefinitely often in all traces,
we expect no states to satisfy this property.

The property $\eglobally{(\mathit{invalid} = 0)}$ for mesi is expected to
be satisfied whenever there is a single processor only in the system; this
is because, in this case, the sole processor comes out of the invalid state
after its first read or write operation, and then never becomes invalid
again. Finally, the property $\eglobally{(\mathit{invalid} \geq 1)}$ is
satisfied by synapse by traces where there are at least two processors, and
one of them is writing in every step (so the other processor becomes invalid). 

\subsection{Other systems and properties}

The non cache coherence systems (Rows~18 onward in Tables~\ref{tbl:results}
and~\ref{tbl:sys-prop-tbl}) are more varied in nature.  Other than
``lift'', ``train'', ``ttp2'', and ``swimmingpool'', the remaining systems
model protocols of various types that mediate between ``jobs'' of some kind
or the other that request ``resources'' and are granted these requests from
time to time. Our $\eglobally{\phi}$ properties generally try to identify
situations where a request is never granted (i.e., states from which
starvation can occur).

For instance, for ``centralserver'' (Rows~18 an~19), the two properties
indicate that resource $C$ (resp. $D$) is continually not allotted to any
job. With regard to system ``consistency'' (Rows~20-23), $S$ and $E$ are
two jobs. \emph{ServeS} indicates whether job $S$ is making a request to be
served, while \emph{GrantS} indicates whether this job is granted the
resource. Job $E$ also has two similar counters. In this system, as per our
understanding of it, any specific job request can face starvation (see the
expected solutions in Rows~20 and~21); however, when a request is pending
from a job, the resource cannot go un-allotted to either job (Rows~22
and~23).

With system ``futurbus'' (Rows~24-25), \emph{pendingR}
and \emph{pendingW} indicate the number of processors waiting to read from
the bus and write to the bus, respectively. With system
``lastinfirstserved'' (Rows~26-28), $I$ indicates the number of
requests in the stack of pending requests; \emph{Ea} (resp. \emph{Eb})
indicates the number of jobs currently having exclusive
access to resource ``a''  (resp. ``b''). With ticket2i, \emph{p1}
and \emph{p2} are two jobs, with `1' standing for ``currently requesting
resource'' and `0' standing for ``currently not requesting resource''. 

The system ``lift'' (Rows~29-32 in the table) models an elevator control
system. The three different values of the counter $a$ indicate whether the
elevator is being commanded to go up, to go down, or stay at its current
level. The counter $c$ represents the current floor where the elevator is
positioned, while the counter $g$ indicates the target floor to move
to. The four properties that we have devised characterize traces along
which the elevator never reaches its target, or keeps moving in a fixed
direction. No such traces are expected to exist.

The system ``train'' (Rows~37-42 in the table) models a train control
system. \emph{stopped}, \emph{late}, \emph{ontime}, and \emph{onbrake} are
control-states of the system; these indicate intuitively the current status
of the running train. The counter $s$ indicates the position where the
train \emph{should} have been at at the current time, while the counter $b$
indicates the position where the train currently is. If $b$ is less than
$s-9$ the train is considered to be late, while if it is greater than $s+9$
it is considered to be too much ahead of schedule.

The system ``ttp2'' models the ``Time Triggered Protocol''. This protocol
is somewhat complex to describe, and is described in greater detail
elsewhere~\cite{bardin04}. We could not understand the system
``swimmingpool'' properly. We chose an arbitrary property for it, and
report the result here for that property only to point it out as an outlier
to which our approach does not scale.

\subsection{Reasons for deviations}

Our objective in this section is to discuss specific issues in the
benchmark systems that came to our notice during the process of comparing
the actual solutions with the expected solutions. In this section we focus
only on properties on which the actual solution differs from the expected
solution, because it is these properties that provide unexpected insights
into the given benchmark systems. 

With system ``firefly'', that is, Rows~1-3 in Table~\ref{tbl:sys-prop-tbl},
the actual solution turned out to be empty for every property. Upon closer
examination of the system, we observed that all traces of the system were
finite traces. Therefore, no state can satisfy any $\eglobally{\phi}$
property. This is somewhat surprising for a reactive system. We feel that
possibly certain transitions could be added to the system that will cause
traces to become of infinite length, while still staying faithful to the
protocol being modeled.

With illinois, Row~5, we found that the property is violated due to a
transition that possibly could be removed from the system. This
transition, named r8, allows a processor in \emph{shared} state to
become \emph{invalid} even though no other processor is writing as part of
the same transition.

The system ``mesi'' appears to have two issues. Let us consider Row~9
first. Whenever there are more than one processors, this property
$\eglobally{(\mathit{invalid}=0)}$ can be
expected to be not satisfied by any state (because writes have to occur
indefinitely often in all traces, and because a write by any processor will
put all other processors into the \emph{invalid} state). However, when
there is a single processor (which the model does admit), then from any
state where the processor is in \emph{modified} state, one should expect to
see a trace along which this processor writes in every step and hence
remains in the \emph{modified} state continuously. In other words, such a
trace should satisfy the property. However, no state actually satisfies
this property. Upon closer investigation, we found that there is no
transition in the system that allows a single-processor system to issue an
infinitely long sequence of writes. We hypothesize that this issue can be
resolved by adding a suitable transition to the system.

Regarding Row~10, this property is violated because there is a transition
in the system (namely, r4) which allows one processor to remain
in \emph{exclusive} state and other processors to remain in \emph{invalid}
state continuously over a trace. That is, no actions are taking place at
all over the trace. We are not sure if this is an issue, or whether it is
acceptable. The unexpected results in Rows~11 and~12 (moesi) are due to a
similar reason (the same transition r4 is present in moesi, too).

With the system ``consistency'', based on our initial study of the system,
we expected to see traces in which a request for a resource was made but
never granted. See the $\eglobally{\phi}$ properties in Rows~20 and~21 in
Table~\ref{tbl:sys-prop-tbl}. But to our surprise, such traces did not
exist as per the computed solutions to the properties. Upon closer
observation, it turned out that the set of initial states specified for
this system was narrower than expected. It is specified as `idle = 1',
whereas we expected it to be `idle $\geq$ 1' (because every job is expected
to be idle in the beginning).

In the system ``futurbus'', from our initial study of the system, we
expected that there would be no traces where pending reads or pending
writes remained in the system continually. See Rows~24 and~25. The former
expectation (i.e., no continuous pending reads) \emph{did not} hold up,
whereas the latter expectation was satisfied. Upon closer inspection, we
identified the cause for the mismatch; there was a transition, namely r8,
which could cause the system to continually stay in the same concrete state
(where there were pending reads) over an infinitely long period. We felt
that this transition needs some form of refinement.

Finally, in the system ``lastinfirstserved'', we observed starvation
scenarios that we did not initially expect. That is, we observed traces
along which the stack of pending requests never becomes empty, and traces
along with a job can hold exclusive access to a resource continually. With
this system, we were not sure whether this starvation is actually a feature
or not of the underlying protocol or not, because we could not find any
separate documentation of the protocol being modeled.

\section{Trace equivalence check}
In this section we give an intuition behind the trace equivalence check used by
the routine $\computeGlobalName$ (Line~\ref{line:checkfl}) and its Variant-X.
This was given by Demri et al~\cite{dem2006}. The trace equivalence check
routine, namely $\checkflattening{M}{N}{\phi}$, takes a counter system $M$, a
flattening $N$ of $M$ and a set of states $\phi$ as the input. It returns true,
if all the traces from $\phi$ in $M$ are present in the flattening $N$. Else it
returns false along with a set of states $\phi^\prime$, such that, $N$ is a trace
flattening of $M$ with respect to $\phi^\prime$. 

The core idea behind the trace flattening check given by Demri et al. is as
follows: Consider a counter system $M$ and a flattening $N$ of $M$. By
definition of flattening, $\traces{N}{\phi}\subseteq\traces{M}{\phi}$. Therefore, if the
traces from $\phi$ are equivalent in both $M$ and $N$, then it is enough if we
show that $\traces{M}{\phi}\subseteq\traces{N}{\phi}$. Note that $N$ is a
flattening of $M$. Therefore each control state in $N$ is a copy of some
control state in $M$. Similarly every transition in $N$ is a copy of some
transition in $M$. Consider any control state
$q_N$ in $N$. A transition $t_1$ is said to be \emph{unabled} in the control state $q_N$
in $N$ if a) there exists a control state $q_M$ in $M$ such that $q_N$ is a
copy of $q_M$ b) $t_1$ is present from $q_M$ but a copy of $t_1$ is not present
from $q_N$. Transition $t_1$ is said to be present in $q_M$ if $q_M$ is the
source control state of $t_1$.
For every unabled transition $t_1$ of a control state $q_N$ in $N$, if
$\transition{\vec{s}}{t_1}{M}{\vec{s^\prime}}$ then there must exist a
transition $t_2$ in $N$ from $q_N$ which is not unabled, such that
$\transition{\vec{s}}{t_2}{N}{\vec{s^\prime}}$. Let $\phi_2$ be the set of
states that do not satisfy the above criterion. The Presburger formula for
$\phi_2$ is complex and is available in ~\cite{dem2006}. These set of states are
guaranteed to have a missing transition and hence a missing trace. The set of
states that have traces in $M$ but not in $N$ is given by
$\preStar{\phi_2}{N}$. Demri et al, give a Presburger formula $\phiteqchk
\equiv \preStar{\phi_2}{N} \wedge \phi$
which represents a set of states that have at least one trace in $M$ but not in
$N$ from a state that satisfy $\phi$. If $\phiteqchk$ is $\fls$, then
$\checkflattening{M}{N}{\phi}$ returns $\tru$ else it returns $\fls$ along with
$\phi-\phiteqchk$ which represents the set of states from which there is a
trace in $M$ but not in $M$.

%
%


\end{document}